\documentclass[apj, twocolumn, numberedappendix]{openjournal}

\usepackage{lipsum}

\usepackage{xcolor}
\usepackage{textgreek}
\usepackage[utf8]{inputenc}
\usepackage[english]{babel}

\usepackage{hyperref}
\hypersetup{
    unicode, 
    colorlinks=true,
    linkcolor=linkcolor,
    citecolor=linkcolor,
    filecolor=linkcolor,
    urlcolor=linkcolor,
}
\usepackage{color,colortbl}
\definecolor{linkcolor}{rgb}{0.0,0.3,0.5}
\usepackage{tensind}
\tensordelimiter{?}
\DeclareGraphicsExtensions{.bmp,.png,.jpg,.pdf}
\usepackage{verbatim}
\usepackage[normalem]{ulem}
\usepackage{orcidlink}
\usepackage{soul}

\usepackage{amsmath}
\usepackage{amssymb}

\DeclareRobustCommand{\ion}[2]{%
\relax\ifmmode
\ifx\testbx\f@series
{\mathbf{#1\,\mathsc{#2}}}\else
{\mathrm{#1\,\mathsc{#2}}}\fi
\else\textup{#1\,{\mdseries\textsc{#2}}}%
\fi}

\usepackage{graphicx}	
\usepackage{subfigure}
\usepackage[figuresright]{rotating}
\usepackage{longtable}

\graphicspath{ {./figures/} }

\shortauthors{Y.~A. Gordon et al.}
\shorttitle{Radio Variables at $3\,$GHz}

\begin{document}
\title{A Census of Variable Radio Sources at $3\,$GHz
\vspace{-4em}
}

\author{Yjan~A. Gordon$^{1,\,4}$\orcidlink{0000-0003-1432-253X}}
\author{Peter~S. Ferguson$^{2}$\orcidlink{0000-0001-6957-1627}}
\author{Michael~N. Martinez$^{1}$\orcidlink{0000-0002-8397-8412}}
\author{Eric~J. Hooper$^{3}$\orcidlink{0000-0003-0713-3300}}

\thanks{Corresponding Author: Yjan~A. Gordon \\ \href{mailto:yjan.a.gordon@gmail.com}{yjan.a.gordon@gmail.com}}

\affiliation{$^1$Department of Physics, University of Wisconsin-Madison, 
1150 University Ave, Madison, WI 53706, USA\\
$^2$Department of Astronomy, University of Washington, 3910 15th Avenue NE, Seattle, WA 98195, USA\\
$^3$Department of Astronomy, University of Wisconsin-Madison, 
475 N. Charter Street, Madison, WI 53703, USA\\
$^4$Department of Physics, University of Wisconsin-Whitewater, 800 W. Main St., Whitewater, WI 53190, USA}

\begin{abstract}
    A wide range of phenomena, from explosive transients to active galactic nuclei, exhibit variability at radio wavelengths on timescales of a few years.
    Characterizing the rate and scale of variability in the radio sky can provide keen insights into dynamic processes in the Universe, such as accretion mechanics, jet propagation, and stellar evolution.
    We use data from the first two epochs of the Very Large Array Sky Survey (VLASS) to conduct a census of the variable radio sky.
    Approximately $3,600$ compact sources are found to significantly vary in brightness during the $\sim2.5\,$ years between observations.
    In this work we focus on sources that are detected in both VLASS epochs, but estimate there may be $>10,000$ additional variable radio sources in VLASS that are only detected in either the first or second epoch.
    For objects detected in both epochs whose mean flux density across the two epochs, $\mu_{S}$, is brighter than $20\,$mJy, $5\,$\% show brightness variations $>30\,$\%, rising to $9\,$\% at $\mu_{S}>300\,$mJy.
    We analyze the redshift distributions, infrared colors, and $\gamma$-ray properties of the variable radio sources, finding that most have multiwavelength characteristics that are consistent with blazars and quasars.
    Blazars in particular are found to be overrepresented among the variable radio sources, and the largest absolute changes in flux density are produced by blazars.
    The largest fractional changes in brightness are exhibited by galactic sources.
    We discuss our results, including some of the more interesting and extreme examples of variable radio sources identified, as well as future research directions.\\
\end{abstract}

\begin{keywords}
{\href{http://astrothesaurus.org/uat/1338}{Radio Astronomy (1338)},
\href{http://astrothesaurus.org/uat/2109}{Time Domain Astronomy (2109)},
\href{http://astrothesaurus.org/uat/508}{Extragalactic radio sources (508)},
\href{http://astrothesaurus.org/uat/16}{Active galactic nuclei (16)},
\href{http://astrothesaurus.org/uat/164}{Blazars (164)}}
\end{keywords}

\maketitle


\section{Introduction} \label{sec:intro}

Radio variability is exhibited by a wide variety of astronomical phenomena on timescales ranging from less than a second to decades or longer \citep{Pietka2015}.
Active galactic nuclei (AGN) dominate the bright radio sky \citep{Novak2018, Condon2019}.
The switch-on of an AGN due to increased accretion onto the supermassive black hole can present as the appearance and brightening of a new radio source on timescales of years to decades \citep{KunertBajraszewska2020, Nyland2020}.
Furthermore, AGN can have shorter term  variations (lasting months to years) in their radio brightness that are typically attributed to stochastic processes such as shocks within the jets or magnetic reconnection events within the central engine \citep{Fiedler1987, Hovatta2007, Hovatta2008, Kadowski2015, Fernandez2022}.
For explosive transients like supernovae (SNe), gamma-ray bursts (GRBs), kilonovae, and tidal disruption events (TDEs), the radio emission from the initial event can be detected within days, and radio afterglows may be observed years or even decades later \citep[e.g.,][]{Weiler1986, Ball1995, Hancock2013, Kathirgamaraju2019, Anderson2020,  Cendes2024}.
Certain types of stars, such as M-dwarfs or RS CVn-type objects that have strong and fluctuating magnetospheres, are known to show strong variability in their radio emission on timescales as short as a few hours \citep[e.g.,][]{Villadsen2019, Andersson2022, Yiu2024, Driessen2024}.
The accretion of matter onto the neutron star or black hole in X-ray binaries can produce microquasars that, like their more massive namesakes, are often variable radio sources \citep[e.g.,][]{Hjellming2000, Gallo2003, Fender2004, Sharma2021}.
The radio variability timescales in microquasars can be much shorter than those of quasars due to the much smaller size of the accretion disk around stellar mass black holes relative to the supermassive black holes that power AGN, with variations possible over periods much shorter than a second \citep{Wang2024}.
Rapidly rotating pulsars and fast radio bursts\textemdash a phenomenon for which the driving mechanism remains unknown, but may result from magnetic reconnection events in neutron stars or magnetars\textemdash can show radio variability on timescales of just a few milliseconds \citep[e.g.,][]{Bhattacharya1991, Petroff2022}.
Furthermore, in addition to intrinsic source variability, interstellar scintillation can result in significant fluctuations in observed radio brightness, particularly at low frequencies \citep[e.g.,][]{Hancock2019, Ross2021, Ross2022}.

Dedicated studies of the dynamic radio sky have historically often been limited to the monitoring of known variable radio sources such as blazars \citep{Richards2011}, wide-area variability surveys sensitive to only the brightest radio sources \citep{Bell2019}, or variability surveys covering a very small on-sky footprint \citep{Hancock2016}.
Such survey designs inherently result in limited sampling of the radio-variable source population, an effect that can be countered by using independent observations from quasi-single-epoch deep and wide area radio surveys.
For instance, the Faint Images of the Radio Sky at Twenty cm survey \citep[FIRST,][]{Becker1995} was used by \citet{deVries2004} and \citet{Thyagarajan2011} to search for radio variability, with the latter finding $>1,600$ sources at $S_{1.4\,\text{GHz}}\gtrsim1\,$mJy that vary in brightness by over $20\,\%$.
By combining FIRST with the NRAO VLA Sky Survey \citep[NVSS,][]{Condon1998}, \citet{Ofek2011} found that $0.1\,\%$ of sources vary significantly on timescales of months to years.

More recently, deep and wide area time-domain radio surveys have started to become more common.
Such projects often have very different survey properties (e.g., depth, observing frequency, cadence etc.), and often have separate approaches to defining what constitutes variability.
Nonetheless, these early studies provide a key benchmark in our understanding of the variable radio sky.
A pilot for the Variables and Slow Transients survey \citep[VAST,][]{Murphy2013} found that only $0.02\,$\% of sources vary significantly at $888\,$MHz \citep{Murphy2021}.
The full VAST survey is currently underway, and will be complemented by the Rapid ASKAP Continuum Survey \citep[RACS,][]{McConnell2020}, which will add additional observing epochs and multi-frequency observations for use with VAST \citep{Duchesne2024}.
\citet{Hajela2019} observed Stripe 82 over multiple epochs at $150\,$MHz, reporting significant variability in $<0.1\,$\% of sources.
All of these are consistent with radio variability being extremely rare at low frequencies \citep{Mooley2013}.
At observing frequencies above a few GHz, radio variability is still rare but more common than at lower frequencies.
\citet{Mooley2016} observed $50\,\text{deg}^{2}$ down to an rms noise level $40\,\mu\text{Jy}\,\text{beam}^{-1}$ at $3\,$GHz over five epochs, identifying $>150$ transients and slow variables, accounting for $\sim4\,$\% of sources.

In this work, we use data from the Karl G. Jansky Very Large Array Sky Survey \citep[VLASS,][]{Lacy2020} to undertake a near-all-sky census of variable radio sources at $\nu\sim3\,$GHz.
The rest of this paper is structured as follows.
In Section \ref{sec:data} we describe the VLASS data used in this work.
Our method for finding variables is detailed in Section \ref{sec:finding-variables}.
The radio properties and statistics of our sample of variables is reported in Section \ref{sec:variability-stats}, and we investigate their multiwavelength attributes in Section \ref{sec:multiwavelength}.
We discuss our results and highlight some interesting examples of objects we identify in Section \ref{sec:discussion}, and present our summary and conclusions in Section \ref{sec:summary}.
The supplementary data products produced in the course of this work, including the catalog of variable radio sources and their multiwavelength identifications where available, are described in Appendix \ref{ap:supdata}.

\section{The Very Large Array Sky Survey} 
\label{sec:data}

Beginning in 2017, VLASS is a multi-epoch, $3\,$GHz survey of the entire sky north of $-40^{\circ}$ in declination.
Each full VLASS epoch consists of observing the entire $\sim34,000\,\text{deg}^{2}$ footprint down to an rms noise level of $\sim 130\,\mu\text{Jy}\,\text{beam}^{-1}$, and requires about $1,840$ hours of observing time \citep{Lacy2019, Lacy2020}.
Observations for VLASS use the S-band ($2\,\text{GHz}< \nu < 4\,\text{GHz}$) with the VLA in B-configuration at $\delta > -8^{\circ}$ and a combination of B- and BnA-configuration\footnote{`BnA' is a hybrid configuration where the north arm antennas of the VLA are in A-configuration while east and west arm antennas are in B-configuration \citep{Lacy2020}.} at more southern declinations, resulting in a synthesized beam that is typically $\lesssim3\arcsec$. 
The high observing frequency is well suited to detecting radio variability that is often absent at lower frequencies, \citep{Mooley2016}, while the high angular resolution enables confident multiwavelength source associations \citep{Gordon2023dragns}.

Three full VLASS epochs have been observed to date, with a fourth epoch covering half the full survey footprint ($\sim17,000\,\text{deg}^{2}$) completed in early 2026 \citep{Nyland2023}.
However, images and catalog data were only readily available for the first two epochs when we were starting this work, and, as such, the analysis and results presented in this paper are based solely on data from the first and second epochs of VLASS. 
The first epoch of VLASS (hereafter referred to as Epoch 1) ran from September 2017 to July 2019, with the second epoch (Epoch 2) running from June 2020 to February 2022.
The median time between the Epoch 1 and Epoch 2 observations of a VLASS field is $974\,$days ($32\,$months), but ranges between $832\,$days ($27\,$months) and $1,120\,$days ($37\,$months), with the $16$th and $84$th percentiles being $923\,$days ($30\,$months) and $1,039\,$days ($34\,$months) respectively. 
The survey design of VLASS is thus well suited to conduct a statistical characterization of the variability of the radio source population on timescales of a few years.

\subsection{VLASS Quick Look Images}
\label{ssec:quicklook}

\textit{Quick Look} images are produced from the raw VLASS visibilities within a few weeks of the observations.
This turnaround time is necessitated by one of the survey's core science goals of identifying new radio transients for follow-up observations \citep{Lacy2020}.
In order to produce images for large areas of the sky rapidly, the VLASS \textit{Quick Look} imaging pipeline does not include self-calibration, undersamples the point-spread function when gridding the images, and does not correct for $w$-terms. 
The result is that the \textit{Quick Look} images have a number of known quality issues to be aware of when using them, including flux density scaling errors of $\lesssim 15\,$\%, and image artifacts (particularly from sidelobes) that can cause spurious source detections \citep{Lacy2019}.

Higher quality imaging that addresses some of the limitations of the \textit{Quick Look} images is underway \citep{Lacy2022}; however, at the time of writing, only a small fraction of the VLASS observations have been processed in this manner.
On the other hand, two full epochs (each covering $80\,$\% of the sky) of \textit{Quick Look} images are already publicly available.
While one needs to take care when using these \textit{Quick Look} images, they have already proven their scientific utility, being used in a wide range of research including studies of radio galaxy evolution, stellar science, identification of gravitational lenses, and analysis of the large scale structure of the Universe \citep[e.g.,][]{Kukreti2023, Gordon2023css, Pelisoli2024, Martinez2025, Darling2022}

\subsection{Catalog Data}

VLASS source catalogs based on the \textit{Quick Look} images and covering the entire survey footprint are available for both Epoch 1 and Epoch 2 via the Canadian Initiative for Radio Astronomy Data Analysis\footnote{\url{https://cirada.ca/}} (\citealp{Gordon2020, Gordon2021}; B. Sebastian et al. in prep).
These catalogs are produced using the source finder \texttt{PyBDSF} \citep{Mohan2015} that fits single- or multi-Gaussian source models to regions of brightness where the peak pixel value is $>5\times$ the local rms noise.
Additional quality assurance metrics are included in these catalogs to flag sources that are more likely to be false-positive detections arising from artifacts in the \textit{Quick Look} images \citep[for a full description of the catalogs, see Sections 2 and 3 of][]{Gordon2021}.
We use version 3 of the Epoch 1 catalog, and version 1 of the Epoch 2 catalog, as both of these catalogs correct for astrometric limitations described in Section 3.3 of \citet{Gordon2021}, resulting in the source positions having an astrometric precision of $\sim0\arcsec.2$ \citep{Lacy2019, Bruzewski2021}.
Following the recommendations laid out in Section 3 of \citet{Gordon2021}, we identify unique sources in the catalogs from both epochs with quality flags consistent with real source detections.
Explicitly, we select sources with:
\begin{itemize}
    \item \texttt{Quality\_flag $==(0|4)$}, meaning the source is likely real and not a spurious detection of an image artifact; 
    \item \texttt{Duplicate\_flag $<2$}, eliminating any duplicate detections within the $\sim30\arcsec$ overlap between adjacent images;
    \item and \texttt{Source\_Code $\neq$`E'} indicating that the source is a distinct detection at $\text{S/N} > 5$ and is well modeled by one or more Gaussians.
\end{itemize}
These criteria identify 1,874,645 sources in Epoch 1 and 1,873,524 in Epoch 2, a difference of 1,121 sources.

\subsection{Obtaining Two-Epoch Source Measurements} 

With most radio sources being stationary, they should have the same coordinates when detected in both epochs\footnote{Radio stars may potentially move between VLASS epochs. However, these are expected to be rare, and thus are not considered here.}.
In practice, the astrometric precision of the cataloged coordinates of a source depends on a number of factors including the point spread function (the VLASS beam size), the pixel size of the image, and the signal to noise of the source.
In order to characterize the relative astrometry between Epoch 1 and Epoch 2, and thus determine a reliable positional tolerance to adopt when matching sources between the two epochs, we analyze the distribution of angular separations between likely point sources (those having \texttt{DC\_Maj}==0) in Epoch 1 and the closest likely point source in Epoch 2.
We limit this analysis to point sources for two reasons: i) the positional uncertainty isn't impacted by the spatial extent of the source, and ii) variable sources are expected to be compact, and so the decisions of how to match sources between the two VLASS epochs based on this analysis shouldn't have a significant impact on our ability to identify variable sources.

\begin{figure}
    \centering
    \subfigure[]{\includegraphics[width=0.99\columnwidth]{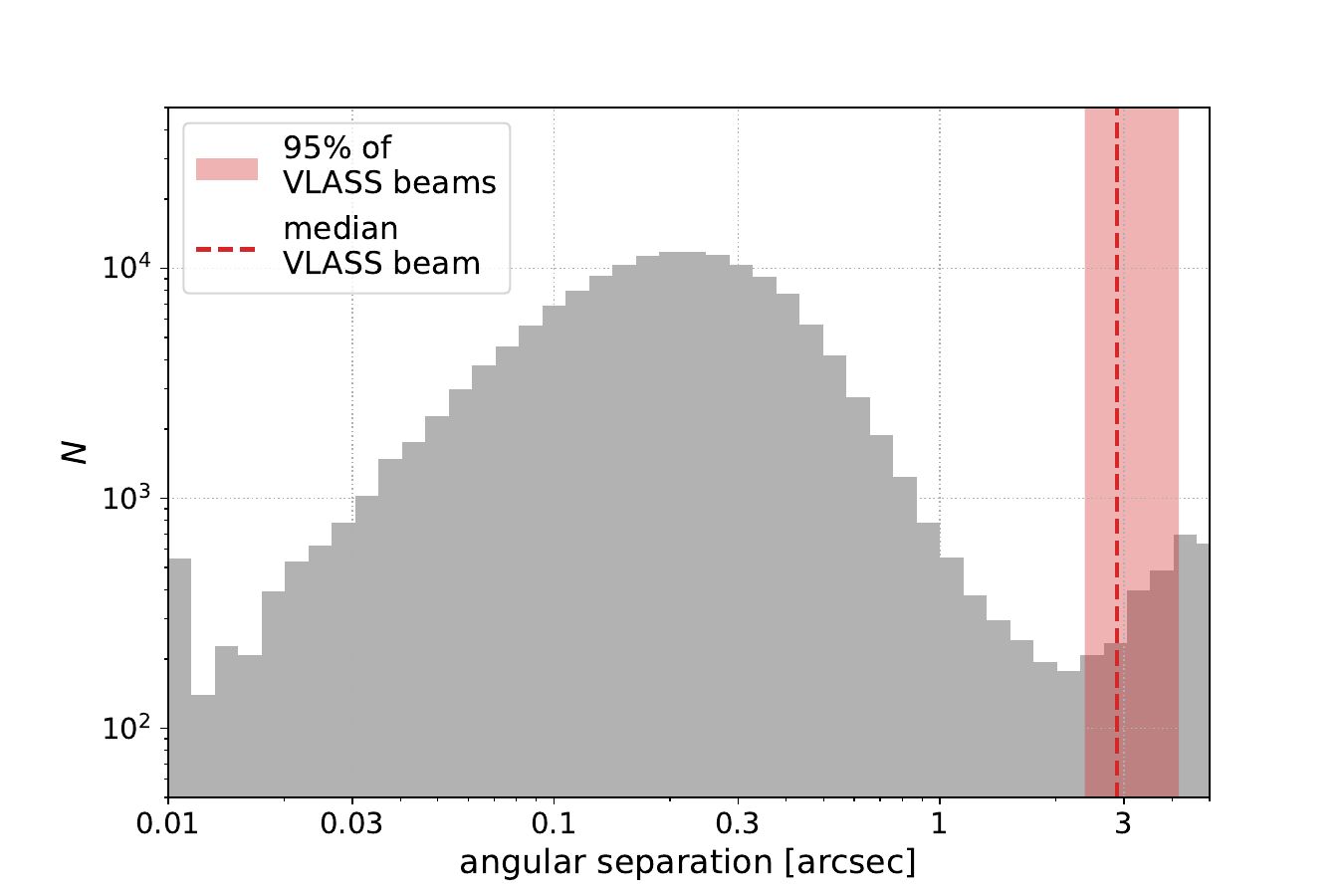}}
    \subfigure[]{\includegraphics[trim={0 0 0 1cm}, clip,width=0.99\columnwidth]{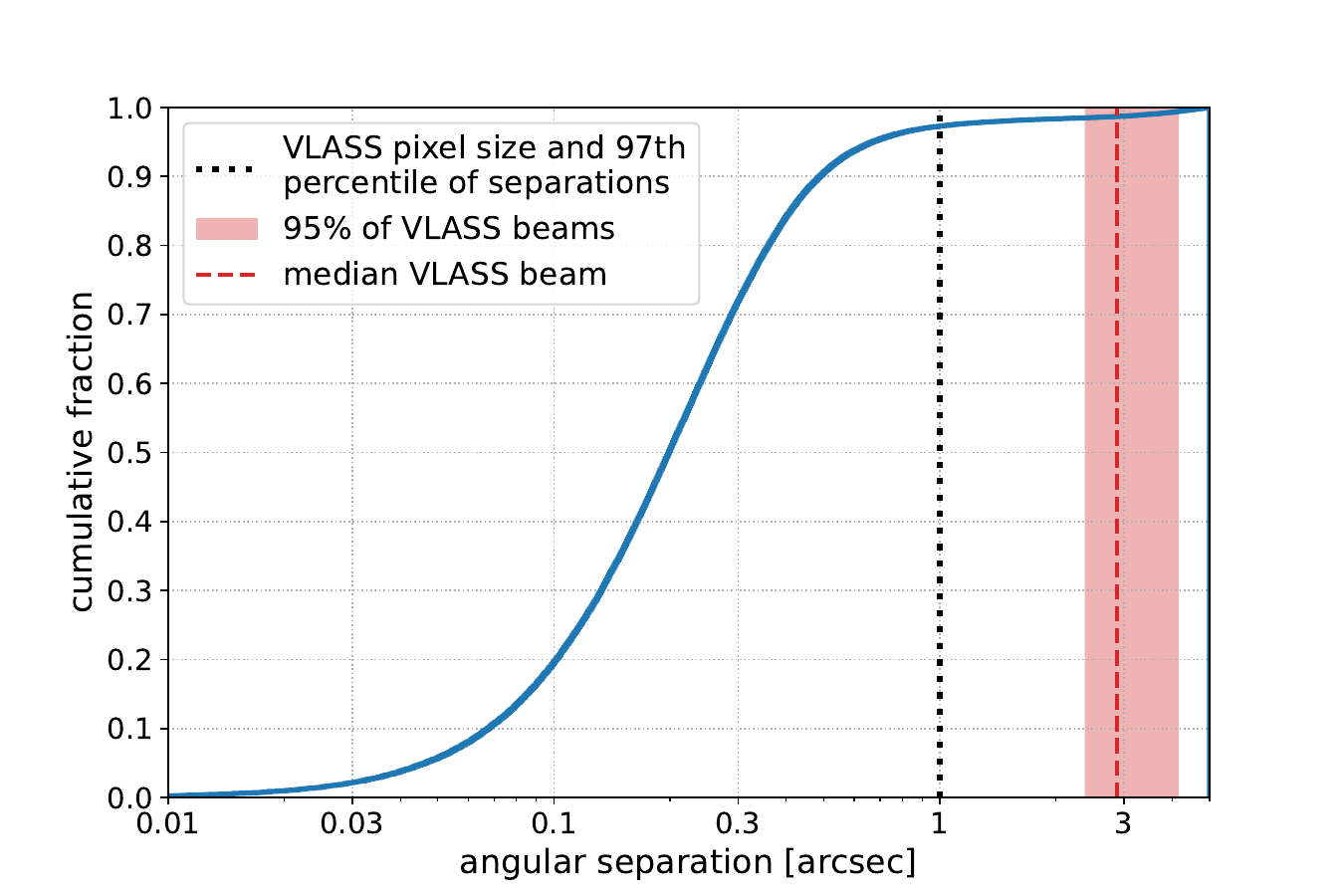}}
    \caption{Panel (a) shows the distribution of angular separations between point sources in VLASS Epoch 1 and the nearest point source in VLASS Epoch 2 out to $5\arcsec$.
    Panel (b) shows the cumulative distribution of angular separations, showing the majority of these to be separated by a few tenths of an arcsecond.
    In both panels the median VLASS beam size ($2\arcsec.9$) is shown by a dashed red vertical line, with the red shaded region showing the $95\,$\% spread in VLASS beam sizes.
    The vertical dotted line shows the pixel size ($1\arcsec$) of the VLASS \textit{Quick Look} images which corresponds to the $97$th percentile of the distributions shown here.}
    \label{fig:matchsepdist}
\end{figure}

More than $99\,$\% of VLASS images have a beam size of \text{$<5\arcsec$}, and in Figure \ref{fig:matchsepdist}a we show a histogram of the angular separations between point sources in one epoch and their closest neighbor within $5\arcsec$ in the other epoch.
The distribution peaks at $\sim0\arcsec.2$, substantially smaller than the median VLASS beam size of $2\arcsec.9$ (red dashed vertical line in Figure \ref{fig:matchsepdist}a), and we expect these to be observations of the same objects in both epochs.
The increase in the counts at separations similar to the VLASS beam size is driven by close pairs of VLASS sources that would be expected from relatively compact examples of double-lobed radio galaxies \citep[see also, e.g.,][]{Gordon2023dragns}.
In Figure \ref{fig:matchsepdist}b, we show the cumulative distribution of the angular separations for those point sources in Epoch 1 within $5\arcsec$ of a point source detection in Epoch 2.
The median separation between these repeat detections is $0\arcsec.2$, and more than $97\,$\% are separated by less than the $1\arcsec$ pixel size of the VLASS \textit{Quick Look} images.

Naturally, these statistics are dominated by fainter sources whose relatively low signal to noise ($S/N>5$ is required for a detection) will impact their positional accuracy.
When only sources with $S/N > 30$ are considered a similar result is found, with $P_{50}=0\arcsec.1$ and $P_{97}=0\arcsec.8$ for sources with a counterpart within $5\arcsec$ in the other epoch's catalog.
If sources have coordinates in Epochs 1 and 2 that are separated by less than $1\arcsec$, we can be assured that they are likely the same source, or at least indistinguishable at the resolution of VLASS. 
At the same time, a tolerance of $1\arcsec$ will allow for a highly complete sample of matched sources.
For the rest of this work, we only consider sources to be detected in both catalogs if the angular separation between the match is less than $1\arcsec$ ($1,414,783$ sources).

\section{Identifying Variable Radio Sources} 
\label{sec:finding-variables}

\subsection{Variability Systematics in the Quick Look Images}
\label{ssec:qlsystematics}

In order to identify variability among radio sources, it is necessary to first understand the self consistency of the measurements made at different times.
In particular, if the flux density measurements of sources are systematically higher in one epoch over the other this must be accounted for prior to characterizing the true variability of the radio sources.
When using measurements based on the VLASS \textit{Quick Look} images, extra care must be taken in such comparisons as the data quality is known to be non-homogeneous across the survey, a feature we demonstrated in the context of how the rms noise of Epoch 1 \textit{Quick Look} images varies over the sky in Figure 4 of \citet{Gordon2021}.
In addition to the main source catalogs, tables of metadata for each \textit{Quick Look} image, including basic statistics on the sources detected within each image (number of detections; mean and standard deviation of rms noise; 0th, 25th, 50th, 75th, and 100th percentiles of the peak flux density distribution), are available for both the Epoch 1 and 2 catalogs \citep{Gordon2020}.

To characterize any inhomogeneity in the systematics of flux measurement repeatability between the two epochs, we analyze how the ratio of the median peak flux density of detected sources in Epoch 2 to the median peak flux density of detected sources in Epoch 1 ($S_{\text{peak,}\,\text{Ep. 2,}\,p50}/S_{\text{peak,}\,\text{Ep. 1,}\,p50}$) varies across the VLASS footprint.
The median brightness of detected sources in each epoch should be robust against the strong variables and radio transients that are expected to make up only a small fraction of radio sources, \citep[e.g.,][]{Hajela2019}.
The VLASS observing strategy consists of observing 899 Tiles per epoch, each typically covering a $4^{\circ} \times 10^{\circ}$ footprint  \citep{Kimball2017} containing $\sim2,100$ sources, with the $1\,\text{deg}^{2}$ \textit{Quick Look} images produced from these Tiles\textemdash each Tile is an independent observation, rather than each \textit{Quick Look} image.
We therefore assess $S_{\text{peak,}\,\text{Ep. 2,}\,p50}/S_{\text{peak,}\,\text{Ep. 1,}\,p50}$ on a per-Tile basis rather than a per-image basis, improving the robustness of the statistic by increasing the number of contributing source measurements by a factor of $40$.

\begin{figure}
    \centering
    \subfigure[]{\includegraphics[trim={1.8cm 0 1.8cm 0}, clip, width=\columnwidth]{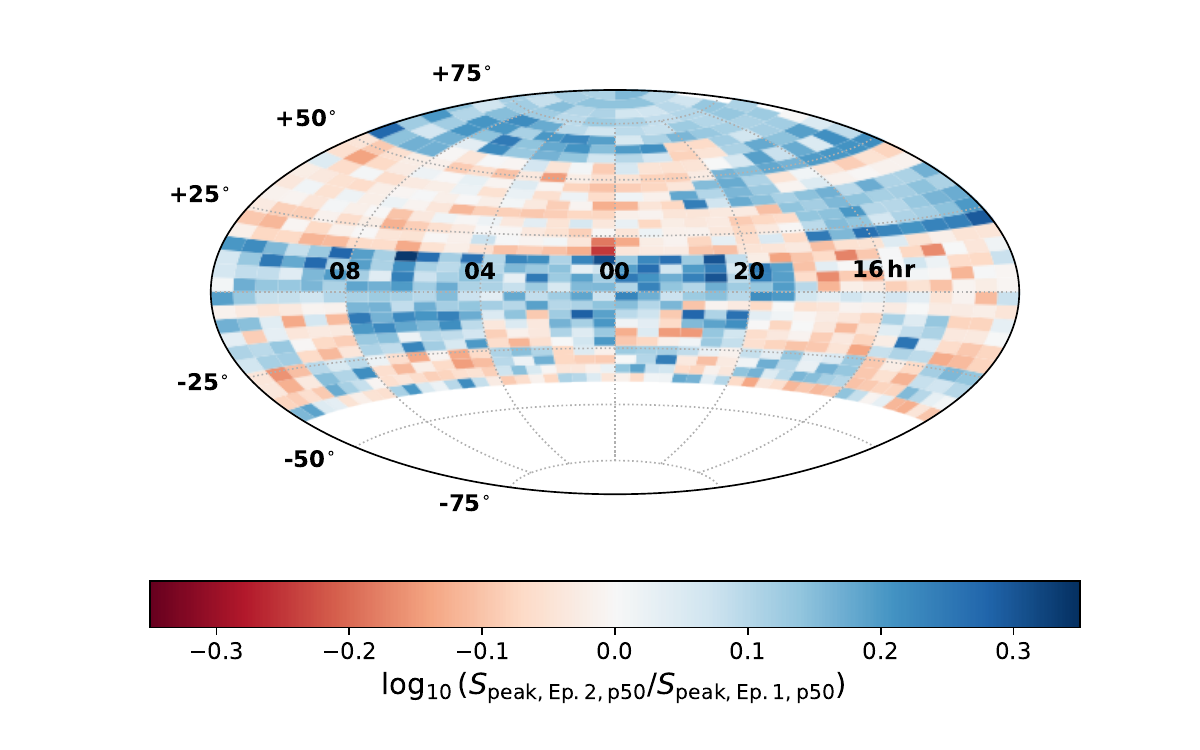}}
    \subfigure[]{\includegraphics[trim={0 0 0 1.5cm}, clip, width=\columnwidth]{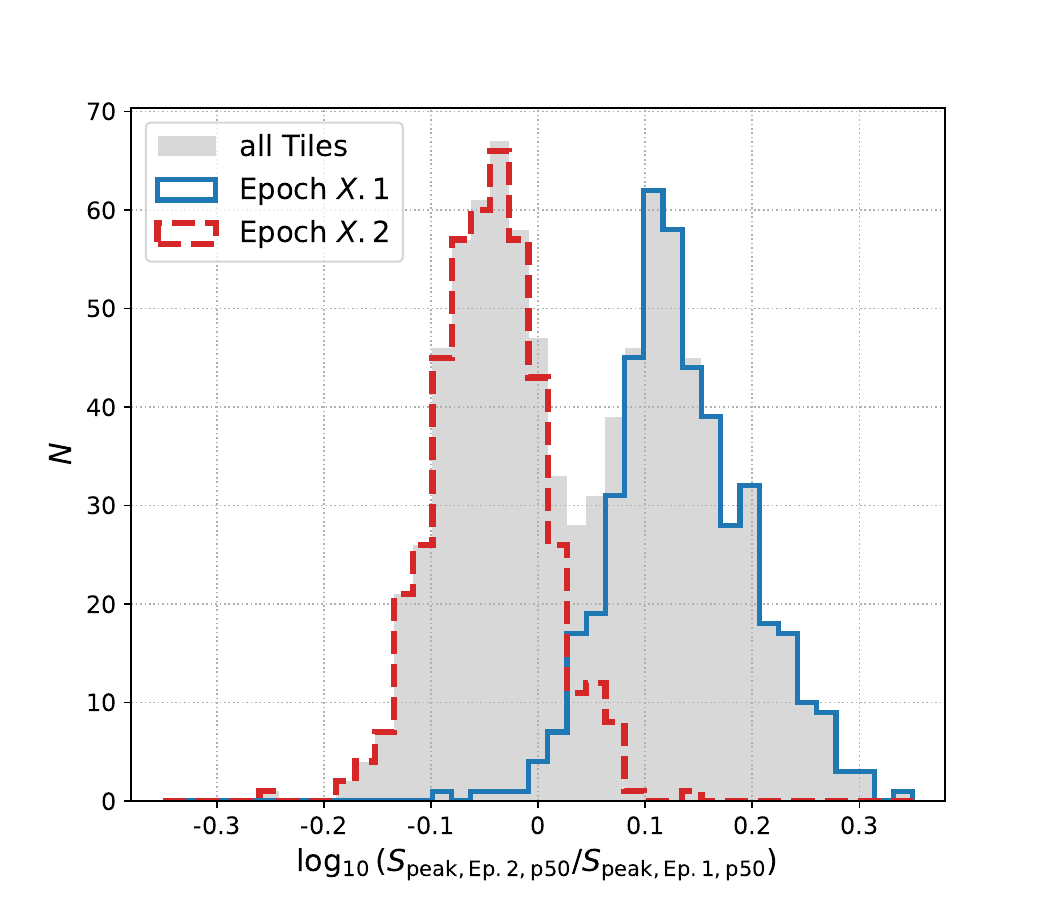}}
    \caption{Panel (a) shows the ratio of the median brightness of sources within a VLASS Tile in Epoch 2 to the median brightness of sources within that Tile in Epoch 1 on an Aitoff projected sky map.
    Panel (b) shows a histogram of the ratio of the median brightness of sources within a Tile in Epoch 2 to Epoch 1, split by whether the tile was observed in the first half (blue solid line) or second half (red dashed line) of a VLASS Epoch.
    A clear difference in image quality is seen between VLASS sub-epochs, with the flux ratios between images in E2.2 and E1.2 typically being lower than between images in E2.1 and E1.1. 
    }
    \label{fig:image_flux_ratio}
\end{figure}

In Figure \ref{fig:image_flux_ratio}a, we show the ratio of median peak flux density of detected sources in a VLASS Tile in Epoch 2 to median peak flux density of detected sources in that tile in Epoch 1 on an Aitoff projected sky map.
Tiles with $S_{\text{peak,}\,\text{Ep. 2,}\,p50}/S_{\text{peak,}\,\text{Ep. 1,}\,p50} > 1$ are shown as progressively darker shades of blue, while Tiles with $S_{\text{peak,}\,\text{Ep. 2,}\,p50}/S_{\text{peak,}\,\text{Ep. 1,}\,p50} < 1$ are shown in progressively darker shades of red. 
Figure \ref{fig:image_flux_ratio}a highlights a clear bimodality in ($S_{\text{peak, Ep. 2, } p50}/S_{\text{peak, Ep. 1, } p50}$) dependent on sky location of the tile.
One potential explanation for this bimodality might lie in improvements made to the \textit{Quick Look} imaging algorithm between the first and second half of Epoch 1 (Epochs 1.1 and 1.2, respectively).
\citet{Lacy2019} note that, as a result of these changes, the peak flux densities of calibrator sources in the \textit{Quick Look} images are systematically $\sim8\,$\% lower than expected in Epoch 1.2 but $\sim15\,$\% lower than expected in Epoch 1.1.
With changes to the imaging routine that impact the flux density scaling being applied half way through Epoch 1, one might expect the typical ratios of flux densities between Epoch 2.1 and 1.1 to systematically differ from ratios based on measurements in Epochs 2.2 and 1.2.

In Figure \ref{fig:image_flux_ratio}b, we show a histogram of $\log_{10}(S_{\text{peak,}\,\text{Ep. 2,}\,p50}/S_{\text{peak,}\,\text{Ep. 1,}\,p50})$
for all VLASS Tiles, split by whether the tile was observed in the first half (Epoch $X.1$) or second half (Epoch $X.2$) of a VLASS Epoch.
For the survey as a whole, the median value of $\log_{10}(S_{\text{peak,}\,\text{Ep. 2,}\,p50}/S_{\text{peak,}\,\text{Ep. 1,}\,p50})$ is $0.04$, i.e., sources appear $~10\,$\% brighter in Epoch 2 than Epoch 1.
The standard deviation in $\log_{10}(S_{\text{peak, Ep. 2, } p50}/S_{\text{peak, Ep. 1, } p50}$) for the survey as a whole is 0.11 (a factor of nearly $30\,$\%).
When considering only the individual half-epochs, the median and standard deviations in $\log_{10}(S_{\text{peak,}\,\text{Ep. 2,}\,p50}/S_{\text{peak,}\,\text{Ep. 1,}\,p50})$ for Epochs $X.1$ and $X.2$ are $0.13\pm0.06$ ($35\pm15\,$\% brighter in Epoch 2) and $-0.05\pm0.05$ ($12\pm12\,$\% brighter in Epoch 1), respectively.
It is thus abundantly clear that the inhomogeneity of the VLASS \textit{Quick Look} data would hinder using statistics based on the survey as a whole when identifying radio variability. 
For instance, the large spread in $S_{\text{peak,}\,\text{Ep. 2,}\,p50}/S_{\text{peak,}\,\text{Ep. 1,}\,p50}$ across the whole survey might prevent the identification of many more weakly variable sources that can be identified cleanly when using statistics based on only using the Tile, or even a half Epoch.
We therefore chose to identify radio variability on a per-Tile basis throughout the rest of this work.
As most radio sources are not expected to be variable, we correct the measured Epoch 2 flux density of sources so that the median value of $S_{\text{Ep.2}}/S_{\text{Ep.1}}$ is $1$ in each Tile.

\subsection{Identifying Unresolved Radio Sources}

For point sources we expect the total and peak flux density measurements to be the same within uncertainties.
As such we adopt the traditional approach of defining point sources based on their position within the parameter space of $S_{\text{total}}/S_{\text{peak}}$ versus source signal-to-noise ratio \citep[$S/N = S_{\text{peak}}/\text{rms}_{\text{local}}$; e.g.,][]{Bondi2008, Huynh2012, Shimwell2019}.
To do this we first model the $95\,$\% spread in $S_{\text{total}}/S_{\text{peak}}$ as a function of S/N by measuring the $2.5$th percentile of $\log_{10}(S_{\text{total}}/S_{\text{peak}})$ in 10 equal width bins of $\log_{10}(S/N)$ for sources that have \texttt{DC\_Maj}==0 and are thus a good first-order approximation for unresolved VLASS sources.
Using a least squares fit, we find that $97.5\,$\% of sources with \texttt{DC\_Maj}==0 have: 
\begin{equation}
    \label{eq:tot2peak-lower}
    \log_{10}\Bigg(\frac{S_{\text{total}}}{S_{\text{peak}}}\Bigg) > 0.018 - 0.147\times \log_{10}\Bigg(\frac{S}{N}\Bigg)^{-1.519}.
\end{equation}
Rather than use the $97.5$th percentile of the $\log_{10}(S/N)$ bins, which will be contaminated by extended objects where the source finder has miscalculated the deconvolved source size, to determine the region containing $95\,$\% of point sources we mirror the $2.5$ percentile model about the median $S_{\text{total}}/S_{\text{peak}}$ value by flipping the sign of the $S/N$ term in Equation \ref{eq:tot2peak-lower}, i.e.,
\begin{equation}
    \label{eq:tot2peak-upper}
    \log_{10}\Bigg(\frac{S_{\text{total}}}{S_{\text{peak}}}\Bigg) < 0.018 + 0.147\times \log_{10}\Bigg(\frac{S}{N}\Bigg)^{-1.519}.
\end{equation}

\begin{figure}
    \centering
    \includegraphics[width=\columnwidth]{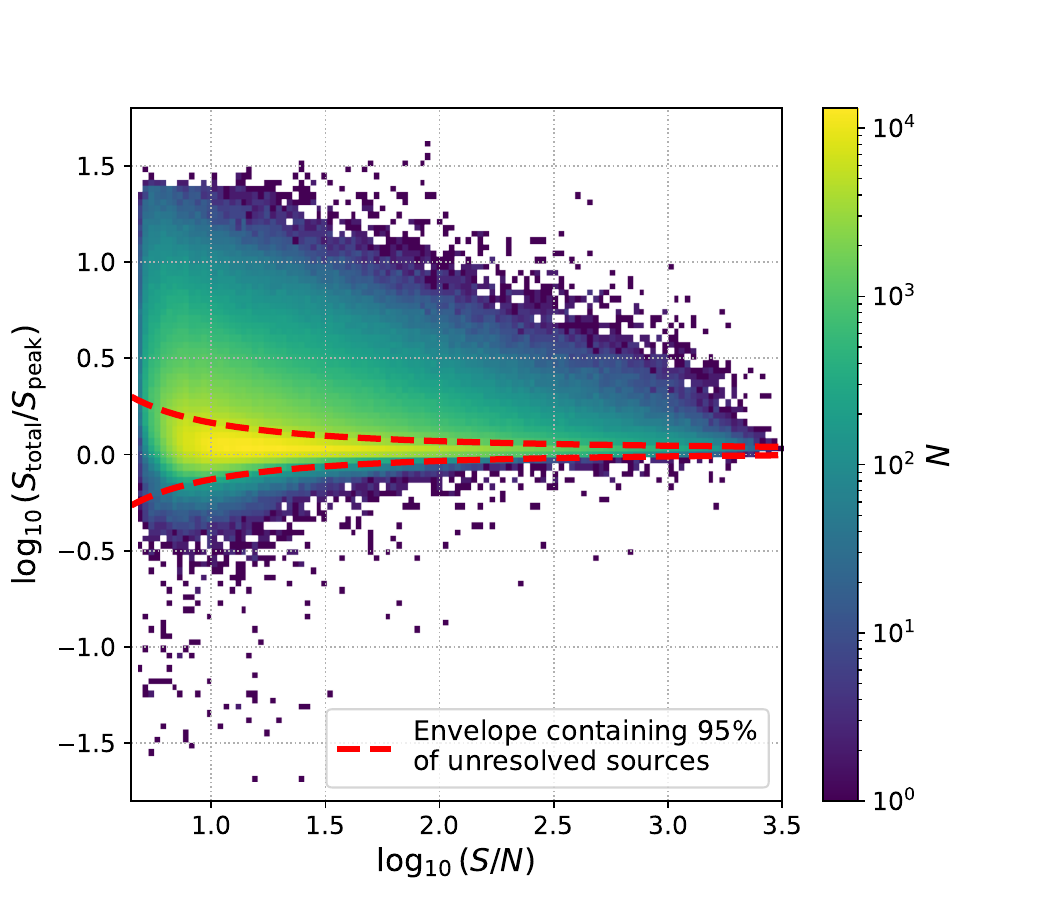}
    \caption{The distribution of the ratio of integrated to peak flux density ($S_{\text{total}}/S_{\text{peak}}$) as a function of source signal to noise.
    The red dashed line encloses the region where $95\,$\% of point sources are expected to lie.
    This figure shows the both the Epoch 1 and Epoch 2 measurements for all sources.}
    \label{fig:tot2peak}
\end{figure}

In Figure \ref{fig:tot2peak} we show the distributions of $\log_{10}(S_{\text{total}}/S_{\text{peak}})$ and $\log_{10}(S/N)$ for our sources with detections in Epochs 1 and 2, with the lower and upper dashed red lines showing the limits of Equations \ref{eq:tot2peak-lower} and \ref{eq:tot2peak-upper} respectively.
For the rest of this work, we consider those sources satisfying Equation \ref{eq:tot2peak-upper} as being likely point sources, while sources not satisfying Equation \ref{eq:tot2peak-upper} are likely resolved.
Notably, sources that are unresolved in one epoch may be resolved in the other.
This effect may be caused by measurement uncertainties and differing beam shapes between the two sets of observations, particularly for sources that are on the cusp of the VLASS resolution limit.
Based on Equation \ref{eq:tot2peak-upper}, $\sim9\,$\% of sources are resolved in one Epoch but not the other, with $130,104$ sources resolved in Epoch 1 but not in Epoch 2 and $126,015$ sources resolved in Epoch 2 but not in Epoch 1.
Approximately $30\,$\% of sources ($N=428,441$) are resolved in both Epoch 1 and Epoch 2.
Of our sources, $52\,$\% ($N=730,223$) are unresolved by VLASS in both Epochs, and for the rest of this work we refer to these as our main sample.

\subsection{Finding Candidate Radio Variables}

\begin{figure}
    \centering
    \subfigure[T13t18]{\includegraphics[trim={0 0 0 0}, clip, width=0.9\columnwidth]{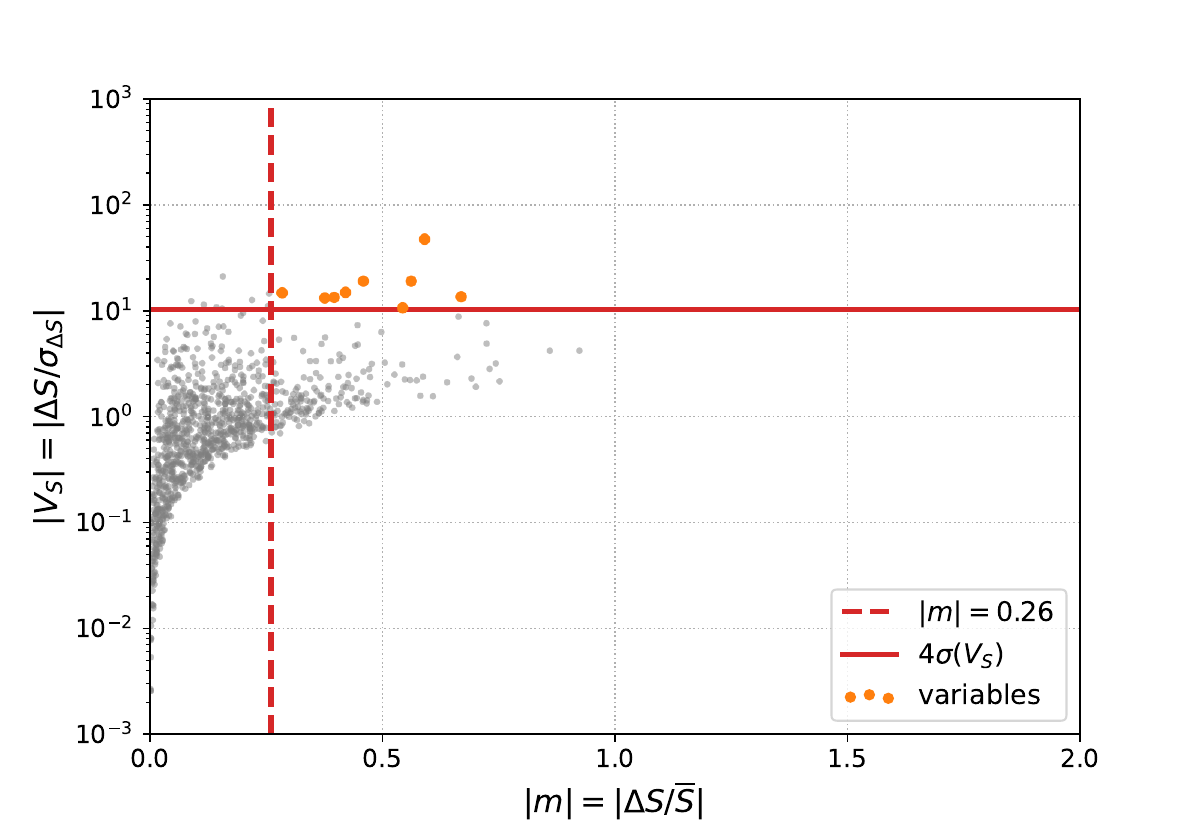}}
    \subfigure[T21t06]{\includegraphics[trim={0 0 0 1cm}, clip, width=0.9\columnwidth]{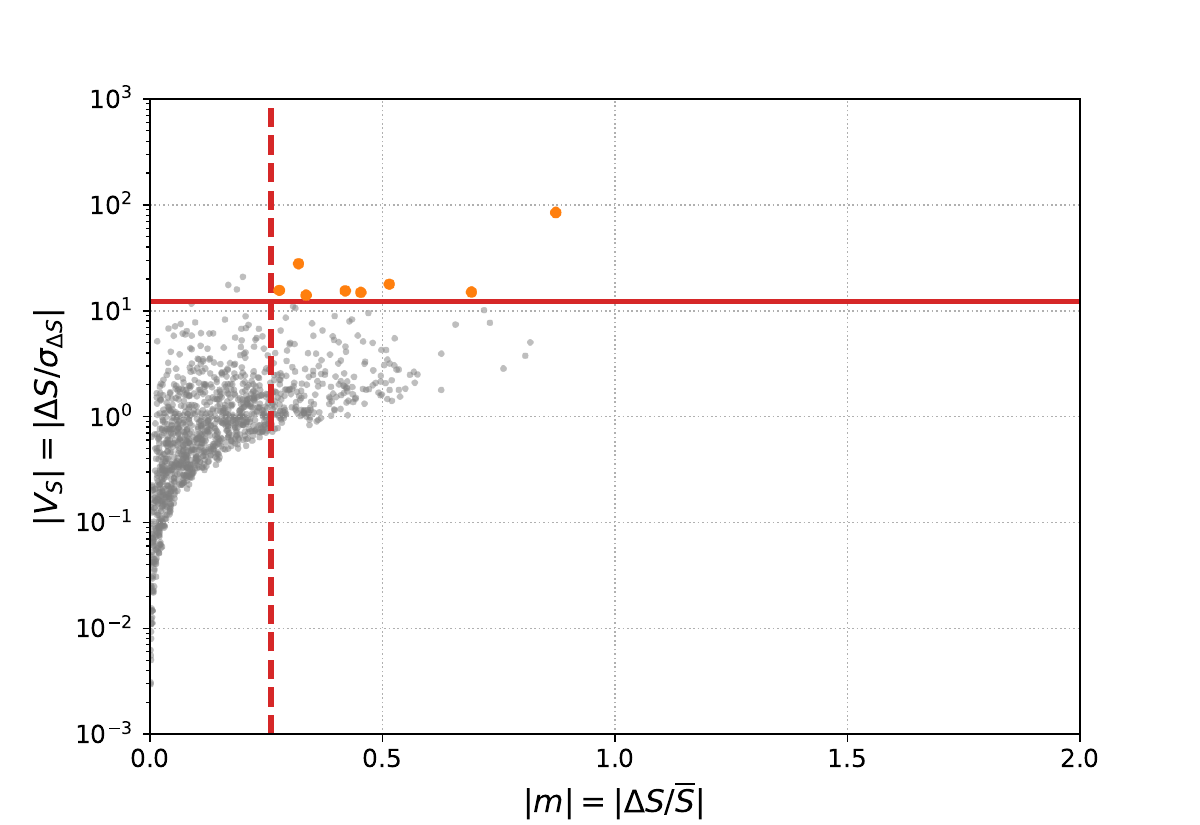}}
    \caption{
    Panels (a) and (b) show the selection of variable radio sources from two example VLASS tiles (Tile ID given below each panel).
    Grey scatter points show individual VLASS sources, with orange circles highlighting the variables.
    The vertical dashed red line shows $|m| = 0.26$, while the horizontal red line marks $4\times$ the standard deviation in $V_{S}$ for the Tile.
    }
    \label{fig:varselect-multipanel}
\end{figure}

Identification of true variability requires that the source have an absolute change in flux that is both significant relative to measurement errors, and large enough to not be attributed to observational data properties such as calibration accuracy.
To that end, in order to identify variable sources within a VLASS Tile we adopt an approach based on two key parameters: i) the Variability Statistic, $V_{s}$, which accounts for the absolute difference in the measured flux densities relative to the measurement uncertainties, and ii) the modulation index, $m$, which measures the fractional change in the flux density.
Explicitly,
\begin{equation}
    \label{eq:vs}
    V_{s} = \frac{\Delta S}{\sigma_{\Delta S}} = \frac{S_{\text{Ep.2}}-S_{\text{Ep.1}}}{\sqrt{\sigma_{S\text{, Ep.2}}^{2} +\sigma_{S\text{, Ep.1}}^{2}}},
\end{equation}
and 
\begin{equation}
    \label{eq:modidx}
    m = \frac{\Delta S}{\overline{S}},
\end{equation}
and consider a source to be variable if it satisfies \text{$V_{s} > 4\sigma_{V_s, \text{Tile}}$}, where $\sigma_{V_s, \text{Tile}}$ is the standard deviation of $V_{s}$ measurements for all unresolved sources within that Tile, and $|m| > 0.26$ \citep{Mooley2016, Hajela2019}.
We show this selection for two example Tiles in Figure \ref{fig:varselect-multipanel}.
Applying these criteria to each Tile and concatenating the results, we identify $4,124$ sources that are likely truly variable.

\begin{figure}
    \centering
    \subfigure[Epoch $X.1$]{\includegraphics[trim={0 0 0 0}, clip, width=0.9\columnwidth]{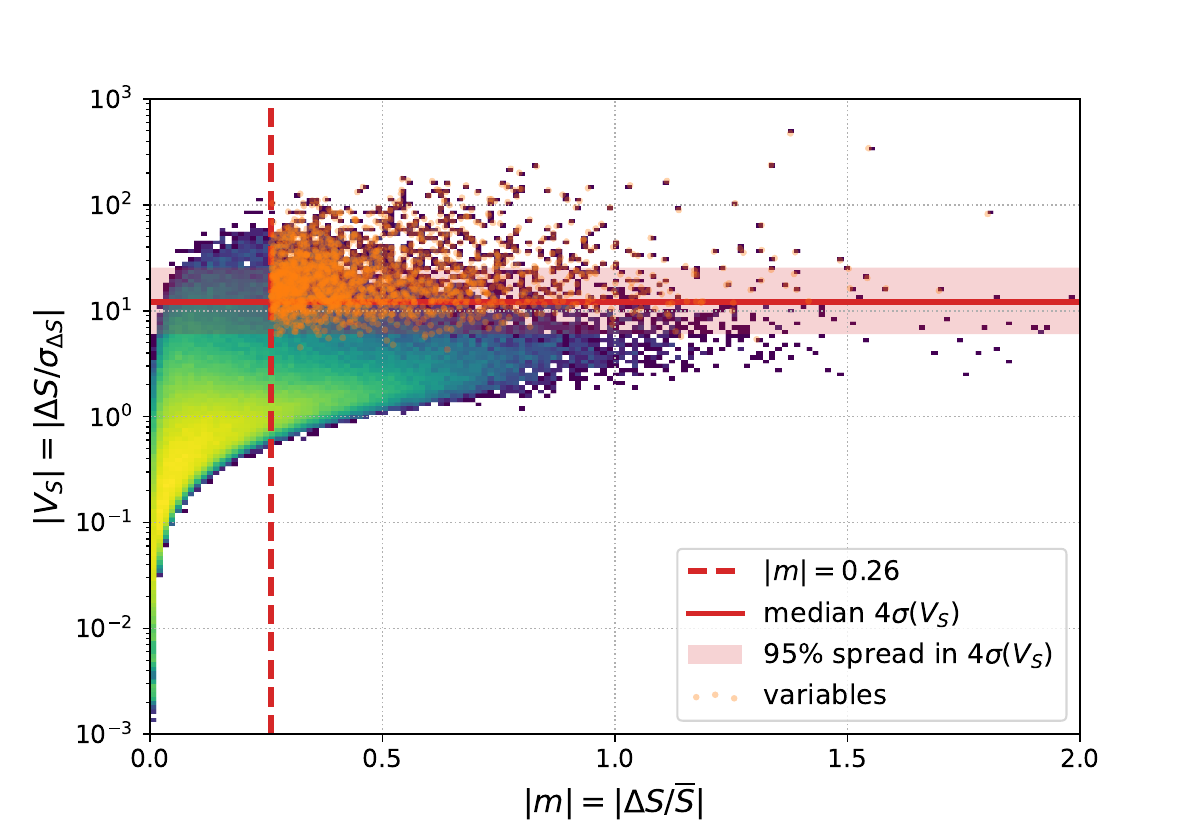}}
    \subfigure[Epoch $X.2$]{\includegraphics[trim={0 0 0 1cm}, clip, width=0.9\columnwidth]{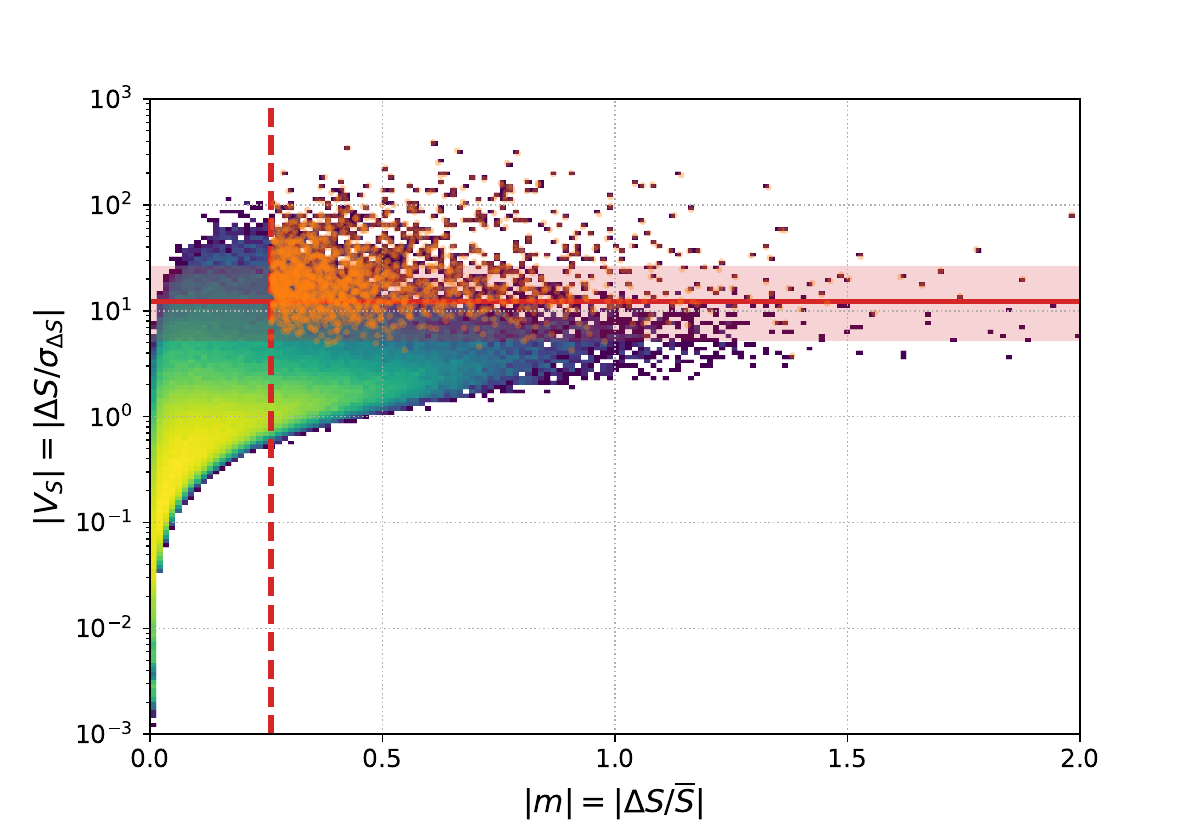}}
    \caption{
    The same variability selection parameter space shown in Figure \ref{fig:varselect-multipanel} but for the entire first (a) and second (b) subepochs.
    The red horizontal line shows the median value of $4\sigma(V_{S})$ for the Tiles within that subepoch, and the pink shaded region shows the spread between the $2.5$th and $97.5$th percentiles of $4\sigma(V_{S})$ from the Tiles within that subepoch.
    The orange dots show sources identified as being variable based on the spread of $V_{s}$ values within their Tile, rather than within the entire subepoch.
    Note that many of the sources identified as variables fall below the red solid line and would not be selected if the global subepoch $V_{s}$ statistics had been used instead of selecting variability on a per-Tile basis.
    }
    \label{fig:varselect-multipanel-subep}
\end{figure}

In Figure \ref{fig:varselect-multipanel-subep}, we show the same variability parameter space as in Figure \ref{fig:varselect-multipanel} but for all the sources from every Tile in each VLASS subepoch.
The solid red horizontal line on both panels in Figure \ref{fig:varselect-multipanel-subep} shows the median value of $4\sigma_{V_{s}\text{, Tile}}$ within that subepoch, while the shaded pink region shows the $95\,$\% spread of  $4\sigma_{V_{s}\text{, Tile}}$ values.
Note how for both panels many sources identified as variable (orange dots) fall below the solid red horizontal line, suggesting that such sources would be missed if our variability selection were done at the sub-epoch level rather than on a per-Tile basis.
If we were to select variables using the $V_{s}$ statistics from an entire sub-epoch rather than on a per-Tile basis, we would find $1,601$ sources in Epoch $X.1$ and $1,432$ in Epoch $X.2$.
Of these $3,033$ variables, $264$ were not selected by the per-Tile method; $147$ from Epoch $X.1$ and $117$ from Epoch $X.2$.
Using the global $V_{s}$ statistics from the entire survey shows a similar result to using the subepoch variability statistics: $3,015$ sources are identified as variables, $266$ of which are not selected by the per-Tile approach.
This highlights that using spread in $V_{s}$ for either an entire subepoch or the entire survey as the variability selection threshold, rather than selecting variables on a per-Tile basis, will a) fail to identify $\sim 1/4$ of variables that are found in higher than average quality images, and b) suffer $\sim 9\,$\% contamination from sources spuriously identified as variable in lower than average quality images.

\subsection{Quality Assurance}
\label{ssec:qa}

\subsubsection{Tile Edge Effects}

The increased noise towards the edge of the field of view (FoV) that is common to many observing setups, including the VLA, has the potential to impact the selection of variable sources from multiple independent observations.
For example, increased variance in comparative flux measurements at the edge of the FoV relative to at the center of the FoV, could cause sources at the edge of the FoV to be more likely to be classified as variables.
To observe a Tile, VLASS uses an on-the-fly mosaicking observing strategy that scans multiple strips of Right Ascension to build up a Tile that typically spans $10^{\circ}$ in RA by $4^{\circ}$ in Decl. \citep{Kimball2017, Lacy2020}.
This approach produces uniform coverage over most of the Tile, but the Tile edges, particularly the east and west edges, are still susceptible to reduced sensitivity \citep[see also Figure 4 of][]{Gordon2021}.

\begin{figure}
    \centering
    \subfigure[]{\includegraphics[trim={0 0 0 0}, clip, width=0.95\columnwidth]{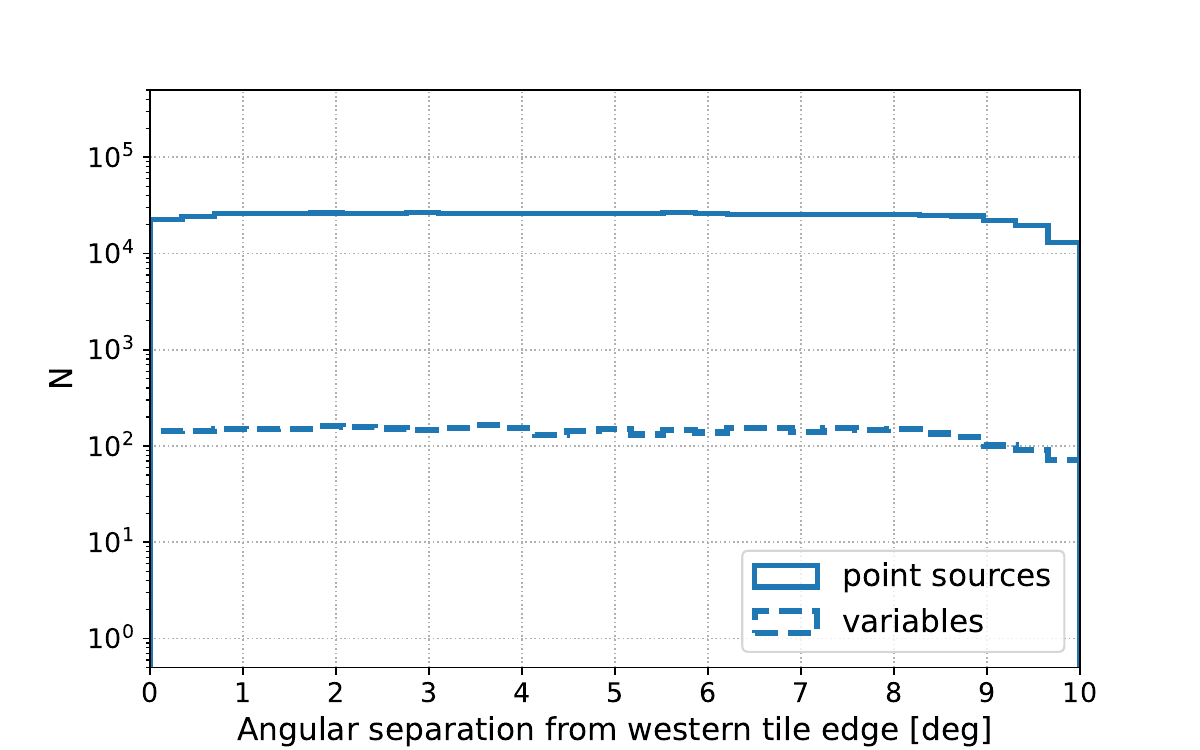}}
    \subfigure[]{\includegraphics[trim={0 0 0 1cm}, clip, width=0.95\columnwidth]{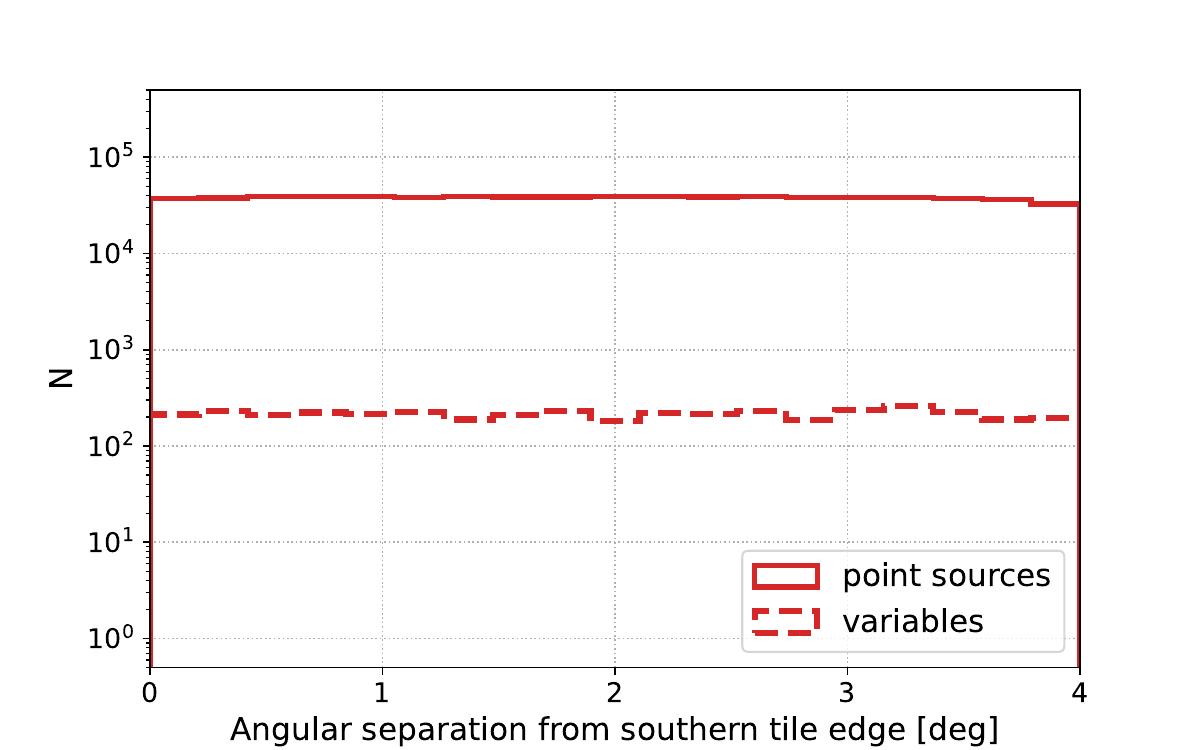}}
    \caption{Angular separation of sources from the western (a) and southern (b) edges of the VLASS Tile they are in.
    A typical VLASS Tile covers $10^{\circ}\times4^{\circ}$.
    Note the similar drop off in counts for point sources and variables that are $\gtrsim9^{\circ}$ from the Tile's western edge, and thus $\lesssim1^{\circ}$ from the Tile's eastern edge.}
    \label{fig:tile_edge_effects}
\end{figure}

To check that we are not preferentially selecting variables at Tile edges, we analyze the typical flux variation of point sources based on how close they are to a Tile edge.
For sources $>1^{\circ}$ from a Tile edge, the median value of $|\Delta S|$ is $0.30\,$mJy.
The median uncertainty in $\Delta S$, $\sigma_{\Delta S}$, for those sources is $0.37\,$mJy.
When considering sources that are $<0.1^{\circ}$ from a Tile edge the median values of $|\Delta S|$ and $\sigma_{\Delta S}$ are $0.36\,$mJy and $0.43\,$mJy, respectively.
While the typical flux variation is larger at the Tile edges, there is a corresponding increase in the uncertainty associated with the $\Delta S$ measurements, suggesting that any increased variability measurements as a result of Tile edge effects may not be significant enough to cause the selection of additional variable sources.

As a further check to any potential impact on our variability selection, we show the distribution of point sources and candidate variables within a Tile in Figure \ref{fig:tile_edge_effects}.
Panel (a) shows the angular separation of sources from the western RA edge of the Tile, and panel (b) shows the angular separation of sources from the southern Decl. edge of the Tile.
The Decl. distribution of point sources and candidate radio variables is flat within a Tile, and the RA distribution is flat $<9^{\circ}$ from the western edge of the Tile.
Within $1^{\circ}$ of the \emph{eastern} Tile edge (i.e., more than $9^{\circ}$ from the western edge), there is a drop off in the counts of both point sources and candidate radio variables.
Both populations drop off at the same rate, with candidate variables accounting for between $0.5\,$\% and $0.6\,$\% of point sources across the entire range of RA within a Tile.
Although this does show that Tile edge effects within VLASS do impact source detectability, such effects do not appear to significantly bias our selection of variables from the sources detected by VLASS.

\subsubsection{Visual Inspection of Candidate Variables}
\label{sssec:qa}

\begin{figure}
    \centering
    \subfigure[]{\includegraphics[trim={3.5cm 0 3.5cm 0}, clip, width=\columnwidth]{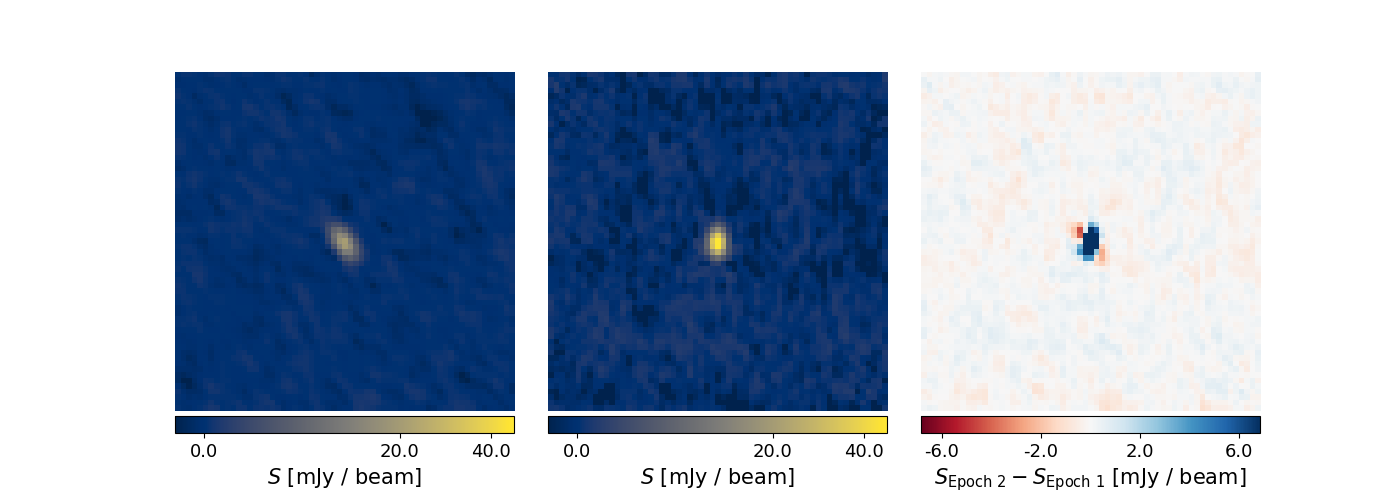}}
    \subfigure[]{\includegraphics[trim={3.5cm 0 3.5cm 1cm}, clip, width=\columnwidth]{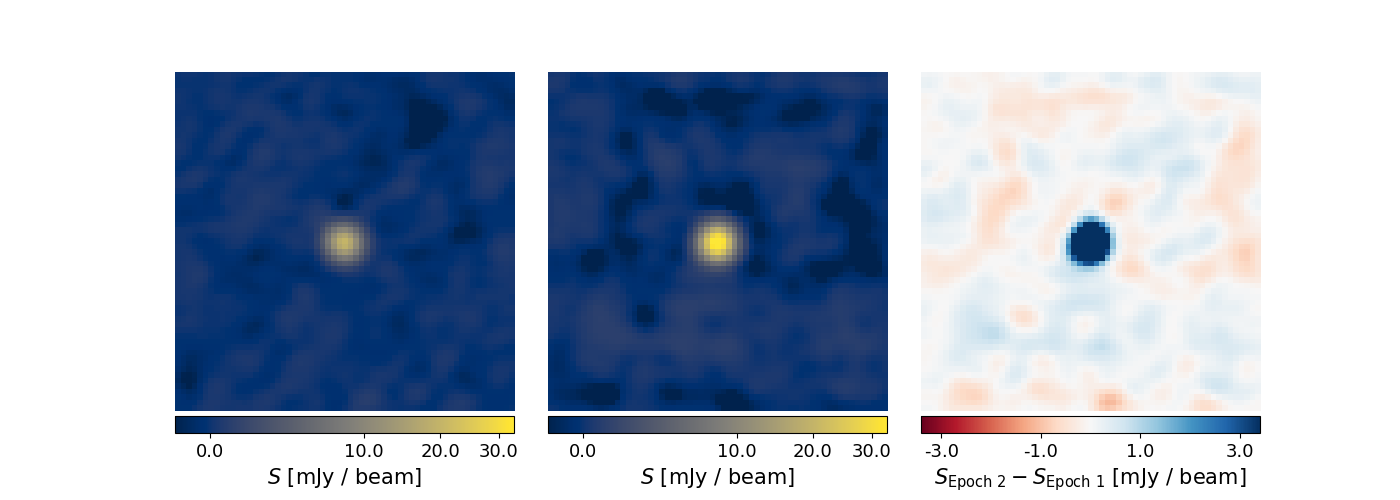}}
    \subfigure[]{\includegraphics[trim={3.5cm 0 3.5cm 1cm}, clip, width=\columnwidth]{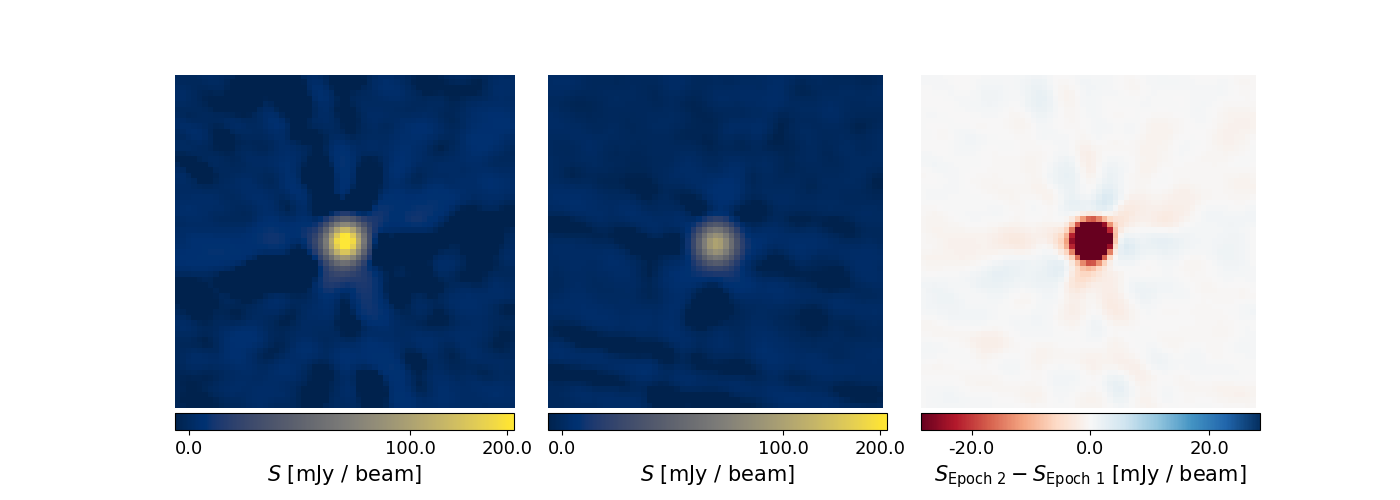}}
    \subfigure[]{\includegraphics[trim={3.5cm 0 3.5cm 1cm}, clip, width=\columnwidth]{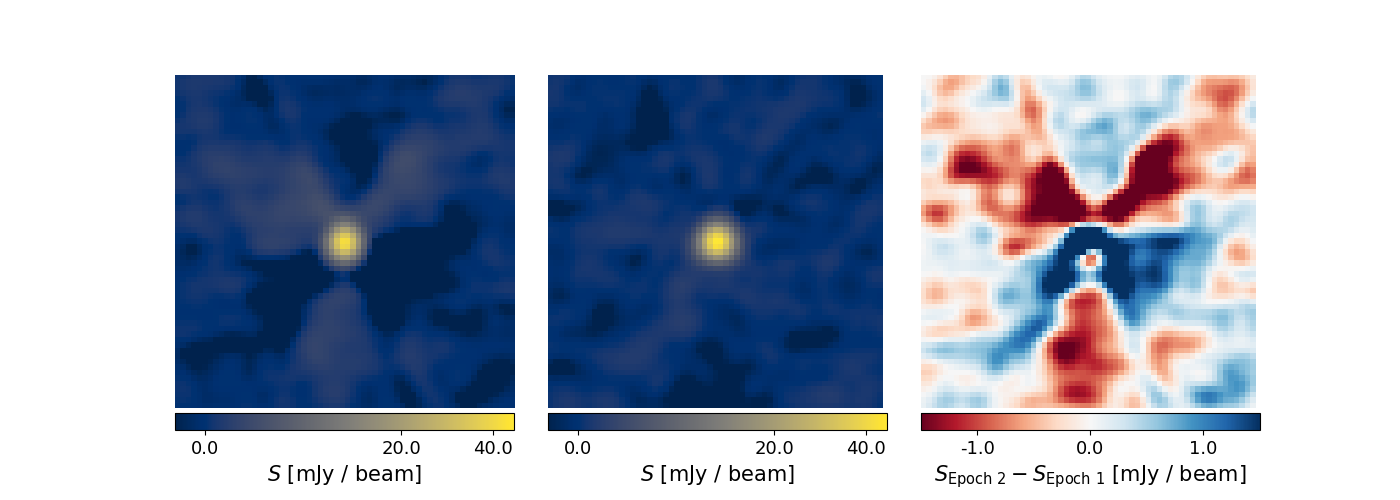}}
    \caption{Example QA images of candidate variable source showing the Epoch 1 (left) and Epoch 2 (middle), and difference (right) images.
    Panel (a) shows the images at native resolution, where the VLASS beam differs between the two images, leading to artificial effects in the difference image.
    In panel (b) the images of the same object as in panel (a) are blurred to a resolution of $4\arcsec.5$, resulting in a cleaner difference image.
    Example variable sources that have brightened and faded between VLASS epochs 1 and 2 are shown in panels (b) and (c), respectively.
    Panel (d) shows an example source that was rejected as a variable on visual inspection. 
    The cataloged flux values used to identify the variability may not be accurate for the example in panel (d) due to artifacts in the \textit{Quick Look} VLASS images.}
    \label{fig:image-qa}
\end{figure}

To ensure that the variables identified by our method are real variable sources rather than the result of bad measurements due to the quality issues in either of the \textit{Quick Look} images of a source, we visually inspect all $4,124$ candidate radio variables identified. 
In order to maximize the efficiency of this process, a three-panel image is produced for each candidate variable source consisting of the $1'\times1'$ postage stamp cutouts from each Epoch as well as difference image showing $S_{\text{Epoch 2}} - S_{\text{Epoch 1}}$ (see Figure \ref{fig:image-qa}).
The VLASS point spread function varies across different observations, which can complicate the difference image. 
For example, in Figure \ref{fig:image-qa}a we show a variable source where the VLASS observations in the Epoch 1 and 2 images have elliptical PSFs with substantially different position angles causing an `X' shaped feature of extreme pixel values to appear in the difference image.
To counter differences in the PSFs, we convolve the cutout images to a common resolution modeled by a circular Gaussian with a FWHM of $4\arcsec.5$.
While this a substantial reduction in resolution (the median FWHM of the VLASS beam is $2\arcsec.9$), we don't expect to require the high native resolution of VLASS here as the variables are expected to be point sources. 
By applying this blurring, however, we clarify the variability presented in the difference image\textemdash Figure \ref{fig:image-qa}b shows the same source as Figure \ref{fig:image-qa}a but blurred to $4\arcsec.5$ resolution\textemdash making it easier to differentiate real variability from noise in the difference images.

The Epoch 1 and 2 cutouts used in the quality assurance process are stretched using a power-law stretch with an exponent of $-0.5$.
Color scale limits are set to $-0.35\,\text{mJy}\,\text{beam}^{-1}$ and the mean of the two maximum pixel values in the Epoch 1 and 2 cutouts.
Having an image stretch that is the same for both the Epoch 1 and 2 images helps to readily visualize fractional variability, while the power-law stretch provides enough dynamic range in the image to see the background noise as well as changes in bright sources.
For the difference image, a linear stretch is used, centered on 0 and ranging between $\pm\Delta x_{\text{max}}/5$, where $\Delta x_{\text{max}}$ is the difference in maximum pixel values in the Epoch 1 and 2 cutouts.
In combination with a diverging colormap (we use the \texttt{matplotlib} colormap `\texttt{RdBu}'), this allows the ready identification of real flux differences by showing the source strongly in a color dependent on the direction of variability as well as the background noise in softer hues of the two colors (e.g., see right-hand cutouts in Figure \ref{fig:image-qa}).
The resulting $4,124$ quality assurance (QA) images we use in the analysis presented here are available as part of the supplementary data to this paper, as described in Appendix \ref{ap:supdata}.

In order to streamline the visual inspection process, we make use of the Zooniverse platform\footnote{\href{https://www.zooniverse.org/}{Zooniverse.org}} \citep{Lintott2008}.
The Zooniverse interface allows us to produce a workflow where each volunteer (YG, PF, and MM) was presented with a three-panel QA image and asked the question \textit{``Does this source look like a real variable?''}
Three options were presented as answers to this question:
\begin{enumerate}
    \item Yes, the source is clearly a real variable.
    \item Unclear from these images.
    \item No, the source is probably not a real variable.
\end{enumerate}
To minimize the inherent subjectivity of this approach, 
each of the three authors involved in the visual inspection
classified all $4,124$ candidate variables independently, presented to them in a random order. 

\begin{figure}
    \centering
    \includegraphics[width=\columnwidth]{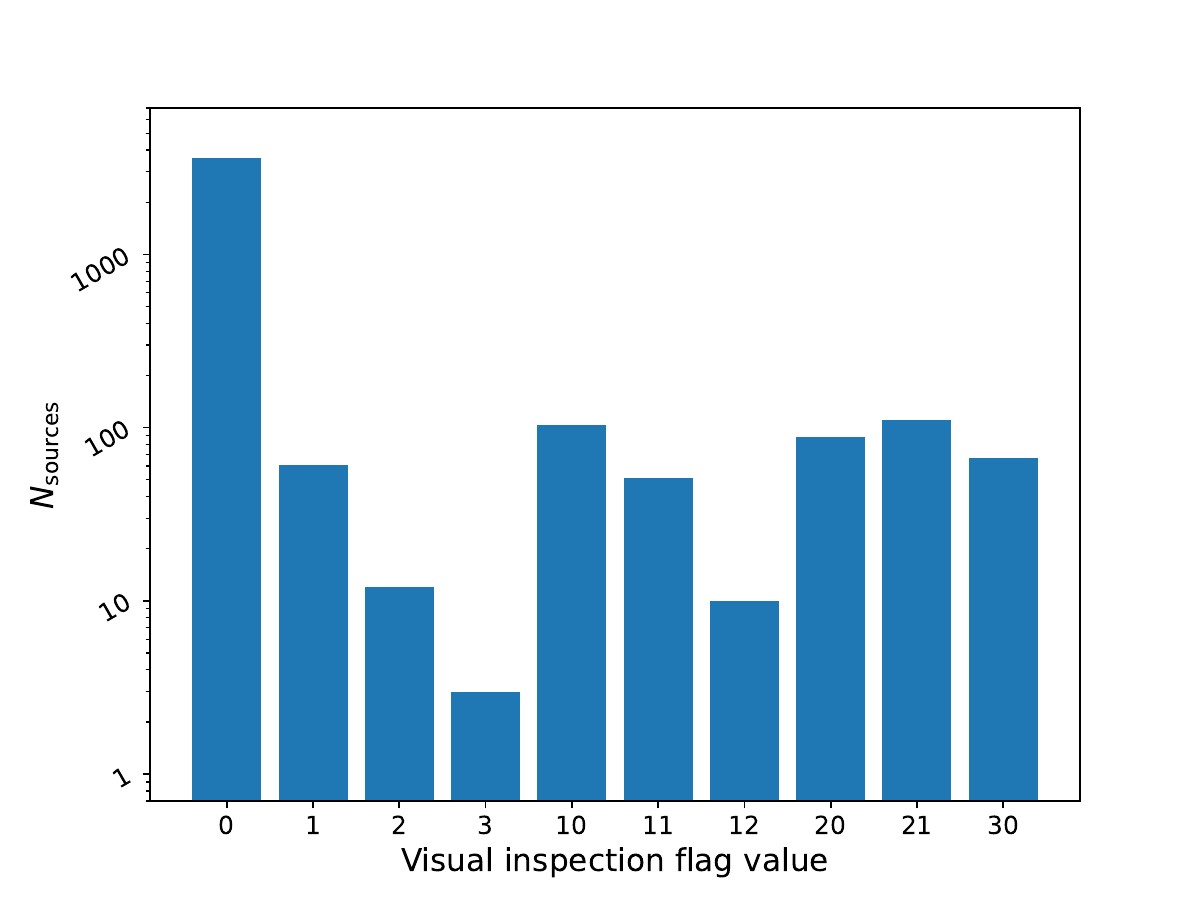}
    \caption{Distribution of the flag values assigned to the candidate variables after visual inspection. Note the logarithmically scaled y-axis, with a flag value of $0$ (meaning all three classifiers agreed the source was a real variable) being applied to $88\,$\% of sources.
    Full details of how to interpret these values are given in the main body text of Section \ref{sssec:qa}.}
    \label{fig:zooflags}
\end{figure}

Between $91\,$\% and $93\,$\% of sources are classified as being likely real variables, depending on exactly who was classifying the images.
However, while the fraction of objects classified as real variables is remarkably consistent among our classifiers, there are cases where there is disagreement over individual objects.
Of our $4,124$ candidate variables, $3,618$ ($88\,$\%) were classified as being real variable sources by all three classifiers.
For $190$ sources ($5\,$\%), all three classifiers either thought the source was clearly not a real variable or were uncertain.
The remaining $316$ sources had a mix of votes including at least one vote in favor of the source being a real variable.
In order to make use of this information, we define a flag, \textit{f\_vi}, for each source defined by the sum of individual votes it received. 
Each source starts with $\text{\textit{f\_vi}}=0$, with the value increased by the following with each visual inspection vote:
\begin{itemize}
    \item $+0$ for \textit{Yes, the source is clearly a real variable};
    \item $+1$ for \textit{Unclear from these images};
    \item $+10$ for \textit{No, The source is probably not a real variable}.
\end{itemize}
For example, should two classifiers vote that the source was not a variable, and the third classifier is uncertain, that source would have $\text{\textit{f\_vi}}=21$.
Each source therefore ends up with a flag value between $0$ (all three classifiers agree the source is a real variable) and $30$ (all three classifiers agree the source is not a real variable), and the distribution of these flag values is presented in Figure \ref{fig:zooflags}.

\renewcommand{\arraystretch}{1.25}
\tabletypesize{\scriptsize}
\begin{deluxetable*}{ccccccccccc}
\tablecaption{Variable Radio Sources in VLASS \label{tab:variables}}
\tablehead{\colhead{VLASS1QLCIR} & \colhead{VLASS2QLCIR} & \colhead{RA} & \colhead{Dec} & \colhead{MJD1} & \colhead{MJD2} & \colhead{dS} & \colhead{Smean} & \colhead{VarSig} & \colhead{modidx} & \colhead{f\_vi\tablenotemark{$\dagger$}}\\ 
\colhead{ } & \colhead{ } & \colhead{[deg]} & \colhead{[deg]} & \colhead{[day]} & \colhead{[day]} & \colhead{[mJy]} & \colhead{[mJy]} & \colhead{ } & \colhead{ } & \colhead{ }\\
\colhead{(1)} & \colhead{(2)} & \colhead{(3)} & \colhead{(4)} & \colhead{(5)} & \colhead{(6)} & \colhead{(7)} & \colhead{(8)} & \colhead{(9)} & \colhead{(10)} & \colhead{(11)}}
    \startdata
    J000014.87$+$275157.5 & J000014.88$+$275157.6 & 0.06199 & 27.86598 & 58557.74 & 59528.27 & $+$13.41$\pm$0.44& 43.35 & 9.299 & 0.309 & 0 \\
    J000038.86$-$174024.8 & J000038.86$-$174024.9 & 0.16195 & -17.67357 & 58166.87 & 59148.20 & $-$7.01$\pm$0.57 & 18.96 & 4.35 & 0.370 & 20 \\
    J000109.53$+$024309.6 & J000109.53$+$024309.7 & 0.28972 & 2.71936 & 58025.23 & 59068.40 & $+$19.71$\pm$0.41 & 40.65 & 14.217 & 0.485 & 0 \\
    J000116.24$+$364402.3 & J000116.25$+$364402.3 & 0.31770 & 36.73398 & 58592.60 & 59526.35 & $-$6.00$\pm$0.42 & 17.79 & 4.676 & 0.337 & 0 \\
    J000121.87$-$170324.6 & J000121.88$-$170324.5 & 0.34115 & -17.05685 & 58166.87 & 59148.20 & $-$6.66$\pm$0.43 & 25.33 & 5.419 & 0.263 & 10 \\
    J000123.69$+$063231.0 & J000123.69$+$063230.9 & 0.34873 & 6.54195 & 58027.31 & 59078.38 & $+$10.60$\pm$0.38 & 28.66 & 9.146 & 0.370 & 0 \\
    J000156.34$+$322813.9 & J000156.35$+$322814.0 & 0.48476 & 32.47055 & 58639.47 & 59523.36 & $+$14.97$\pm$0.39 & 40.35 & 11.123 & 0.371 & 0 \\
    J000224.14$+$142043.8 & J000224.13$+$142043.9 & 0.60059 & 14.34552 & 58031.35 & 59070.37 & $-$32.62$\pm$0.53 & 64.05 & 15.94 & 0.509 & 0 \\
    J000230.62$-$033140.3 & J000230.61$-$033140.3 & 0.62759 & -3.52788 & 58025.35 & 59047.43 & $+$8.04$\pm$0.36 & 20.03 & 5.858 & 0.401 & 0 \\
    J000231.12$+$745947.1 & J000231.08$+$745947.0 & 0.62967 & 74.99643 & 58067.18 & 59095.31 & $+$62.36$\pm$0.61 & 116.08 & 21.382 & 0.537 & 0 \\
    ... & ... & ... & ... & ... & ... & ... & ... & ... & ... & ...\\
    ... & ... & ... & ... & ... & ... & ... & ... & ... & ... & ...
    \enddata
\tablecomments{VLASS source name in Epoch 1 catalog (1); source name in the Epoch 2 catalog (2); Right Ascension from the Epoch 1 catalog (3); Declination from the Epoch 1 catalog (4); Modified Julian Date of the Epoch 1 observation (5); Modified Julian Date of the Epoch 2 observation (6); change in brightness, $S_{\text{Epoch 2}}-S_{\text{Epoch 1}}$, and associated uncertainty (7); mean flux density of the source over both Epochs (8); significance of variability, $V_{s}/\sigma_{V_{s},\text{Tile}}$ (9); modulation index, $|m|$ (10); and the quality flag from visual inspection (11).
The first 10 rows are shown here for references, and the full table consisting of $4,124$ rows is available at \url{https://zenodo.org/records/18010746}, and will be made available on \href{https://cds.unistra.fr/}{CDS} after the paper is accepted for publication.}
\tablenotetext{$\dagger$}{set to 0 if marked as reliable by all classifiers in the visual inspection step (see Section \ref{ssec:qa}).}
\end{deluxetable*}

\begin{figure}
    \centering
    \includegraphics[trim={1.8cm 0 1.8cm 0}, clip, width=\columnwidth]{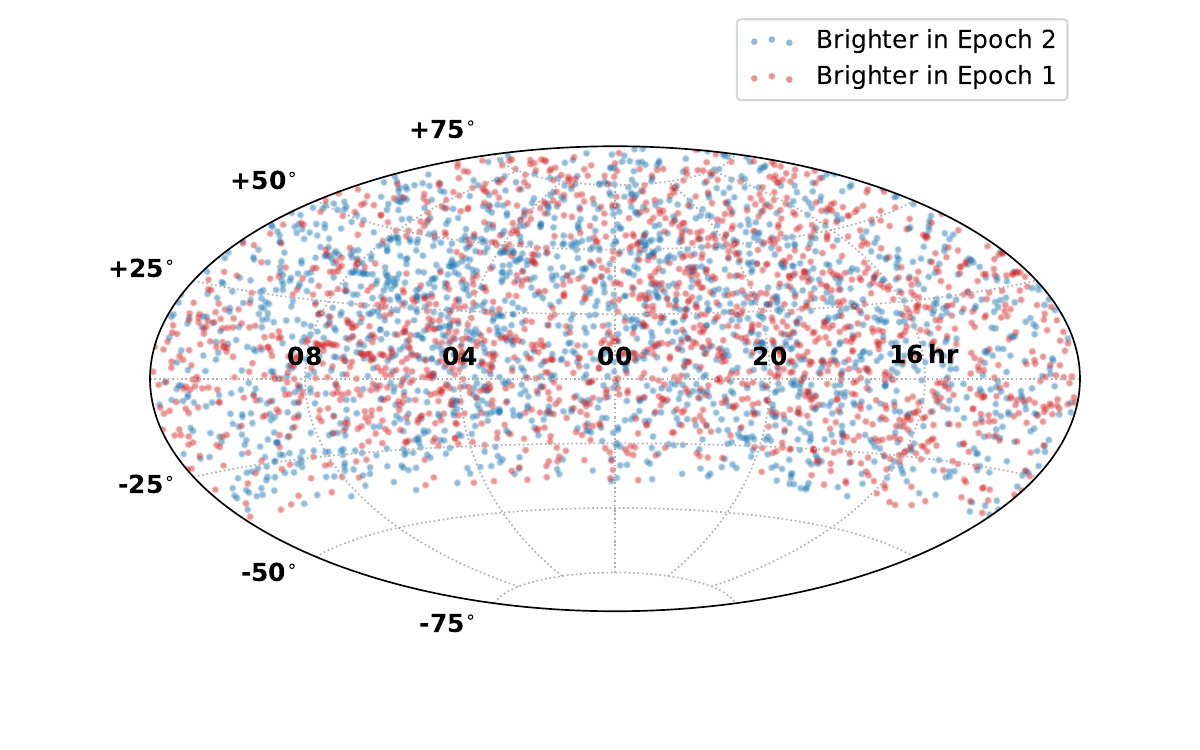}
    \caption{Aitoff projection showing the on-sky distribution of variable radio sources identified in this work.
    Red points show sources that are brighter in Epoch 1 than in Epoch 2, and blue points show sources that are brighter in Epoch 2 than in Epoch 1.}
    \label{fig:variable-aitoff}
\end{figure}

In Table \ref{tab:variables} we list the candidate variables identified in this work, their identifiers in the VLASS catalogs for Epoch 1 and Epoch 2, their variability statistics, and visual inspection flags.
Note that while the flux density measurements and variability statistics provided in Table \ref{tab:variables} account for the relative systematic flux biases between the two VLASS epochs as described in Section \ref{ssec:qlsystematics}, no external-to-VLASS flux calibration has been applied.
Care should therefore be taken when comparing the flux densities in Table \ref{tab:variables} to radio measurements not obtained from VLASS epochs 1 or 2.
The first 10 rows of Table \ref{tab:variables} are shown here as an example of the table format.
The full table of $4,124$ rows is available at \url{https://zenodo.org/records/18010746}, and will be made available on \href{https://cds.unistra.fr/}{CDS} after final publication of this article.
Throughout the rest of this work, we limit our analysis to the $3,618$ sources with $\text{\textit{f\_vi}}=0$, as this is the most robust selection of radio variables. 
In Figure \ref{fig:variable-aitoff} we show the distribution of these $3,618$ sources on an Aitoff projected sky map colored by whether they are brighter in Epoch 1 (red) or Epoch 2 (blue).
Notably, the distribution of variables appears homogeneous across the sky, indicating that the steps taken in Section \ref{ssec:qlsystematics} do a good job of mitigating potential selection biases due to systematics in the \textit{Quick Look} images. 

\section{Radio Variability Statistics}
\label{sec:variability-stats}

\subsection{Variable Sources Detected in Both Epochs}

\begin{figure}
    \centering
    \includegraphics[width=0.95\columnwidth]{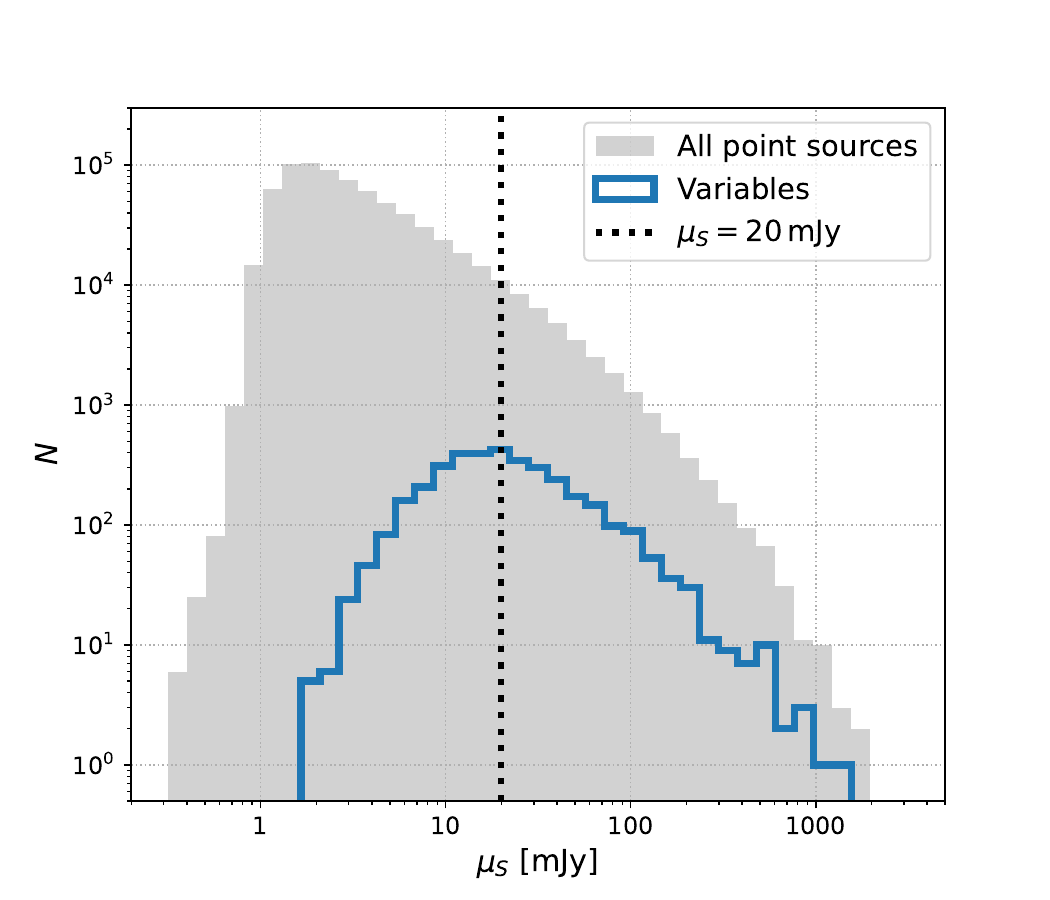}
    \caption{The distribution of the mean flux densities, $\mu_{S}$, for all point sources detected in both epochs (gray solid histogram) and those sources identified as variables (blue line). 
    The black dotted line shows $\mu_{S}=20\,$mJy, below which we lose completeness in our sample of variables as sources fade below the survey detection limits.}
    \label{fig:smeandist}
\end{figure}

Approximately $0.5\,$\% of the $\sim730,000$ point sources that are detected in the VLASS Epoch 1 and 2 catalogs show significant variability.
However, this number may not be truly representative of the fraction of variable sources.
At lower brightnesses, variables may drop below the sensitivity of VLASS, resulting in them not being detected in one of the epochs.
We assess this effect by comparing the distributions of mean flux density across Epochs 1 and 2, $\mu_{S}$, for all point sources with detections in both epochs and variable sources. 
From Figure \ref{fig:smeandist}, it is clear that the completeness of our variable source selection (blue line) decreases at $\mu_{S} \lesssim 20\,$mJy.
The number counts for point sources (gray solid histogram) increase steadily down to $\mu_{S}\sim 2\,$mJy, before dropping off sharply as sources fall below the survey noise floor.
For the variable sources the drop off is far more gradual as the number of sources detected in both epochs is determined by both the mean flux density and the scale of variability of the source.
The $\mu_{S}=20\,$mJy limit is broadly consistent with our expectations given data quality of VLASS.
Consider a model source detected in Epoch 1 at $2\pm0.4\,$mJy (a $S/N=5$ source at about the completeness limit for non-variable sources from Figure \ref{fig:smeandist}). 
Assuming the same noise level in Epoch 2, resulting in $\sigma_{\Delta S} = 0.57\,$mJy, then by combining Equations \ref{eq:vs} and \ref{eq:modidx}, $\mu_{S} \sim 2.2\,V_{S}$ at the variability selection limit of $|m|=0.26$.
The median value of $\sigma_{V_{s}, \text{Tile}}$ in our data is $3.03$, and requiring $V_{s}>4\sigma_{V_{s}, \text{Tile}}$ thus suggests that the variables we detect in this work are likely to have $\mu_{S}\gtrsim 26\,$mJy.
At $\mu_{S} > 20\,$mJy, $1,744$ objects, accounting for $4.9\pm0.1\,$\% of VLASS point sources, exhibit significant variability over two VLASS epochs.
The split of variable sources between those flaring and those fading is approximately equal, with $51\pm1\,$\% of variables being brighter in Epoch 2 than in Epoch 1.

\begin{figure}
    \centering
    \subfigure[]{\includegraphics[trim={0 0 0 0}, clip, width=0.95\columnwidth]{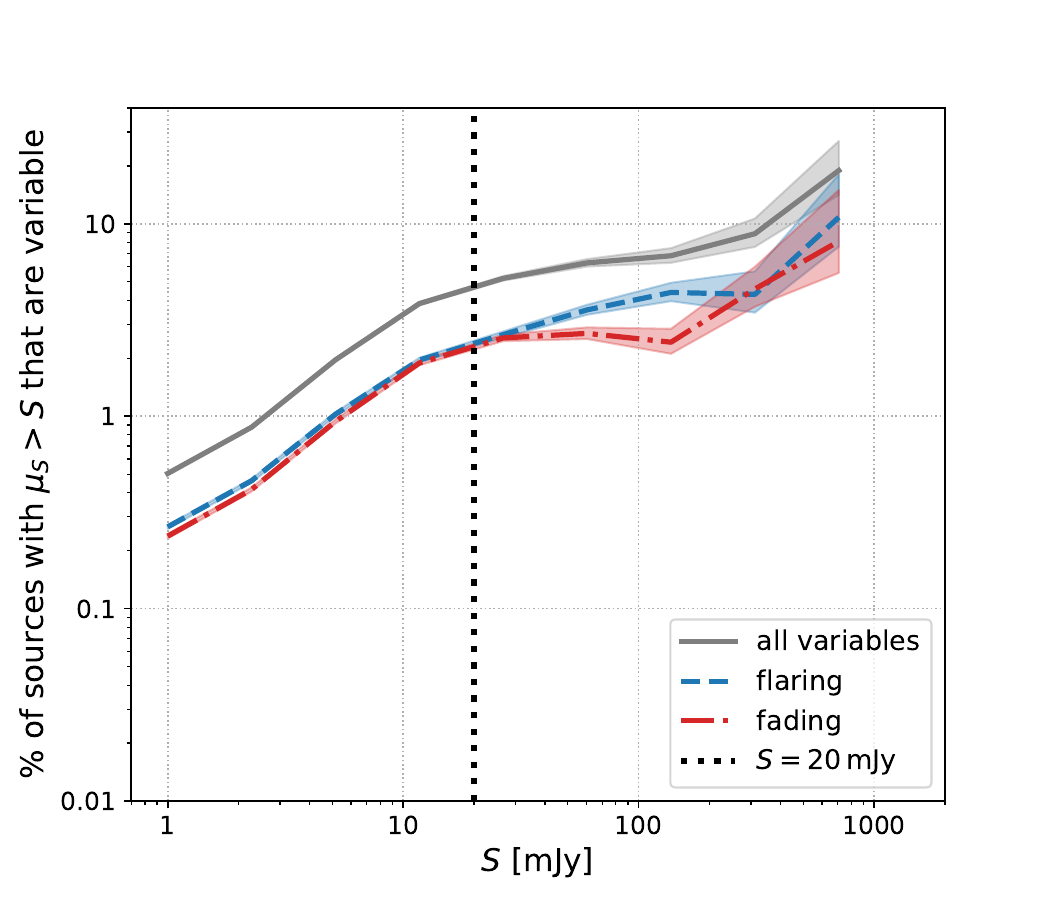}}
    \subfigure[]{\includegraphics[trim={0 0 0 1cm}, clip, width=0.95\columnwidth]{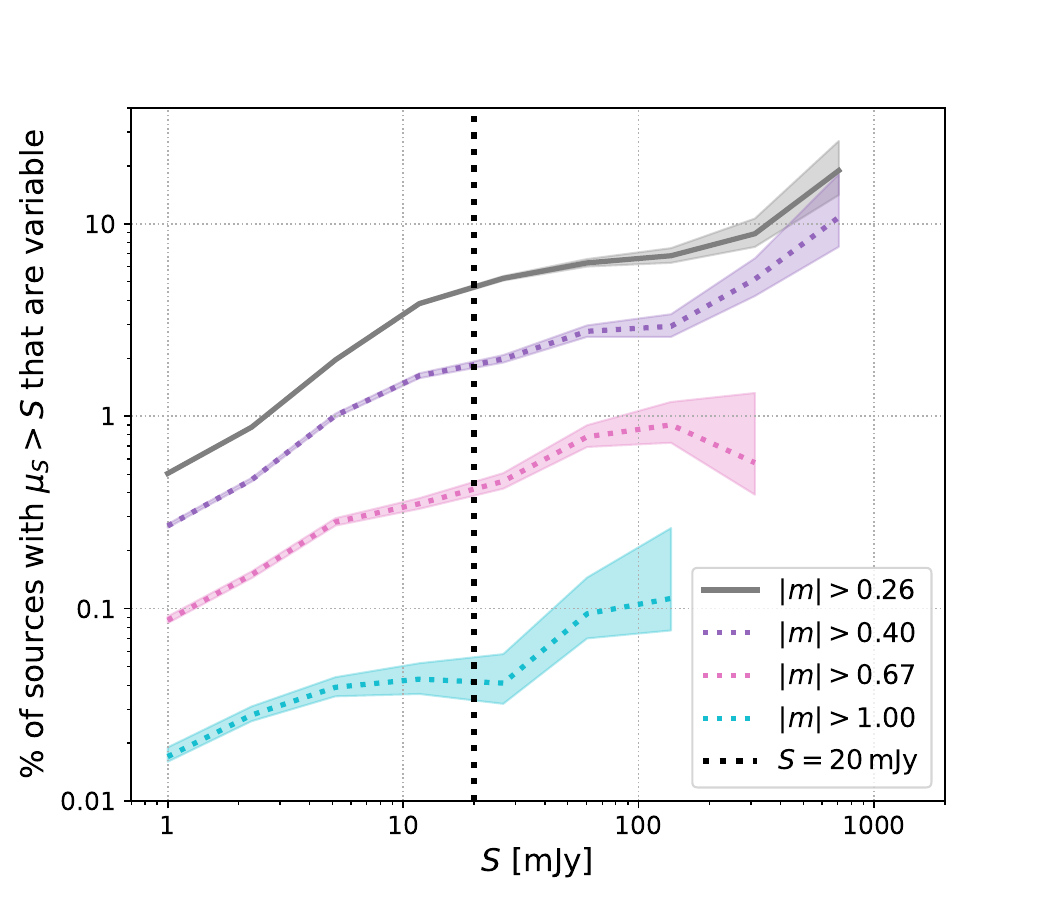}}
    \caption{The fraction of unresolved radio sources that are variables as a function of mean flux density.
    In both panels the grey solid line shows all variable sources identified in this work, and the black dotted line represents a flux density of $20\,$mJy below which we expect to have an incomplete variable selection.
    In Panel (a) the blue dashed and red dot-dashed lines show the fractions of sources that have brightened and faded respectively between Epochs 1 and 2 of VLASS.
    Panel (b) shows the percentage of sources that vary in brightness by at least $30\,$\% (grey solid line), $50\,$\% (purple dotted line), $100\,$\% (pink dotted line), and $200\,$\% (cyan dotted line). 
}
    \label{fig:varyfrac}
\end{figure}


In Figure \ref{fig:varyfrac}, we show the fraction of sources that are detected in both epochs and showing variability as a function of limiting mean flux density. 
Panel (a) splits variables by whether they are flaring (blue dashed line) or fading (red dot-dashed line).
Panel (b) shows variables split by the scale of their fractional variability, with the purple dotted line showing sources with $>50\,$\% fractional variability ($|m| > 0.4$), the pink dotted line showing those with $>100\,$\% fractional variability ($|m| > 0.67$) and the cyan dotted line representing sources with $>200\,$\% fractional variability ($|m| > 1$).
In both panels of Figure \ref{fig:varyfrac} the gray solid line represents all identified variables, and the vertical black dotted line shows $20\,$mJy, below which we know our variable sample lacks completeness.
Here, and throughout, the uncertainties on population fractions are estimated using a binomial $68\,$\% confidence interval \citep{Cameron2011}, shown in Figure \ref{fig:varyfrac} by the shaded regions around the lines.
In general, the fraction of variable source increases with source brightness, rising to $9.0_{-1.3}^{+1.7}\,$\% at $\mu_{S} > 300\,$mJy.
Moreover, this is a trend that shows no dependence on whether the source is flaring or fading, or on the scale of fractional variability.
Such a rise in the fraction of radio sources that are variable with increasing $\mu_{S}$ may be driven by changes in the make up of the radio source population.
For instance, blazars are generally bright radio sources due to Doppler boosting, a point we consider further in Section \ref{sssec:nontransient-variables}.

\begin{figure*}
    \centering
    \subfigure[]{\includegraphics[trim={0 0 0 0}, clip, width=0.95\columnwidth]{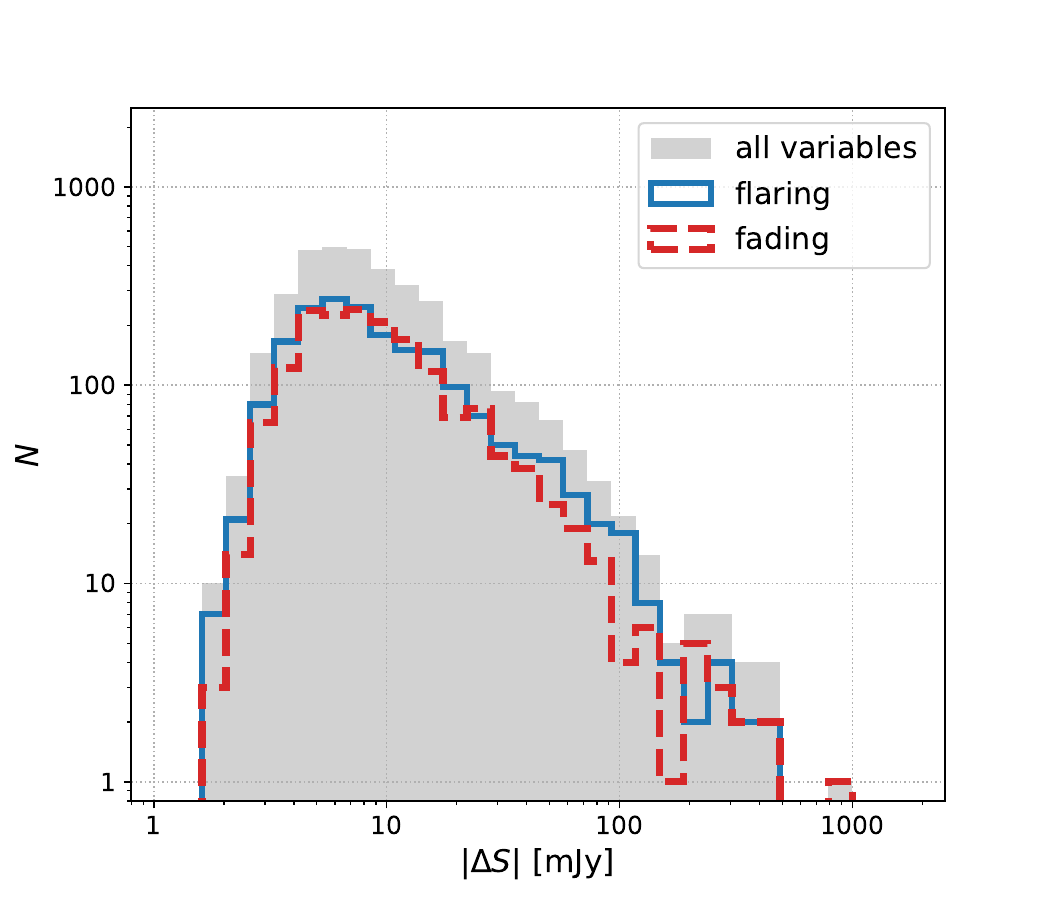}}
    \subfigure[]{\includegraphics[trim={0 0 0 0}, clip, width=0.95\columnwidth]{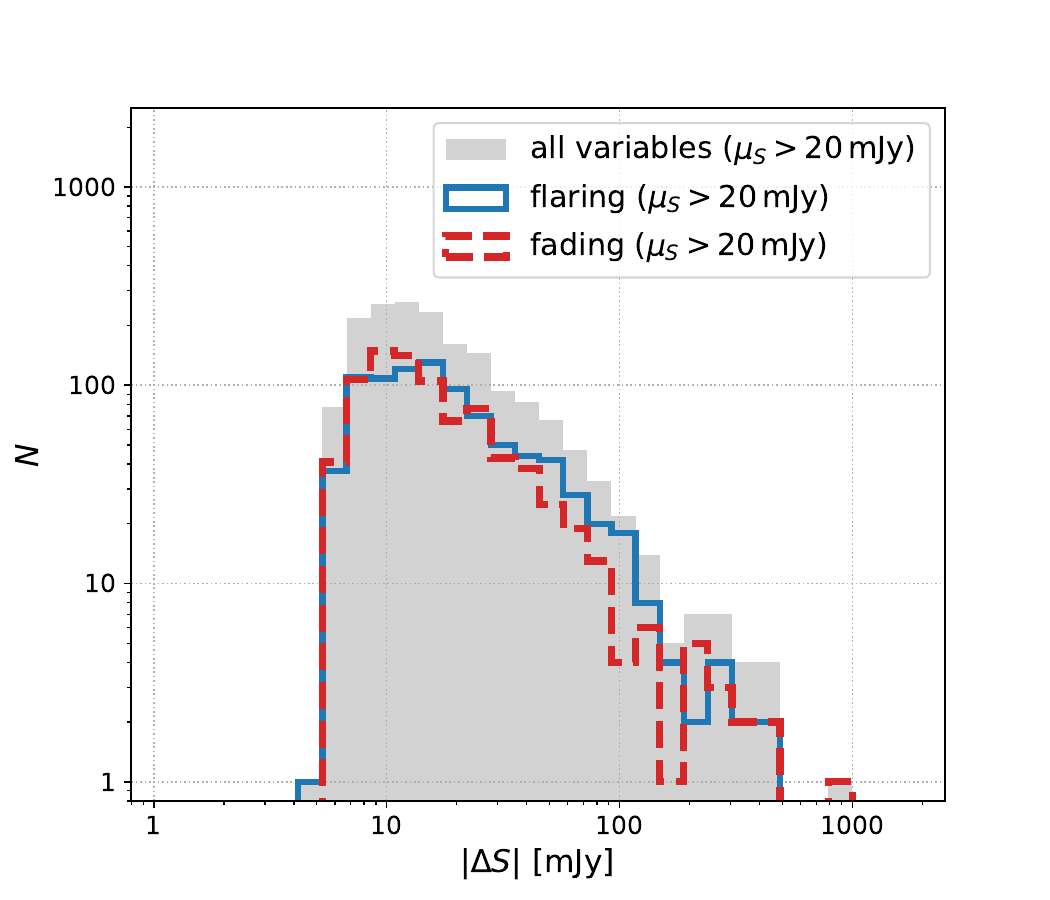}}
    \subfigure[]{\includegraphics[trim={0 0 0 1.5cm}, clip, width=0.95\columnwidth]{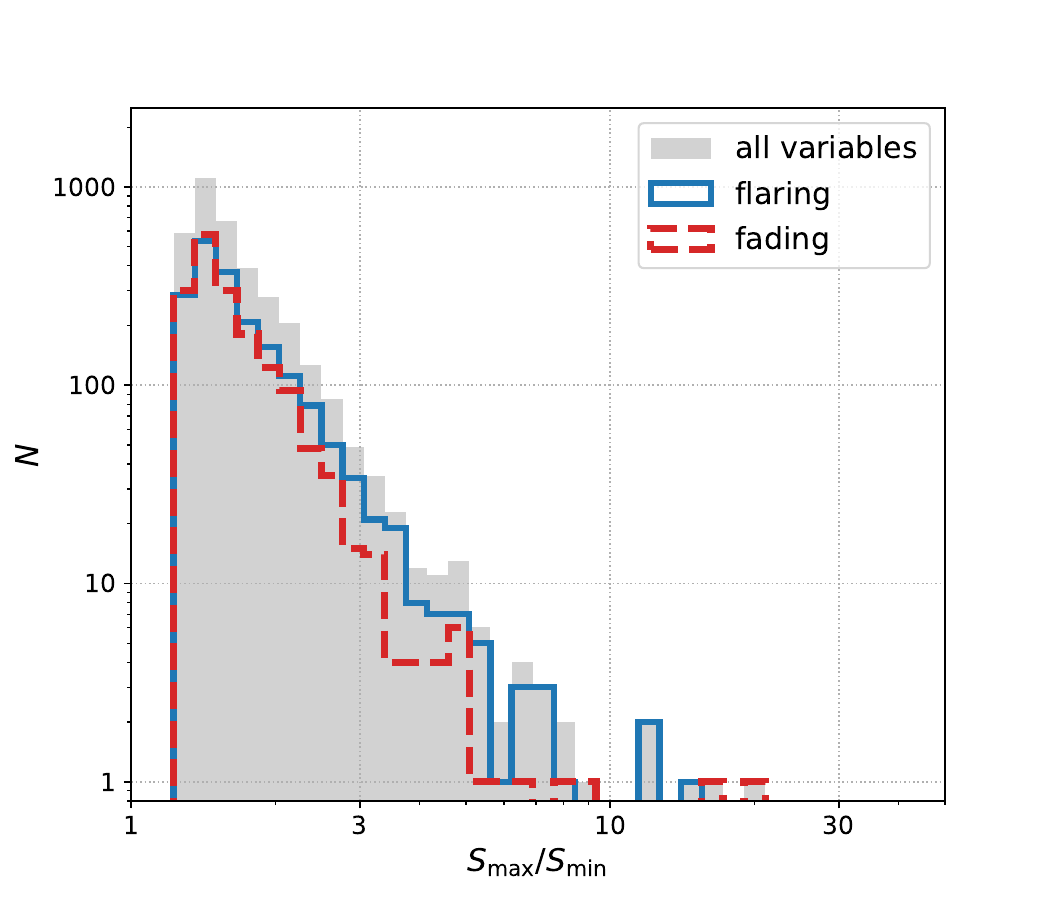}}
    \subfigure[]{\includegraphics[trim={0 0 0 1.5cm}, clip, width=0.95\columnwidth]{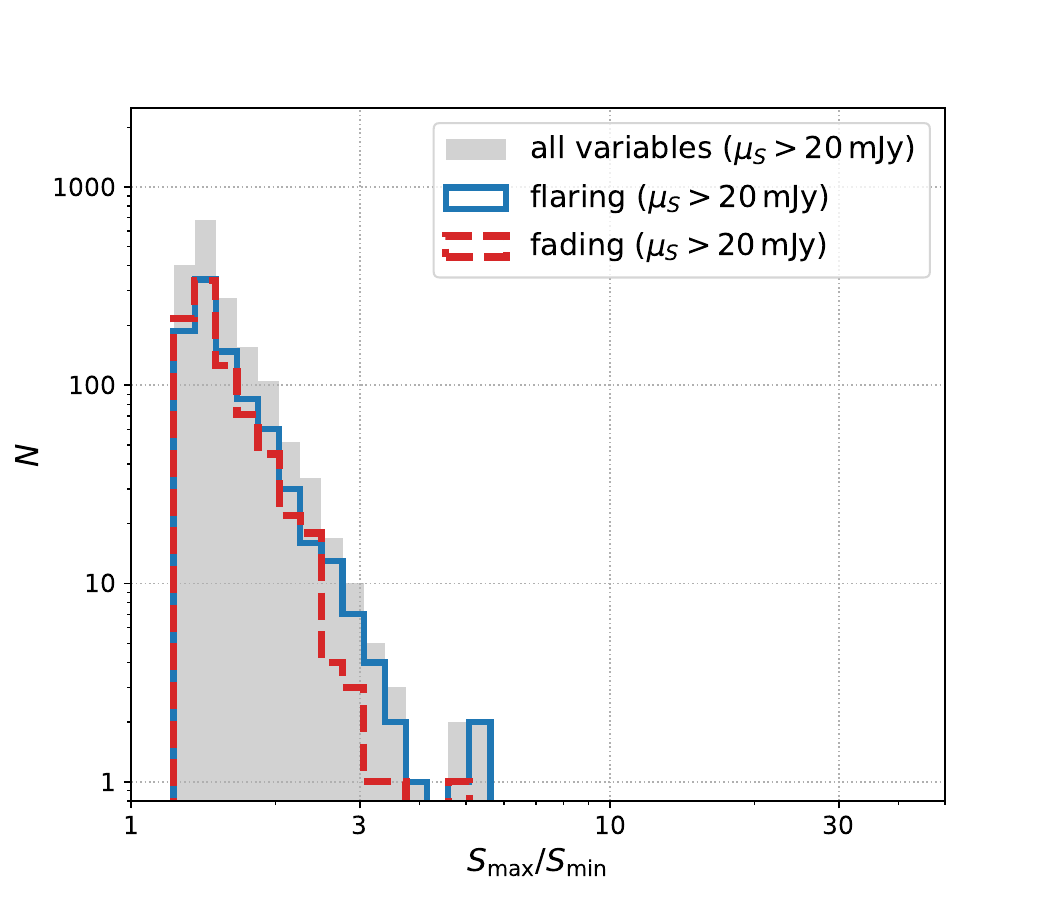}}
    \caption{Distributions of absolute change in flux density, $|\Delta S|$, and fractional change in flux density $S_{\text{max}}/S_{\text{min}}$, for our variable radio sources. Panels (b) and (d) show only those objects with $\mu_{S}>20\,$mJy, while panels (a) and (c) have no cut applied to the source mean brightness.
    In all panels the blue solid line shows flaring sources, the red dashed line shows fading sources, and the gray solid histogram represents all variable sources.}
    \label{fig:ds_srat_dists}
\end{figure*}

In Figure \ref{fig:ds_srat_dists}, we show the distributions of absolute change in brightness, $|\Delta S|$, for all variables (Panel a) and variables with $\mu_{S} > 20\,$mJy (Panel b), and the fractional change in brightness, $S_\text{max}/S_{\text{min}}$, for all variables (Panel c) and variables having $\mu_{S}>20\,$mJy (panel d).
In all panels the solid gray histogram gives the distribution for all variable radio sources, the blue solid line represents flaring sources, and the red dashed line shows fading sources.
When considering all variables without a flux density cut there is no significant difference in $|\Delta S|$ distributions of flaring and fading sources. 
However, for sources with a mean brightness above $20\,$mJy, fading sources appear to have smaller changes in brightness than flaring sources, having median $|\Delta S|$ values of $13.4\,$mJy and $15.6\,$mJy respectively.
Performing a Kolmogorov-Smirnov (KS) test on the $|\Delta S|$ distributions for flaring and fading sources with $\mu_{S}>20\,$mJy returns a $p$-value of $4\times10^{-4}$, suggesting a statistically significant difference in absolute changes in brightness for flaring and fading radio sources.

A weaker difference is seen in the distributions of fractional variability for these sources with the median $S_{\text{max}}/S_{\text{min}}$ being $1.45$ for flarers and $1.43$ for faders, and a KS-test giving $p=0.007$.
If the difference in $|\Delta S|$ distributions is in fact real, these differences likely only represent relatively small fractional variations in bright sources, which would be consistent with the more subtle differences seen in the $S_{\text{max}}/S_{\text{min}}$ for flaring and fading radio sources.
A plausible explanation for such differences between flaring and fading radio sources may be that the average rise time of the radio variability light curves in our sample is shorter than the typical fall time, i.e., radio sources brighten quicker than they fade.
Such light curve asymmetry is expected for, e.g., shocks in blazar jets \citep{Fromm2015}, and synchrotron transients \citep{Metzger2015}.

\subsection{The Missing Variable Radio Sources}

In this work we are focusing primarily on the variability of radio sources that have detections in both the Epoch 1 and Epoch 2 VLASS catalogs.
This approach will likely miss a large number of variable sources that are detected in one epoch, but not the other.
The existing cataloged information such as brightness of sources in the epoch they are detected in, and the typical noise in the epoch in which they are not detected, can be used to place constraints on the fraction of radio sources that are variable but only detected in one epoch.
Consider a toy model source that is detected in Epoch 2 at a given flux density, $S_{2}$, with uncertainty $\sigma_{S_{2}}$, that is not detected in Epoch 1.
We can estimate the expected image noise in Epoch 1 by taking the mean of the \texttt{Isl\_rms} values of sources detected within that Tile in the \text{Epoch 1} catalog as $\sigma_{S_{1}}$.
As the source detection threshold in the VLASS catalogs is $5\times$ the local rms noise, we would expect the flux density of the source in Epoch 1 to be $S_{1} < 5\sigma_{S_{1}}$ for the source not to be detected, and by making this assumption we can place lower limits on $|\Delta S|$, $V_{s}$, and $|m|$ for the source.

In each epoch, there are $\sim225,000$ sources that do not have a match in the other epoch, i.e., detected in Epoch 1 but not Epoch 2, or vice versa.
Of these, $\sim6,000$ in each epoch have $V_{s}> 4\sigma_{V_{s}\text{, Tile}}$ and $|m|>0.26$ based on the upper flux density limits required for a non-detection in the other epoch.
Assuming a similar selection accuracy of $88\,\%$ as determined in Section \ref{sssec:qa}, we expect $\sim 5,000$ non-detections in each epoch are likely variable sources.
In reality, many more of sources will be true variables, with their real flux density in the epoch in which they were not detected being fainter than $5\times$ the local noise that we assume here. 
Nonetheless, these numbers suggest that $\gtrsim1/50$ sources that are only detected in one VLASS epoch are likely variable, consistent with previous studies of the faint $3\,$GHz sky \citep{Mooley2016}.

The statistics of sources that are detected in one epoch but not the other can also be used to estimate a lower bound on the completeness of our flux-limited sample of variables.
Assuming again that the non-detected counterpart is at the survey sensitivity limit, non-detections that would satisfy our flux limit of $\mu_{S}>20\,$mJy are mostly found within the typical VLASS beam size ($2\arcsec.9$) of another VLASS source.
In such cases their seeming variability is typically due to a larger source being modeled differently by the source-finder in the two different images, e.g., being modeled as a single larger source in one epoch, but as two separate sources in the other epoch. 
These sources would likely be flagged on visual inspection (see Section \ref{sssec:qa}).
Considering only those non-detections that satisfy $\mu_{S}>20\,$mJy, and that are $>2\arcsec.9$ from another VLASS source, $\sim 140$ are found in each epoch.
Many of these $280$ objects will likely have true values of $\mu_{S}$ that are below $20\,$mJy, but can still be considered an upper limit on the number of sources missed by our flux-limited variable sample, suggesting that at $\mu_{S}>20\,$mJy we detect $>86\,$\% of radio sources that exhibit significant variability in VLASS between Epoch 1 and Epoch 2.


\section{Multiwavelength Properties of Variable Radio Sources}
\label{sec:multiwavelength}

Radio observations alone provide a limited amount of information about the objects exhibiting radio variability. 
Analyzing the multiwavelength properties of radio sources has the potential to provide insights into the type of object producing the radio emission and the mechanisms that may be responsible for the radio variability.
In this Section, we make use of multiwavelength archival data to characterize the populations of sources that exhibit radio variability.
The multiwavelength cross identifications that are used in this Section are made available in the supplementary data to this paper, as detailed in Appendix \ref{ap:supdata}.
Given the incompleteness of our radio variable selection at $\mu_{S}<20\,$mJy (see Section \ref{sec:variability-stats} and Figure \ref{fig:smeandist} for details), we limit our analyses to sources with $\mu_{S}>20\,$mJy when comparing radio variables to the parent population of radio sources, unless otherwise stated.
For brevity throughout the rest of this paper, we use the terminology \textit{flux-limited parent} sample to refer to the $35,781$ point radio sources with $\mu_{S}>20\,$mJy, and \textit{flux-limited variables} for the $1,744$ objects with $\mu_{S}>20\,$mJy that are also identified as being variable radio sources.

\subsection{Redshifts}
\label{ssec:redshifts}

Most radio sources are extragalactic in nature, and obtaining redshift measurements, $z$, for these allows us to quantify their distribution throughout the history of the Universe.
Moreover, it is interesting to consider whether variable radio sources exist at the same redshifts as the wider radio source population; if their redshift distributions differ it may provide insights into particular epochs of AGN activity for instance.
Using a $1\arcsec$ cross match radius, we obtain spectroscopic redshifts (spec-$z$s) for our radio sources from the 16th data release of the Sloan Digital Sky Survey \citep[SDSS,][]{York2000, Blanton2017, Ahumada2020} and photometric redshifts (photo-$z$s) from the DESI Legacy Surveys 8th data release \citep[LS DR8,][]{Dey2019, Duncan2022}.
SDSS provides spec-$z$s for galaxies and quasars down to $r\lesssim 22\,$mag across $\sim 30\,$\% of the VLASS footprint, while photo-$z$s from LS DR8 are available for sources down to $r\lesssim 24\,$mag for $\sim 55\,$\% of the VLASS footprint. 
To exclude highly uncertain redshift measurements, we only include redshifts where the uncertainty on the redshift, $\sigma_{z}$, satisfies $\sigma_{z}/(1+z) < 0.2$.

\begin{figure}
    \centering
    \includegraphics[width=\columnwidth]{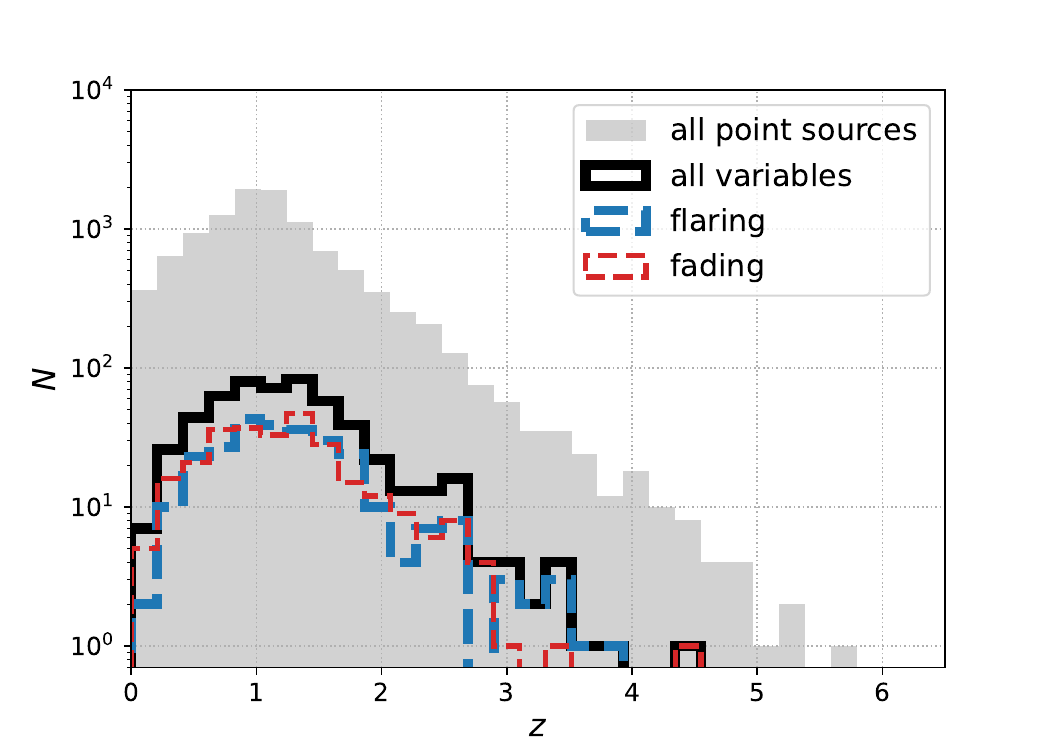}
    \caption{Distribution of the redshifts ($z$) found in SDSS and LS-DR8 for point radio sources (gray solid histogram), variables (black line), flaring variables (blue dashed line), and fading variables (red dashed line).
    Only radio sources with $\mu_{S}>20\,$mJy are included on this plot.
    }
    \label{fig:zdist}
\end{figure}

In Figure \ref{fig:zdist}, we show the redshift distributions for our flux-limited parent sample (gray solid histogram; $N\sim11,000$), and flux-limited variables sample (thick black line; $N\sim 500$).
For the variables, we also highlight the redshift distribution of flaring sources (blue dashed line), and fading sources (thin red line) in Figure \ref{fig:zdist}.
The shapes of all these distributions are qualitatively similar\textemdash the only differences being in the size of the samples\textemdash suggesting that variable radio sources are found at same times throughout the history of the Universe as non-variable radio sources.
The distributions shown in Figure \ref{fig:zdist} include both spec-$z$s and photo-$z$s so as to maximize the sample size.
The majority ($86\,$\%) of the redshifts we obtain are photometric, but the results do not change if the spec-$z$s or photo-$z$s are considered independently, with similar distributions for all populations within these subsamples.

\subsection{Infrared Colors}
\label{ssec:wise}

\begin{figure}
    \centering
    \subfigure[]{\includegraphics[trim={0 0 0 0}, clip, width=\columnwidth]{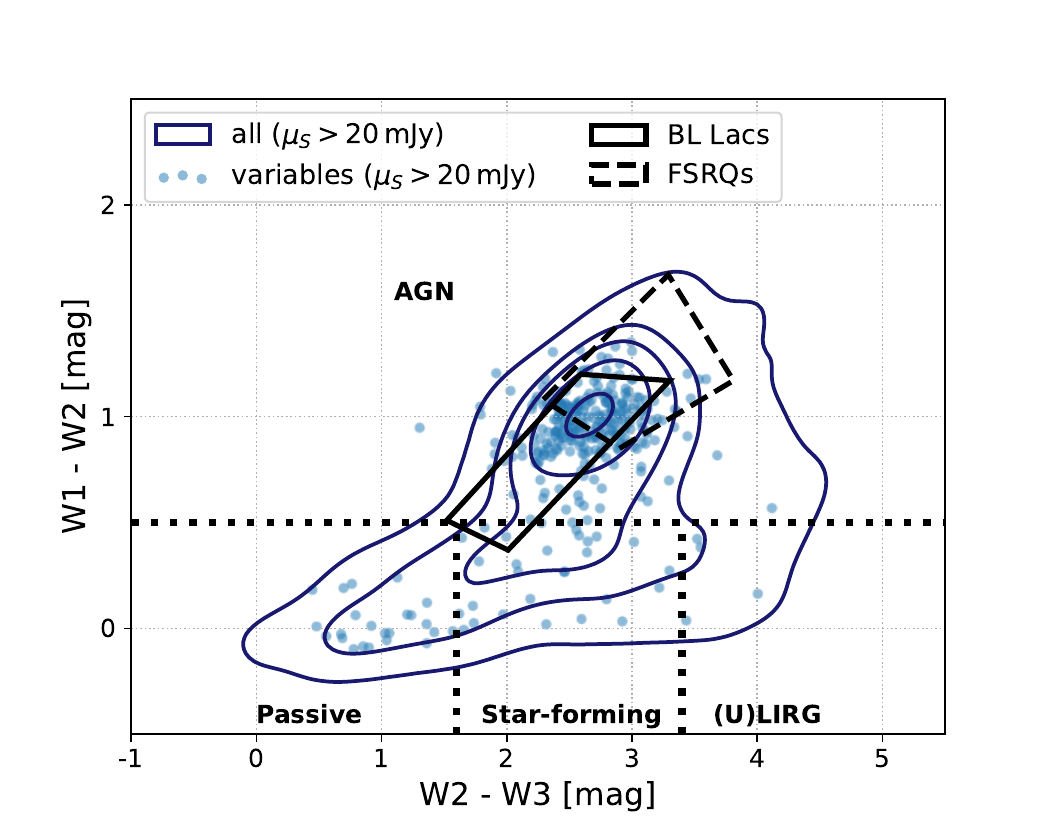}}
    \subfigure[]{\includegraphics[trim={0 0 0 0}, clip, width=\columnwidth]{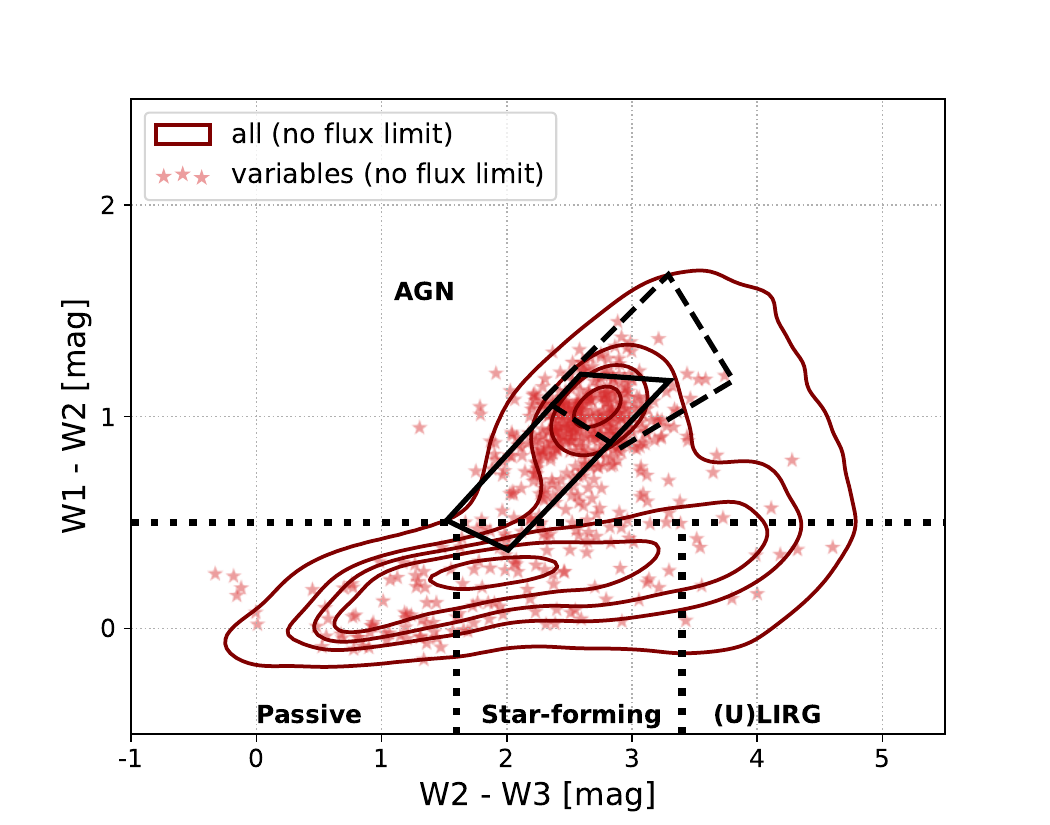}}
    \caption{WISE color/color diagrams for our radio sources with $\mu_{S}>20\,$mJy (a), and without a lower flux limit (b).
    In both panels the variables are shown by scatter points (blue circles in panel a, and red stars in panel b), while the contour lines enclose $5\,$\%, $25\,$\%, $50\,$\%, $75\,$\%, and $95\,$\% of the parent samples of point-like VLASS sources.
    The AGN, passive, star-forming, and (U)LIRG populations are annotated and separated by black dotted lines, while the regions occupied by BL Lacs and FSRQs are shown by black solid and black dashed lines respectively.
    }
    \label{fig:wise-colors}
\end{figure}

Extragalactic radio sources are often relatively bright at infrared (IR) wavelengths, usually resulting from either active star-formation or an AGN \citep[e.g.,][]{vanderKruit1971, Condon1992, Jarrett2011}, and the IR colors of radio sources are frequently used to identify which type of object they are \citep[e.g.,][]{Banfield2015, Norris2021, Gordon2023dragns, Hardcastle2023, Wong2025}.
We cross match our sources to the AllWISE catalog \citep{Cutri2014}, which covers the entire sky down to $\text{W}1 (3.4\,\mu\text{m}) \lesssim 17\,$mag based on observations from the Wide Field Survey Explorer mid-infrared telescope \citep[WISE,][]{Wright2010}, finding an AllWISE counterpart to $\sim410,000$ ($56\,$\%) of our $\sim730,000$ radio sources.
The WISE color/color diagram compares the $\text{W}2(4.3\,\mu\text{m})-\text{W}3(12\,\mu\text{m})$ color of a source to its $\text{W}1 - \text{W}2$ color (e.g., see Figure \ref{fig:wise-colors}), and we therefore limit our analysis to sources that have high quality measurements in all three of these bands.
Explicitly, we require that the source is detected with $S/N > 2$, is not a lower limit on the magnitude, and has no image quality flags.
Furthermore, the WISE colors of sources at high redshift ($z>1$) are less reliable as the WISE bands start to trace different regions of the galaxy spectral energy distribution \citep{Assef2013}.
To account for this we only include sources with $z<1$ where we have a reliable redshift measurement (see also Section \ref{ssec:redshifts}).
For sources where we don't have a redshift measurement, we only include those with $\text{W}1 < 14\,$mag, as $99\,$\% of our sources with $\text{W}1 < 14\,$mag and a redshift measurement have $z<1$.
This leaves us with $54,344$ sources, of which $570$ are variable in VLASS.

In Figure \ref{fig:wise-colors}a, we show the WISE color/color diagram for our objects in our flux-limited variables sample (blue circles) and flux-limited parent sample (dark blue contour lines). 
Taking the criteria of \citet{Mingo2016}, galaxies with $\text{W}1-\text{W}2 > 0.5\,$mag usually host AGN.
For galaxies with $\text{W}1-\text{W}2 < 0.5\,$mag, they are considered to be passive (not actively forming stars) if $\text{W}2-\text{W}3 < 1.7\,$mag, star-forming if $1.7 < \text{W}2-\text{W}3 < 3.4\,$mag, and experiencing a rapid burst of star-formation associated with (ultra) luminous infrared galaxies ((U)LIRGs) if $\text{W}2-\text{W}3 > 3.4\,$mag.
AGN-like colors dominate both samples but are more common among the variables than the parent sample, making up $83\pm2\,$\% and $62\pm1\,$\% of these populations respectively.
The difference between variables and the wider population can mostly be attributed to a lower fraction of star-forming galaxies (SFGs) among the variables, where only $9_{-1}^{+2}\,$\% have WISE colors consistent with SFGs compared to $21\pm1\,$\% of the wider radio source population.
Other IR populations only contribute a minority of the flux-limited variable and flux-limited parent samples, and a full breakdown is given in Table \ref{tab:irpops}.
The uncertainties in the percentages presented here and in Table \ref{tab:irpops} are estimated using the $68\,$\% confidence interval for binomial population proportions as laid out in \citet{Cameron2011}.

\begin{deluxetable}{rcc}
    \tabletypesize{\scriptsize}
    \tablecaption{The distribution of different types of object among our flux-limited variables and flux-limited parent samples of VLASS sources.
    \label{tab:irpops}}
    \tablehead{
    \colhead{IR Population} & \colhead{Flux-limited variables} & \colhead{Flux-limited parent}\\
    \colhead{(1)} & \colhead{(2)} & \colhead{(3)}}
    \startdata
        Passive [\%] & $7.3_{-1.2}^{+1.7}$ & $12.2_{-0.5}^{+0.6}$\\
        SFGs [\%] & $8.8_{-1.3}^{+1.9}$ & $20.9_{-0.6}^{+0.7}$\\
        (U)LIRGs [\%] & $1.3_{-0.4}^{+1.0}$ & $4.7_{-0.3}^{+0.4}$\\
        AGN [\%] & $82.6_{-2.3}^{+1.9}$ & $62.2\pm0.8$\vspace{0.5em}\\
        \hline
        \vspace{-0.5em}\\
        Blazars [\%] & $67.5_{-2.7}^{+2.5}$ & $41.6\pm0.8$\\
        BL Lacs [\%] & $55.8_{-2.8}^{+2.7}$ & $31.2_{-0.7}^{+0.8}$\\
        FSRQs [\%] & $40.4_{-2.7}^{+2.8}$ & $25.3\pm0.7$
    \enddata
    \tablecomments{IR population based on WISE colors (1) for variables with $\mu_{S}>20\,$mJy (2), and all point sources (including non-variables) with $\mu_{S}>20\,$mJy (3).
    Due to the overlap in IR color distributions exhibited by BL Lacs and FSRQs, many blazars have colors consistent with both these populations and are counted in both the BL Lac and FSRQ rows here. 
    }
\end{deluxetable}

Among AGN, blazars in particular are well known to exhibit strong radio variability \citep[e.g.][]{Hufnagel1992, Urry1995, Hovatta2019}.
To investigate whether our variable radio sources are predominantly blazars, we make use of the WISE $\gamma$-ray strip of the WISE color/color diagram as defined by \citet{Massaro2012}.
This `strip' consists of two overlapping regions that contain known, $\gamma$-ray selected, BL Lacertae objects (BL Lacs) and flat-spectrum radio quasars (FSRQs). 
We show the region bounding the BL Lac population as a black solid line and the FSRQ region as a black dashed line in Figure \ref{fig:wise-colors}.
For our flux-limited variables, $68\pm3\,$\% of sources with WISE colors have WISE colors consistent with blazars, $56\pm3\,$\% similar to BL Lacs and $40\pm3\,$\% consistent with FSRQs\footnote{Due to the overlap in the BL Lac and FSRQ regions, sources\\can have WISE colors consistent with both types of blazar.}.
This is significantly higher than for the flux-limited parent population where only $42\pm1\,$\% have blazar-like colors ($31\pm1\,$\% BL Lacs, $25\pm1\,$\% FSRQs).
These results (also shown in Table \ref{tab:irpops}) suggest that the majority of radio sources that exhibit variability at $\nu\sim3\,$GHz on timescales of $\sim 2.5\,$ years may be blazars, highlighting the potential for deep radio-time domain sky surveys to identify faint blazars, enabling the study of these AGN at low-luminosities and/or high redshifts.
While selecting variable radio sources preferentially samples blazars compared to the wider radio population, this doesn't appear to be driven by selecting a particular blazar subpopulation, with BL Lacs and FSRQs making up similar proportions of the WISE blazars in both of the radio selections presented here.
Considering the flaring and fading sources separately from one another has no substantive impact on the results presented here.

In addition to extragalactic sources, the WISE color/color diagram can be used to identify stars within the Milky Way, which often have $\text{W}1-\text{W}2 \sim \text{W}2-\text{W}3 \sim 0$ \citep{Nikutta2014}.
To investigate the potential of VLASS variability as an aid for identifying galactic radio sources, we plot the WISE color/color diagram without the $\mu_{S}>20\,$mJy flux limit in Figure \ref{fig:wise-colors}b.
A distinct grouping of $6$ variable radio sources is present at $\text{W}1-\text{W}2 \sim \text{W}2-\text{W}3 \sim 0$, seemingly separated from the rest of the passive galaxy population.
None of these sources have redshift measurements, and the IR associations for all of these sources are bright ($4.5\,\text{mag} < \text{W}1 < 6.9\,\text{mag}$), suggesting that these objects are radio variable stars.
Indeed, all of these objects are classified as stars in the SIMBAD database \citep{Wenger2000}, and we list these six objects with their SIMBAD identifiers and classification in Table \ref{tab:candidate-stars}.
Five of these six objects are known to be radio stars, appearing in the Sydney Radio Stars Catalog \citep{Driessen2024}.
Only one of our candidate radio-variable stars, BD$+$18 1512, is not a previously known radio star.
When querying the CDS VizieR service \citep{Ochsenbein2000}, the only radio catalogs that list detections for BD$+$18 1512 are VLASS and RACS.
In VLASS Epoch 1, BD$+$18 1512 had a flux density of $S_{3\,\text{GHz}}=9.13\pm0.29\,$mJy observed on March 30, 2019, before fading to $S_{3\,\text{GHz}}=0.98\pm0.26\,$mJy in VLASS Epoch 2, observed on November 21, 2021.
In RACS, BD$+$18 1512 was observed to have $S_{1.4\,\text{GHz}}=2.4\pm0.6\,$mJy on January 13, 2021 \citep{Duchesne2024}.
The absence of this object in earlier radio catalogs including NVSS and past RACS epochs may be due to a combination of its intrinsic variability and the more limited sensitivity of previous generations of radio sky surveys. 
The case of BD$+$18 1512 thus highlights how radio variability surveys can help identify radio stars.

\begin{deluxetable}{lccc}
    \tabletypesize{\scriptsize}
    \tablecaption{Candidate radio-variable stars based on WISE colors \label{tab:candidate-stars}}
    \tablehead{
    \colhead{Name} & \colhead{VLASS1QLCIR} & \colhead{W1} & \colhead{Type}\\
    \colhead{} & \colhead{} & \colhead{[mag]} & \colhead{}\\
    \colhead{(1)} & \colhead{(2)} & \colhead{(3)} & \colhead{(4)}}
    \startdata
        HD 9770C & J$013501.24-295435.2$ & $4.52\pm0.25$ & HighPM$*$\\
        BD$+$18 1512 & J$071011.01+182620.3$ & $6.94\pm0.06$ & HighPM$*$\\
        V$*$ Fl Cnc & J$083217.31+291909.0$ & $4.97\pm0.22$ & RotV$*$\\
        V$*$ DM UMa & J$105543.44+602809.5$ & $6.34\pm0.08$ & RS CVn V$*$\\
        HD 115247 & J$131602.99-054008.2$ & $5.42\pm0.16$ & HighPM$*$\\
        HD 192785 & J$201449.07+450142.6$ & $4.96\pm0.22$ & SB$*$
    \enddata
    \tablecomments{Main object name in SIMBAD (1), identifier in the VLASS Epoch 1 catalog (2), magnitude in the WISE $\text{W}1$-band (3), object type from SIMBAD (4; see also \url{https://www.ivoa.net/rdf/object-type/2020-10-06/object-type.html}).}
\end{deluxetable}

\subsection{$\gamma$-ray Counterparts}
\label{ssec:yrays}

\subsubsection{Fermi Large Area Telescope Detections}

Blazars are known to be highly variable producers of non-thermal radiation across the electromagnetic spectrum \citep{Urry1995}.
Moreover, with the axis of the jet aligned toward Earth, blazars often present a relatively unobscured view of the AGN central engine and inner jet that is enhanced by Doppler boosting, allowing better opportunities to observe high-energy emission from the AGN.
Indeed, blazars are a dominant source of extragalactic $\gamma$-rays \citep{DiMauro2018}.
Given the prevalence of blazar-like IR colors among our radio variables, we search for $\gamma$-ray counterparts to our radio sources in the fourth Fermi Large Area Telescope catalog \citep[4FGL,][]{Abdollahi2020}.
The 4FGL catalog covers the entire sky over the energy range $50\,\text{MeV} < E < 1\,\text{TeV}$, and is sensitive to sources with $S_{1-100\,\text{GeV}}\gtrsim 10^{-10}\,\text{cm}^{-2}\,\text{s}^{-1}$.
We use the fourth data release (DR4) of 4FGL here, which contains measurements for $7,195$ sources from $14\,$years of observations \citep{Abdollahi2022, Ballet2023}.

\begin{figure}
    \centering
    \subfigure[]{\includegraphics[trim={0 0 0 0}, clip, width=\columnwidth]{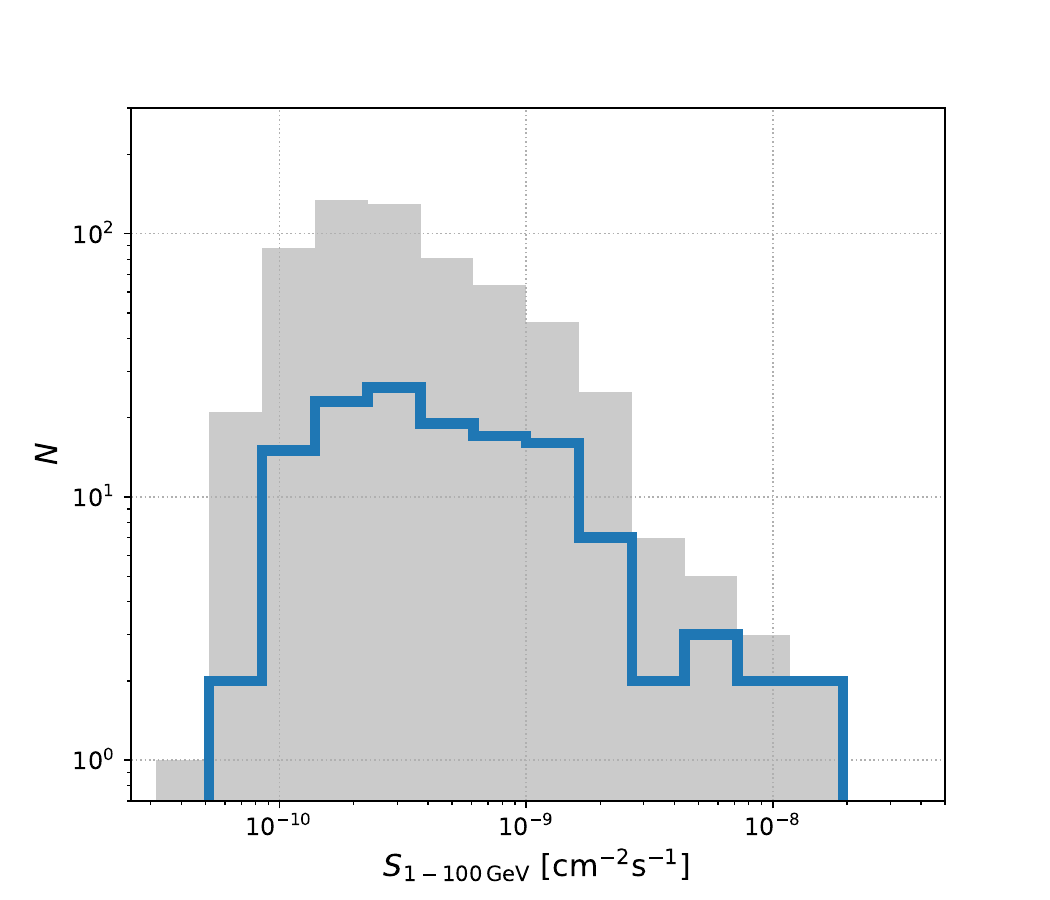}}
    \subfigure[]{\includegraphics[trim={0 0 0 1.6cm}, clip, width=\columnwidth]{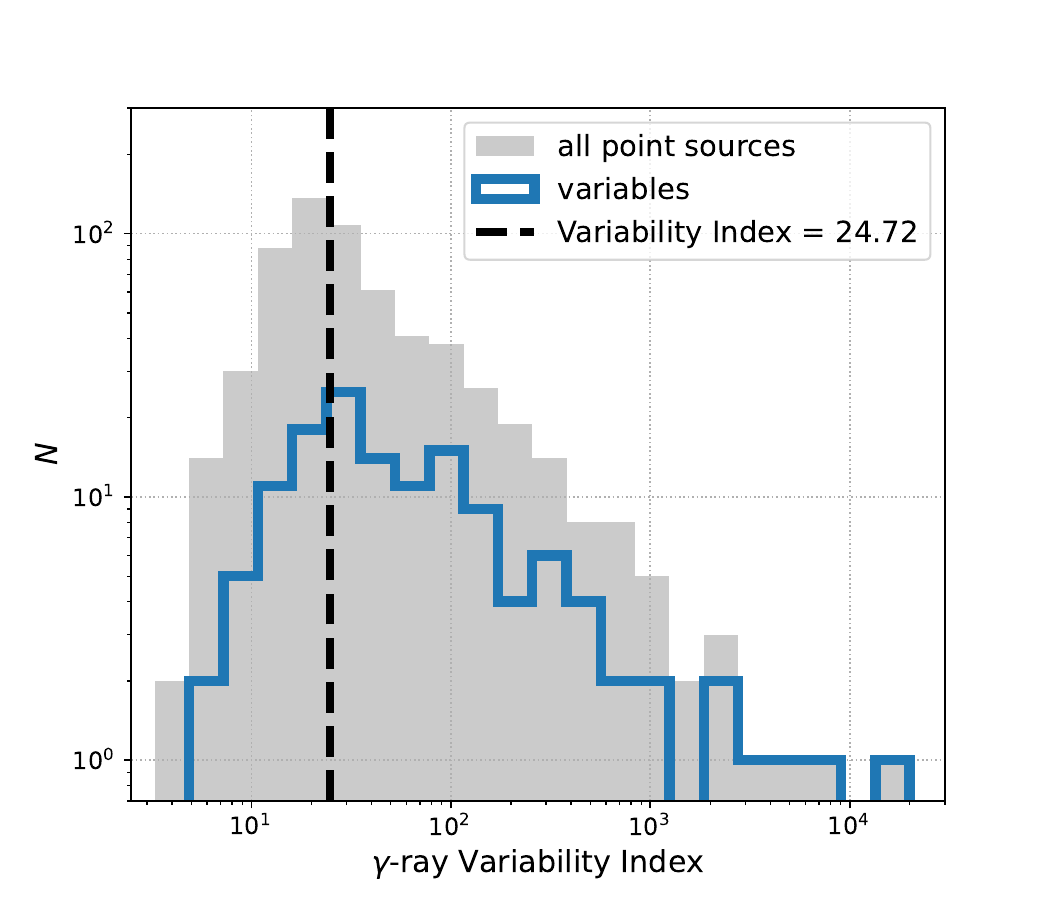}}
    \caption{Distributions of the $1\,\text{GeV}<E<100\,\text{GeV}$ integrated flux density (a) and the $\gamma$-ray variability index (b) of 4FGL counterparts to VLASS sources.
    Only sources with $\mu_{S}>20\,$mJy are included in this Figure, and in both panels the filled gray histogram shows all VLASS point sources while the solid blue line shows variable radio sources.
    The black dashed line in panel (b) shows $\text{Variability Index}=24.72$, above which 4FGL sources are considered to be variable.}
    \label{fig:yray-properties}
\end{figure}

In comparison to the radio, optical, and infrared data used so far in this paper, the astrometric precision of the $\gamma$-ray sources is relatively low.
The cataloged 4FGL sources are constrained to $95\,$\% confidence error ellipses, which have a median semi-major axis of $4'.3$, and range between $0.'4$ and $52'$.
Our radio sources are matched to 4FGL
sources should they lie within the $95\,$\% confidence positional error ellipse.
Furthermore, we only consider $\gamma$-ray sources with measurements considered to be reliable by requiring that \texttt{Flags} $=0$ in the 4FGL catalog.
We find 4FGL matches for $607$ ($2\,$\%) of our flux-limited parent sample, and $134$ ($8\,$\%) of our flux-limited variables sample.
The fraction of false positive associations is estimated by shifting the declinations of the radio sources by half a degree then redoing the cross match, and is found to be $7\,$\% for the variables, and $14\,$\% for the larger sample of radio point sources (due to the higher on sky density of non-variable radio sources).
In Figure \ref{fig:yray-properties}a we show the distributions of $1\,\text{GeV}-100\,\text{GeV}$ fluxes for the 4FGL sources matched to our VLASS sources.
The gray histogram shows our flux-limited parent sample while the flux-limited variables sample is shown by the blue solid line.
Performing a KS test shows no significant differences between these two distributions ($p=0.04$), suggesting that the radio variability of a source is likely not correlated with its $\gamma$-ray flux.

Previous works have demonstrated that radio variability in blazars is correlated with $\gamma$-ray variability \citep[e.g.,][]{Rani2013, Richards2014, MaxMoerbeck2014}.
As such, we might expect our radio-variable sources to also show variability at the high energies observed by Fermi.
The 4FGL catalog includes a measure of source variability (the \texttt{Variability\_Index} column), with 4FGL sources having \texttt{Variability\_Index} $>24.72$ considered to exhibit significant variability in their $\gamma$-ray brightness \citep{Abdollahi2022}.
In panel (b) of Figure \ref{fig:yray-properties}, we show the distributions of the 4FGL \texttt{Variability\_Index} for our samples of VLASS sources, using the same color scheme as in panel (a).
A KS test shows that these distributions are likely significantly different, with a $p$-value of $8\times10^{-4}$.
The radio-variable sources typically have a higher $\gamma$-ray \texttt{Variability\_Index} with a median value of $42.88$, compared to a median for VLASS point sources of $26.59$.
Of the $134$ VLASS variables with a Fermi counterpart, $96$ ($72\pm4\,$\%) have \texttt{Variability\_Index} $>24.72$ and are considered to be variable in $\gamma$-rays, compared to $328$ ($54\pm2\,$\%) of the VLASS point source parent sample.
These results suggest that radio-variable sources in VLASS a) are more likely to have a $\gamma$-ray detection than the wider radio population, b) trace a similar $\gamma$-ray brightness distribution to the non-variable radio sources that are detected in $\gamma$-rays, and c) are more likely to be variable in $\gamma$-rays than sources that don't exhibit radio variability.

\subsubsection{Missing $\gamma$-rays from Sources at High Redshift}

\begin{figure}
    \centering
    \includegraphics[width=\columnwidth]{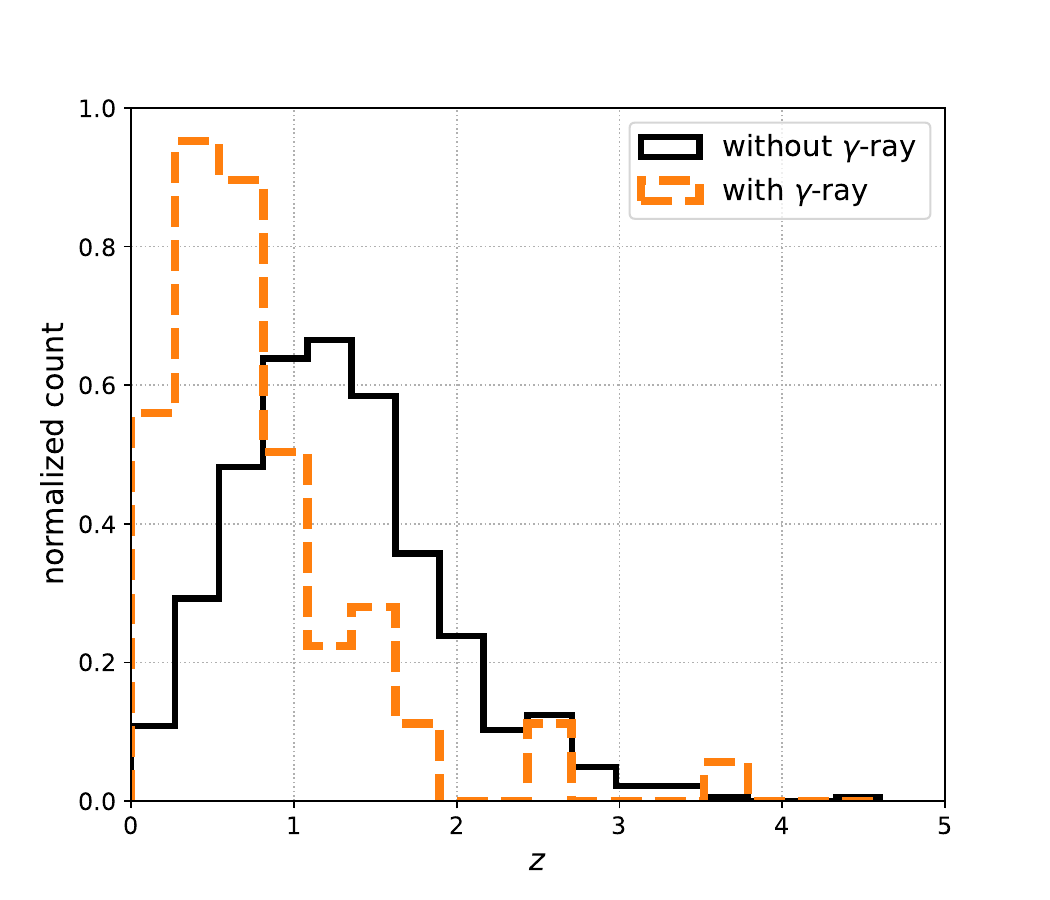}
    \caption{Normalized distributions of redshifts (when available) for variable radio sources at $\mu_{S}>20\,$mJy with (orange dashed line) and without (black solid line) a $\gamma$-ray counterpart in the 4FGL catalog.}
    \label{fig:yray_redshifts}
\end{figure}

In Section \ref{ssec:wise} we showed that approximately $2/3$ of the radio variable sources at $\mu_{S}>20\,$mJy have IR colors consistent with blazars.
Interestingly, $\lesssim 1/12$ radio variable sources have a $\gamma$-ray counterpart in 4FGL, suggesting that the majority of blazars are not detected by the current generation of wide-area $\gamma$-ray surveys.
In the \textit{Roma-BZCat}, a multi-frequency catalog of $>3,500$ blazars, $28\,\%$ of blazars have $\gamma$-ray detections from Fermi \citep{Massaro2009, Massaro2015}.
This is a higher fraction of blazars with $\gamma$-ray detections than found in this work, and is likely due to our sampling fainter radio sources than in \textit{Roma-BZCat}.
The median $\mu_{S}$ of our flux-limited variables sample is $37\,$mJy, while the median $1.4\,$GHz flux density of the \textit{Roma-BZCat} blazars is $180\,$mJy.
A likely explanation is that many of the variable radio sources we identify are simply too faint to be detected by Fermi, either as a result of intrinsically low luminosity or because they are at too high a redshift.
In Figure \ref{fig:yray_redshifts}, we show the redshift distributions for our flux-limited variables with (orange dashed line) and without (black solid line) a 4FGL counterpart.
It is clear that those sources in 4FGL are typically found at lower redshifts than those not in 4FGL.
The median redshift of radio variables with a $\gamma$-ray detection is $0.64$, compared to $1.22$ for those without.
For reference, the median redshift of \textit{Roma BZCat} sources is $0.93$.
Considering only flux-limited variables at $z<1$ ($N=301$), we find $53$ ($18\,$\%) have a 4FGL counterpart.
If we were to assume that $68\,$\% of our radio variables are blazars based on the WISE color distribution of our sample (see Section \ref{ssec:wise}), then $18\,$\% of radio variables having a 4FGL counterpart at $z<1$ could be accounted for by $\gamma$-ray emission from $\sim26\,$\% of blazars, similar to the \textit{Roma BZCat} numbers.

\section{Discussion} 
\label{sec:discussion}

\subsection{The Sources Powering the Variable Radio Sky}
\label{ssec:variableobjecttypes}
\subsubsection{Persistent Sources Showing Radio Variability}
\label{sssec:nontransient-variables}


The flux-limited variable radio sources identified in this work account for a total of $\sum |\Delta S| = 46.0\,$Jy change in flux density between Epoch 1 and Epoch 2 of VLASS.
Normalizing by the $34,000\,\text{deg}^{2}$ footprint of the survey and the $977\,$day mean time between observations, VLASS detects $1.7\,\text{Jy}\,\text{sr}^{-1}\,\text{yr}^{-1}$ of radio variability down to $\mu_{S}>20\,$mJy.
In Section \ref{ssec:wise} we used the IR colors to identify the types of object showing variability, with the majority appearing to be blazars.
While only $317$ ($18\,$\%) of the variables at $\mu_{S}>20\,$mJy have WISE photometry robust enough to use in IR color/color diagnostics, these few sources account for a disproportionately large ($28\,$\%) proportion of the change in flux density between the two epochs, with $\sum |\Delta S| = 13.0\,$Jy, $10.2\,$Jy ($78\,$\%) of which seemingly originates from blazars.

Blazars are relatively rare objects, with only on the order of a few thousand robustly identified to date \citep{Massaro2015}, and it is prudent to ask whether it makes sense that blazars would dominate our selection.
To address this, we consider our selection in the context of the expected radio source population.
Our input sample consists of unresolved sources in VLASS, a survey with a typical beam size of $<3\arcsec$.
We replicate such a selection of radio sources in the Tiered Extragalactic Radio Continuum Simulation \citep[T-RECS,][]{Bonaldi2019}, a data set that provides expected observational and physical properties, including angular size ($\psi$), flux density at $3\,$GHz ($S_{3\,\text{GHz}}$), and angle between the axis of the AGN jet and the line of sight to the source ($\theta$), for a simulated radio source population.
Requiring $\psi<3\arcsec$ (i.e., smaller than the typical VLASS beam) and $S_{3\,\text{GHz}}>20\,$mJy (consistent with our flux limit of $\mu_{S}>20\,$mJy), we find that $66\,$\% of the selected T-RECS sources have $\theta < 10^{\circ}$, as would be expected of blazars \citep{Urry1995}.
This is very similar to the fraction of flux-limited variables having blazar-like IR colors (see Table \ref{tab:irpops}), supporting our interpretation of blazars dominating the radio variable population, based on their WISE colors.
Moreover, blazars are expected to be increasingly common with increasing radio brightness.
In T-RECS, the fraction of sources with $\psi < 3\arcsec$ and $\theta < 10^{\circ}$ increases from $66\,$\% at $S_{3\,\text{GHz}}>20\,$mJy to $84\,$\% at $S_{3\,\text{GHz}}>300\,$mJy.
This likely contributes to the rising fraction of radio sources that are variable with increasing $\mu_{S}$ seen in Figure \ref{fig:varyfrac}.

For radio sources with blazar-like IR colors that have optical spectra, $\sim 1/3$ have spectroscopic properties that confirmed them as blazars \citep{deMenezes2019}.
In the majority of those cases where the optical spectrum can not confirm the blazar nature of the source, \citet{deMenezes2019} find that the broader characteristics are still consistent with the object being a quasar.
It is likely that some of these quasars may indeed be blazars that are not yet identified as such due to limited observational data preventing a confident classification.
Moreover, contaminating sources with blazar-like WISE colors tend to be edge-cases in the WISE $\gamma$-ray strip, either as a result of large uncertainties in their measurements or colors that locate the source close to the edge of the blazar region boundaries in the WISE color/color diagram \citep{Massaro2012}.
The majority of our sources with blazar-like WISE colors are well centralized within the blazar region, with their distribution peaking in the overlap between the BL-Lac and FSRQ bounding boxes (see Figure \ref{fig:wise-colors}).
Even if the majority of our sources with blazar-like IR colors turn out to be non-blazar quasars, random contamination would not explain the $\sim60\,$\% increase in blazar-like colors among variable radio sources over the parent population of compact radio sources.
Further support for blazars dominating our sample of radio variable sources can be drawn from their $\gamma$-ray properties.
When accounting for the redshift distributions of the sources, we find that similar fractions of our radio variable selection and the existing population of known blazars have counterparts in the Fermi 4FGL catalog. 

\begin{figure*}
    \centering
    \includegraphics[trim={0 0 0 0}, clip, width=1.8\columnwidth]{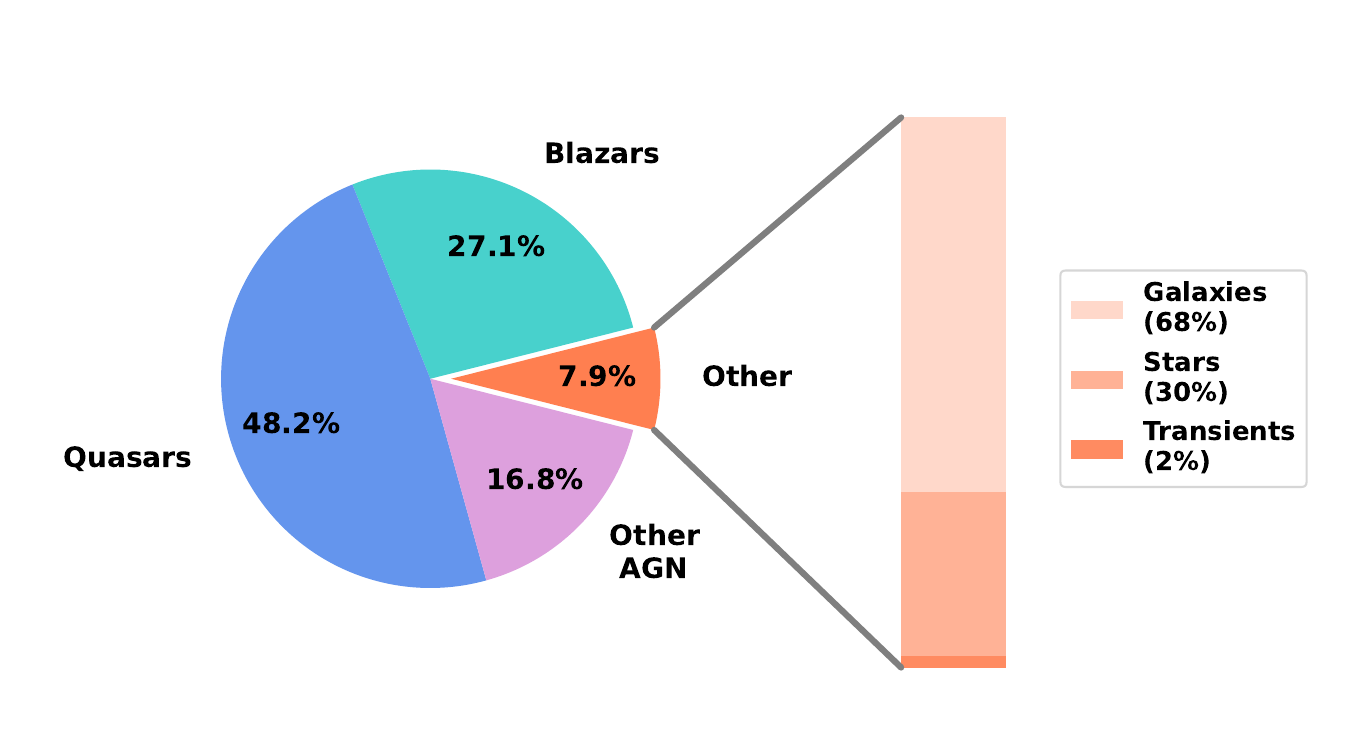}
    \caption{Breakdown of the SIMBAD object classifications for radio variables at $\mu_{S}>20\,$mJy with a SIMBAD association.
    }
    \label{fig:simbadtypes}
\end{figure*}

Although the majority of radio variables identified in this work do not have robust IR colors and/or $\gamma$-ray detections, should those that do be representative of the broader sample, then it suggests that blazars may dominate the variable radio sky.
To expand on this a little further, we cross-match our flux-limited variables to SIMBAD, finding matches to $1,135$ ($65\,\%$) sources.
Of these, $598$ have a classification consistent with a physical type of object (e.g., `galaxy' or `star' instead of just `radio source' or `X-ray source'), and we show the breakdown of these object types in Figure \ref{fig:simbadtypes}.
For clarity on Figure \ref{fig:simbadtypes}, we group subpopulations into a parent population, e.g., `BL Lacs' and `Blazars' are grouped together, `Seyferts' and `Radio galaxies' that are not quasars are grouped as `other AGN', etc.
Over $90\,$\% of the radio variables with a SIMBAD classification are AGN, mostly ($48\,$\%) quasars, followed by blazars ($27\,$\%) and other AGN ($17\,$\%).

Figure \ref{fig:simbadtypes} shows that $8\,$\% of our flux-limited variable sources in SIMBAD are not classified as AGN.
Approximately $2/3$ of these other sources are galaxies.
These are likely host to low-luminosity radio AGN, such as low-excitation radio galaxies where an AGN jet is produced by a radiatively inefficient accretion flow \citep[e.g., see][and references therein]{Tadhunter2016}.
Most ($14$) of the remaining sources are stellar, including three X-ray binaries and six variables stars.
While the SIMBAD data comes from a multitude of different sources, and as such the classifications may be approximate and/or incomplete, the dominance of AGN-like SIMBAD classifications\textemdash particularly quasars and blazars\textemdash among our radio variables supports the picture given by our analysis of the IR and $\gamma$-ray properties of our sample.

\subsubsection{Transient Sources}

In this work we have focused on characterizing radio variable sources, requiring a detection in both Epoch 1 and Epoch 2 of VLASS, and limiting most of our analysis to a more complete sample of variables with mean flux densities across the two epochs brighter than $20\,$mJy.
A natural consequence of this selection is that we are biased against finding transient events.
Transients are typically faint in VLASS ($\lesssim$ a few mJy), and are likely to only be detected in one epoch \citep{Dong2021, Cendes2024, Sharma2025}.
Thus, different methodologies to the one we employ are usually required in order to identify transients, e.g., looking specifically for appearing/disappearing sources or performing forced photometry on the VLASS images at the locations of known optical transients.
Indeed, by adopting such techniques VLASS has already found a swath of transients in recent years, including SNe, GRBs, and TDEs, \citep{Law2018, Dong2021, Somalwar2025a, Somalwar2025b}. 

\begin{figure}
    \centering
    \includegraphics[width=\columnwidth]{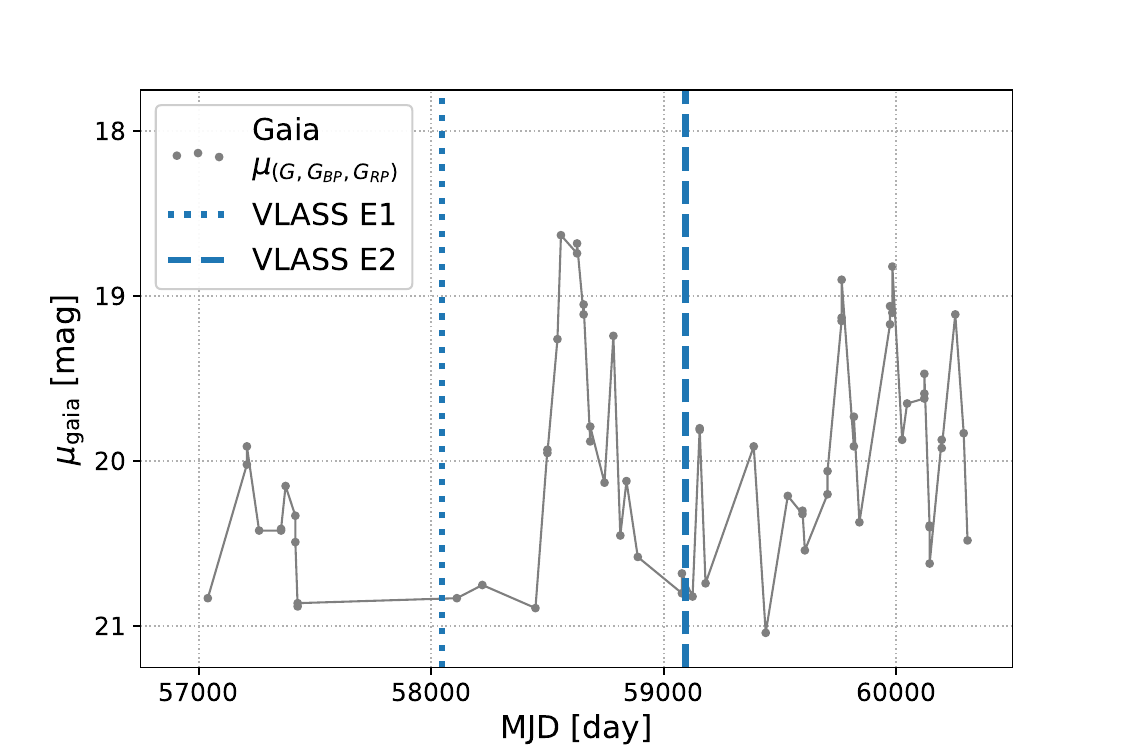}
    \caption{\textit{Gaia} light curve of AT 2019bvr, identified as a transient by SIMBAD.
    The gray points and line show the magnitude in \textit{Gaia} (averaged across the three passbands $G$, $G_{BP}$, and $G_{RP}$), the blue dotted line shows observation date of VLASS Epoch 1 ($S_{3\,\text{GHz}} = 26.5\pm0.2\,$mJy), and the blue dashed line shows the observation date of VLASS Epoch 2 ($S_{3\,\text{GHz}} = 71.1\pm0.4\,$mJy).
    The big peak in the middle of the light curve corresponds to the candidate transient event.}
    \label{fig:AT2019bvr-lc}
\end{figure}

Interestingly, one of our flux-limited radio variables is classified as a transient object in SIMBAD.
VLASS1QLCIR J105533.58$+$650955.9 is spatially consistent with AT 2019bvr, a candidate supernova that brightened by more than two magnitudes in March 2019 \citep{Prokhorov2021}.
The VLASS brightness in October 2017 was $26.5\pm0.2\,$mJy, rising to $71.1\pm0.4\,$mJy in September 2020.
One possibility is that the radio brightening between the two VLASS epochs is, at least in part, due to a supernova.
In this case the radio emission already present in the Epoch 1 observation ($17\,$months prior to AT 2019bvr) may be from the host galaxy.
Another possibility is that AT 2019bvr was not a supernova, and was actually an extreme optical outburst associated with an AGN. 
Not only would this second hypothesis explain the radio emission prior to March 2019, but would also explain the repeated optical variability visible in the optical light curve of the source obtained from \textit{Gaia} \citep{Gaia2016, Gaia2023} shown in Figure \ref{fig:AT2019bvr-lc}.
Furthermore, at the time of the Epoch 2 observation, the optical flare brightness had returned nearly to its baseline magnitude, and in blazars radio variability is known to lag behind variability at shorter wavelengths by hundreds of days \citep{Hufnagel1992, MaxMoerbeck2014}.
Should the radio variability in AT 2019bvr in fact be from a blazar or other AGN rather than a potential supernova, then there are no true transient events among our $598$ VLASS variables at $\mu_{S}>20\,$mJy with a physical classification in SIMBAD.
Taking there to be no transient sources in our sample, assuming a beta distribution as per \citet{Cameron2011} suggests that transients account for $<0.6\,$\% (upper limit at $95\,$\% confidence) of radio variable sources at this brightness.
At first glance such a low rate of radio transients might seem surprising.
However, it is important to consider that our selection criteria\textemdash namely requiring a detection in two VLASS epochs separated by $\sim2.5\,$ years\textemdash actively biases our sample against radio outbursts that can fade significantly over timescales as short as a few weeks \citep{Milisavljevic2013}.

\subsection{Case Studies of Interesting Examples of Radio Variability}

\subsubsection{The Most Extreme Variables}

We discuss here the most extreme cases of radio variability among our flux-limited variables.
In Table \ref{tab:extremes} we list the sources with the largest absolute and fractional increases and decreases in brightness, and discuss these below.
The largest increase in brightness is $410\pm2\,$mJy, exhibited by LB 2449 between September 2017 and June 2020.
Having flux densities of $102.2\pm0.6\,$mJy and $511.8\pm1.6\,$mJy in Epochs 1 and 2 respectively, this represents a five-fold increase in brightness over $33\,$months, and is also one of the most extreme cases of fractional variability in our sample.
LB 2449 is known to be a BL Lac \citep{Fleming1993}, a class of AGN often defined by their extreme variability that likely results from parsec-scale changes within the jet and Doppler boosting of the emission \citep[e.g][]{Aller1999, Richards2011}.

The largest decrease in brightness is seen in the Flat Spectrum Radio Quasar PKS 1502+106.
This object faded from $1,817\pm2\,$mJy in March 2019 to $971\pm1\,$mJy in November 2021, a change in flux density of $-846\pm2\,$mJy that represents a nearly $50\,$\% drop in brightness over 32 months.
Between 2014 and 2020, PKS 1502+106 experienced a substantial outburst of radio activity, becoming $\sim 3$ times brighter than its previously observed baseline observations \citep{Kouch2024}.
The VLASS Epoch 1 observation occurred during that outburst, while the Epoch 2 observation was made after the period of increased activity had ended.
\text{PKS 1502+106} is a particularly interesting source as its radio variability has been tentatively associated with high-energy neutrino detections \citep{Kun2021, Hovatta2021}.
With VLASS capable of detecting similar scales of fractional variability in much fainter sources, PKS 1502+106 highlights the potential for VLASS to identify additional candidate neutrino sources \citep[e.g., see][]{Filipovic2025, Gordon_icecube}.

\begin{deluxetable}{lccc}
    \tabletypesize{\scriptsize}
    \tablecaption{The most extreme cases of radio variability identified in this work
    \label{tab:extremes}}
    \tablehead{
    \colhead{Name} & \colhead{VLASS1QLCIR} & \colhead{$S_{\text{Epoch 1}}$} & \colhead{$S_{\text{Epoch 2}}$}\\
    \colhead{} & \colhead{} & \colhead{[mJy]} & \colhead{[mJy]}\\
    \colhead{(1)} & \colhead{(2)} & \colhead{(3)} & \colhead{(4)}}
    \startdata
        * $\sigma$ Gem & J074318.81$+$285256.2 & $55.7\pm0.2$ & $11.3\pm0.2$\\
        LB 2449 & J124818.79$+$582028.5 & $102.2\pm0.6$ & $511.8\pm1.6$\\
        PKS 1502$+$106 & J150424.99$+$102939.3 & $1816.9\pm2.0$ & $970.6\pm1.0$\\
        Cyg X-3 & J203225.77$+$405727.8 & $59.6\pm0.2$ & $322.6\pm0.5$
    \enddata
    \tablecomments{The common identifier of the extreme variable (1); the name of the object in the VLASS 1 catalog (2); the VLASS flux density in Epoch 1 (3); and the VLASS flux density in Epoch 2 (4).}
\end{deluxetable}

In terms of fractional change in flux, the most extreme case of a flaring source in our sample is Cygnus X-3.
A high-mass X-ray binary consisting of a Wolf-Rayet star orbiting a compact object thought to be a stellar-mass black hole, Cygnus X-3 is in effect a low mass analog of the quasars that dominate the radio sky \citep{Koljonen2017}.
This object is well known to be highly variable at radio wavelengths, typically having a brightness around $S\sim100\,$mJy, but prone to outbursts of activity as bright as a few tens of Jy \citep[e.g.,][]{Waltman1996, MillerJones2004, Egron2021}.
In VLASS, Cygnus X-3 is observed to be 5.4 times brighter in Epoch 2 than in Epoch 1, increasing in flux density from $59.6\pm0.2\,$mJy in October 2017 to $322.5\pm0.5\,$mJy in September 2020.
RACS also observed Cygnus X-3 in December 2020, measuring a flux density of $S=62.4\pm 4.1\,$mJy at $\nu \sim 1.4\,$GHz \citep{Duchesne2024}, suggesting that the outburst we see in VLASS Epoch 2 had faded within three months.

The object with the largest fractional decrease in flux in our sample is $*\ \sigma$ Gem, an RS CVn type variable, and a known radio star \citep{Yiu2024, Driessen2024}.
RS CVn objects are close binary systems with highly active magnetic fields \citep{Feldman1978, Mutel1985}.
The radio emission from these sources is thought to be driven by the acceleration of electrons within the stellar chromosphere(s) or the electrodynamic interactions of the two stars \citep{Toet2021}.
In VLASS, $*\ \sigma$ Gem is only $\sim20\,$\% as bright in Epoch 2 as in Epoch 1, fading from $55.7\pm0.2\,$mJy in April 2019 to $11.3\pm0.2\,$mJy by November 2021.
Of particular note, RACS observed $*\ \sigma$ Gem in April 2019, just one week after VLASS Epoch 1, measuring a flux density of $1.51\pm0.03\,$mJy at $\nu\sim888\,$MHz.
The extreme discrepancy in between the RACS and VLASS flux measurements suggests that the outburst in radio activity happened within the timescale of the week preceding the VLASS Epoch 1 observation.
Indeed, RS CVn variables are known to vary wildly on timescales of a few hours \citep{Feldman1978}. 
Notably, the most extreme cases of fractional radio variability discussed here are both objects within the Milky Way, highlighting the usefulness of radio time domain surveys in studying galactic radio sources.

\subsubsection{Interesting Objects Identified During Visual Inspection}

\begin{figure}
    \centering
    \subfigure[]{\includegraphics[trim={3.4cm 0 3.4cm 0}, clip, width=\columnwidth]{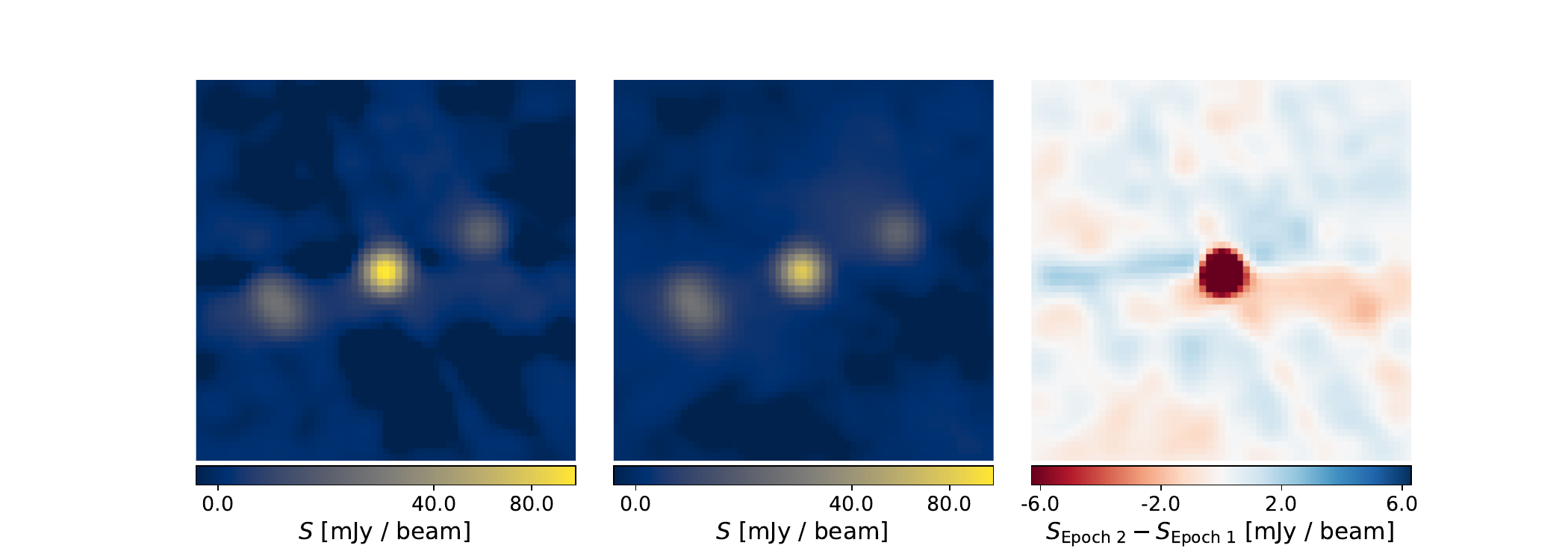}}
    \subfigure[]{\includegraphics[trim={0 0 0 0}, clip, width=\columnwidth]{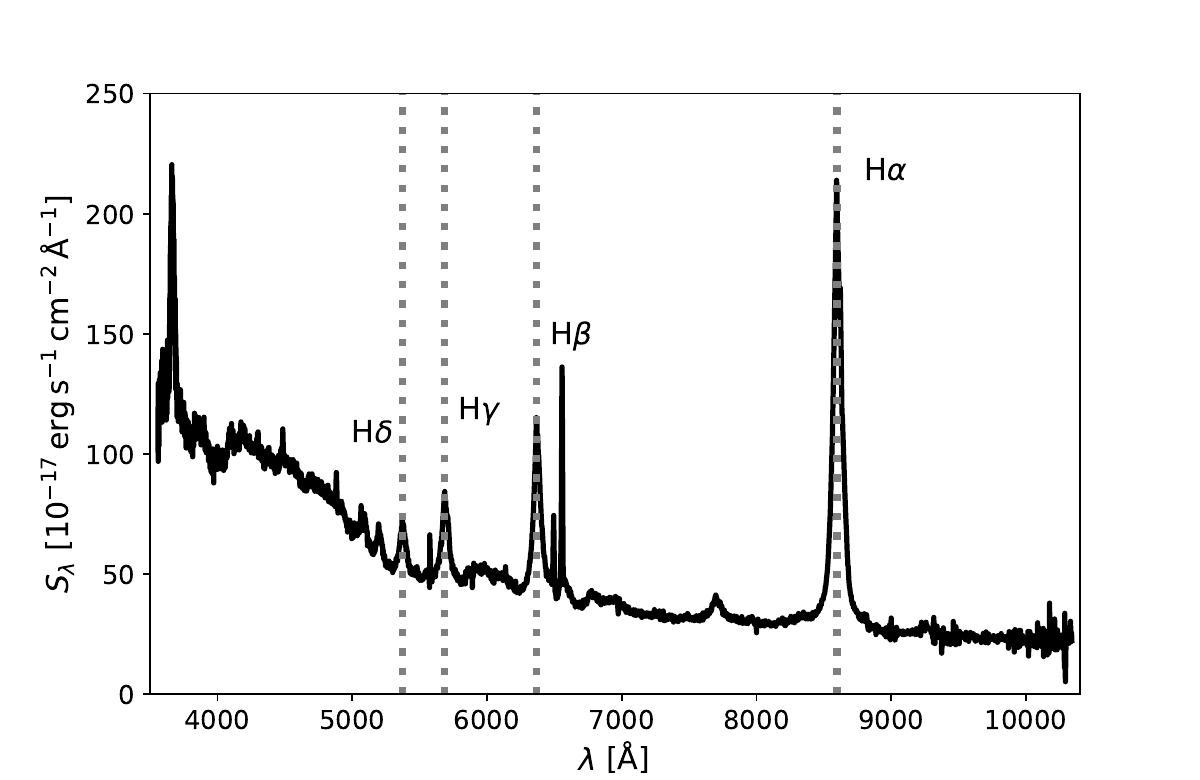}}
    \subfigure[]{\includegraphics[trim={0 0 0 0}, clip, width=\columnwidth]{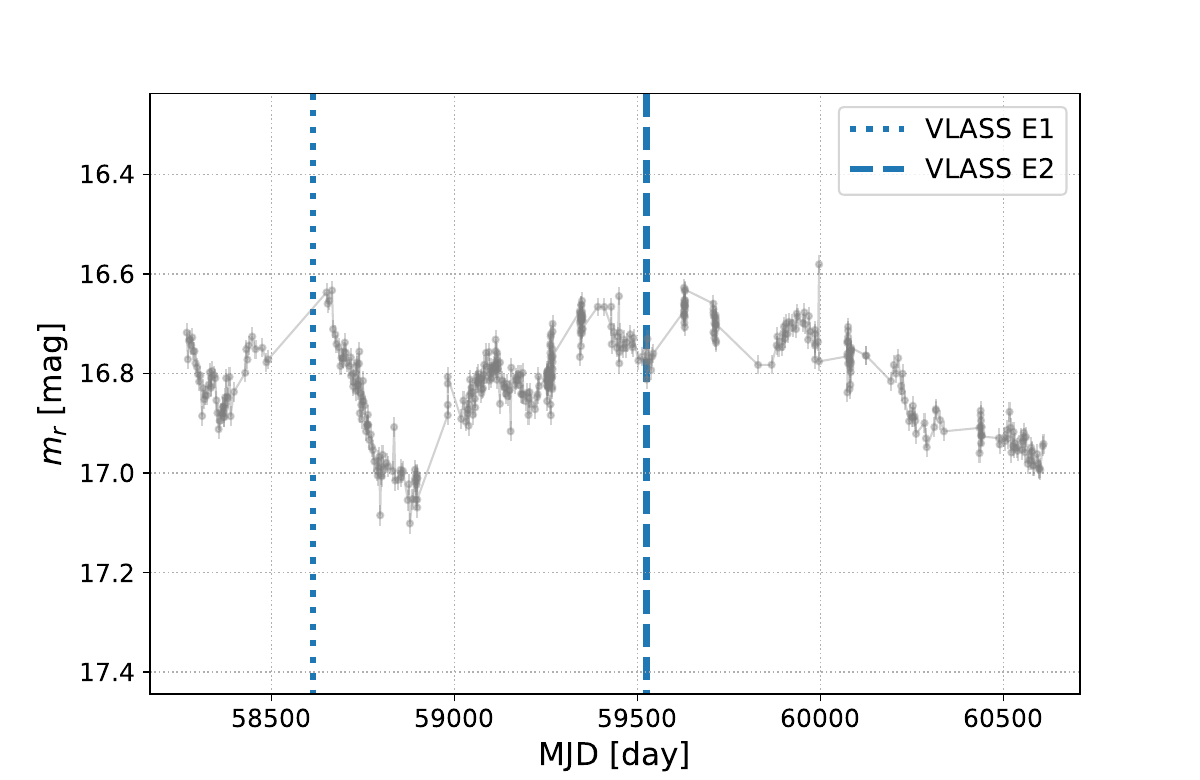}}
    \caption{An example of an extended radio galaxy with a fading core.
    Panel (a) shows the Epoch 1, Epoch 2 and Epoch 2 - Epoch 1 flux density maps from VLASS.
    Panel (b) shows the optical spectrum of the host galaxy from SDSS, revealing a broad line AGN.
    Panel (c) shows the $r$-band light curve of the host galaxy from ZTF, demonstrating the source is optically variable.}
    \label{fig:eg_fadingcore}
\end{figure}

While most radio variable sources are isolated point sources, some are embedded within the cores of larger radio galaxies.
Here we discuss a couple of such examples identified by the authors during the visual inspection step of our sample selection (Section \ref{sssec:qa}).
In Figure \ref{fig:eg_fadingcore}a, we show the VLASS images (Epoch 1, 2 and difference) for J000934.9+180343, a triple radio galaxy with a dominant, quasar-like core that has faded by $30\,$mJy, from $114.1\pm0.6\,$mJy in May 2019 to $84.1\pm0.4\,$mJy in November 2021.
An optical spectrum for the host galaxy was obtained by SDSS, which clearly shows a broad-line AGN at $z=0.31$ (Figure \ref{fig:eg_fadingcore}b).
Optical variability is apparent in the $r$-band light curve from the Zwicky Transient Facility \citep[ZTF,][]{Bellm2019} shown in Figure \ref{fig:eg_fadingcore}c, varying between $16.6\pm0.1\,\text{mag} < m_{r} < 17.1\pm0.1\,\text{mag}$.
In AllWISE, the galaxy has $\text{W1}=12.58\pm0.02\,$mag, $\text{W2}=11.62\pm0.02\,$mag, and $\text{W3}=9.16\pm0.03\,$mag.
The optical variability and IR colors are consistent with J000934.9+180343 being a blazar, and the presence of strong broad emission lines would make it an FSRQ.
There is no Fermi source associated with the galaxy.

\begin{figure}
    \centering
    \subfigure[]{\includegraphics[trim={3.4cm 0 3.4cm 0}, clip, width=\columnwidth]{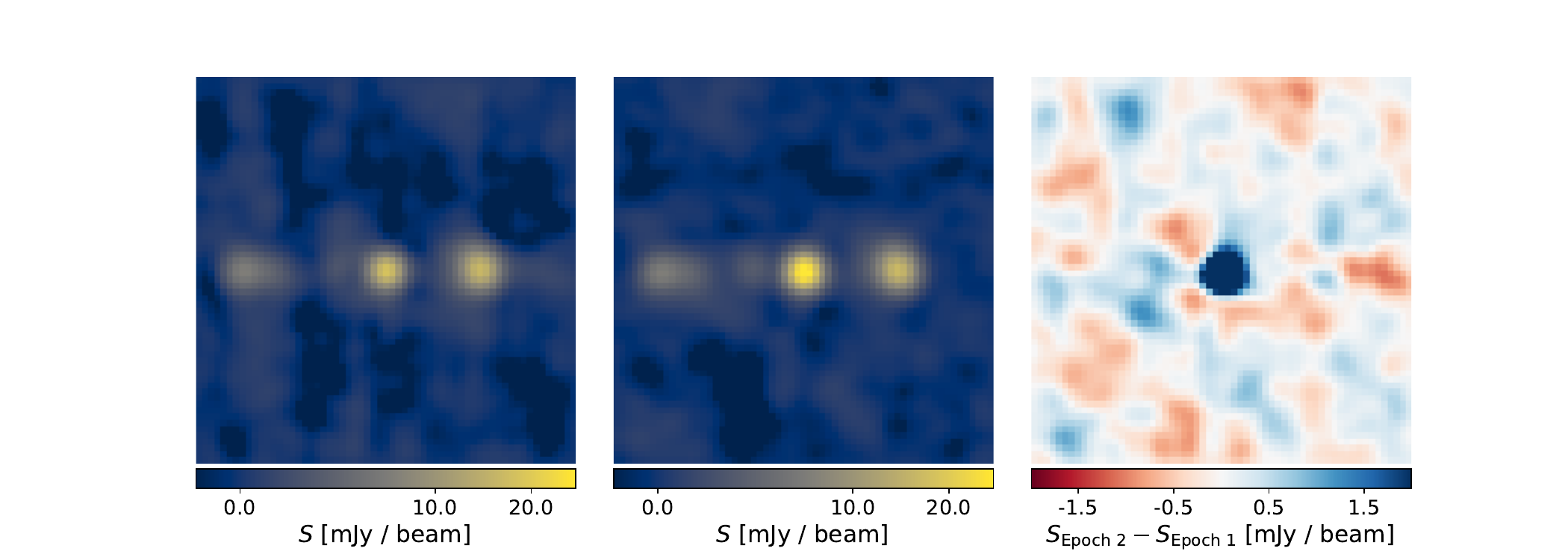}}
    \subfigure[]{\includegraphics[trim={0 0 0 0}, clip, width=\columnwidth]{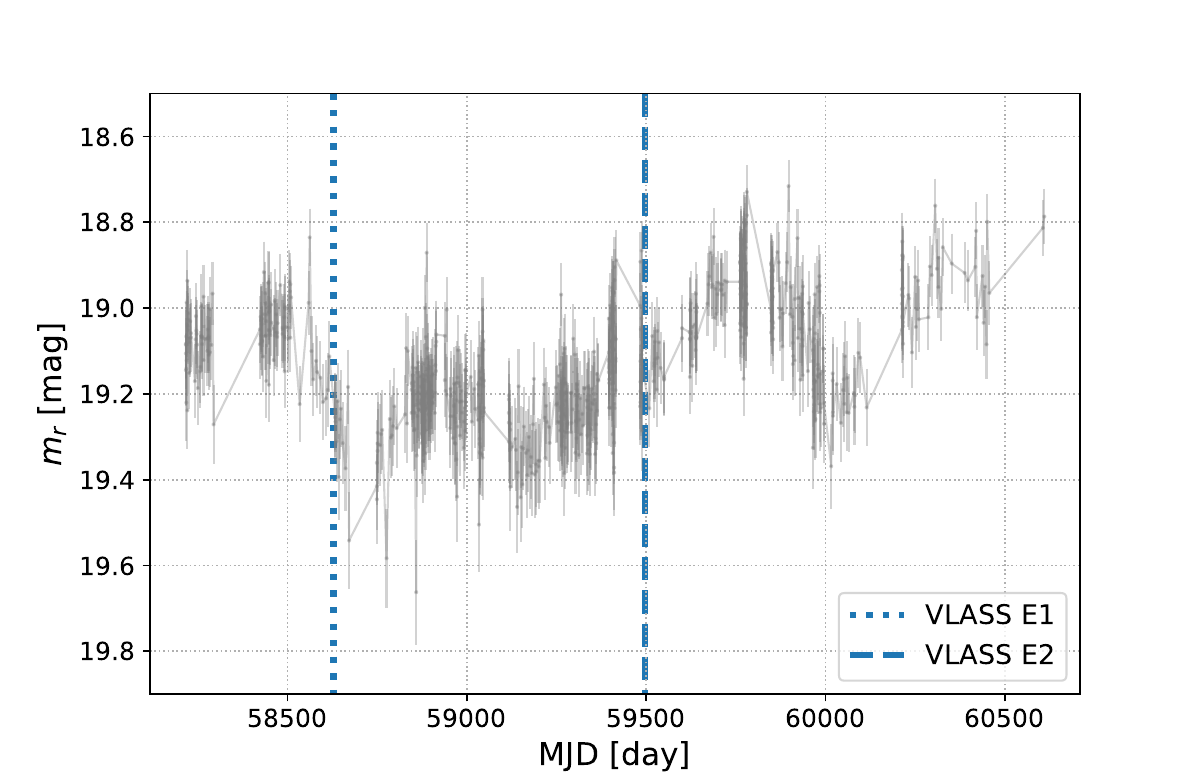}}
    \caption{An example of an extended radio galaxy with a flaring core.
    Panel (a) shows the Epoch 1, Epoch 2 and Epoch 2 - Epoch 1 flux density maps from VLASS.
    Panel (b) shows the $r$-band light curve of the host galaxy from ZTF}
    \label{fig:eg_flaringcore}
\end{figure}

In Figure \ref{fig:eg_flaringcore}a, we show the VLASS images for \text{7C 1025$+$4321}, a quasar at $z=0.72$ \citep{Wold2000} that brightened from $19.2\pm0.3\,$mJy in VLASS Epoch 1 to $28.4\pm0.2\,$mJy in Epoch 2.
In ZTF the host galaxy $r$-band magnitude varies between $18.7\pm0.1\,\text{mag} < m_{r} < 19.7\pm0.1\,\text{mag}$, and in AllWISE it is observed to have  $\text{W1}=15.52\pm0.04\,$mag, $\text{W2}=14.31\pm0.04\,$mag, and $\text{W3}=12.03\pm0.26\,$mag.
These IR magnitudes place the host galaxy in the AGN region of the WISE color-color diagram (see Figure \ref{fig:wise-colors}), although not in the blazar strip.
Like the previous example, 7C 1025$+$4321 does not have a counterpart in the 4FGL catalog.

\subsection{Data Limitations}

Identification of radio-variables in this work was made using catalogs based on the VLASS \textit{Quick Look} images, a data product that is known to suffer from quality limitations \citep[see][as well as Section \ref{ssec:quicklook} of this work]{Lacy2019}.
The reliability of the flux density measurements presents the biggest concern when using the \textit{Quick Look} images to identify variable radio sources.
\citet{Lacy2019} demonstrate that the peak flux density measurements of sources in the \textit{Quick Look} images have a scatter of $15\,$\%, particularly in Epoch 1.1, with later epochs being less severely affected.
This is consistent with the $15\,$\% and $12\,$\% scatter we see in ratio flux densities measured respectively in Epoch 2.1 / Epoch 1.1 and Epoch 2.2 / Epoch 1.2 in Section \ref{ssec:qlsystematics}.
Therefore, the minimum $30\,$\% change in flux density ($|m| > 0.26$) required by variable source selection should only be identifying real variables, and not sources that simply appear to be variable as a result of poor flux density measurements from the \textit{Quick Look} images.

As an additional test to confirm that $|m|>0.26$ is indeed a suitable choice to identify variables in this work, we repeat all of the cross match statistics presented in Section \ref{sec:multiwavelength} using only variable sources with at least a $50\,$\% change in flux density between Epoch 1 and Epoch 2 ($|m|>0.4$), accounting for approximately half of variables.
The results are qualitatively similar to those presented in Section \ref{sec:multiwavelength}: about $2/3$ of sources with an IR match are likely blazars, and at $z<1$ $\gamma$-ray counterparts are found for $\sim1$ in $5$ radio variables.
Similarly, variables with $|m|>0.4$ have a similar breakdown of source types in SIMBAD as those with $|m|>0.26$, with fewer than $1$ in $10$ not being associated with AGN, and the AGN being dominated by quasars and blazars.
These results imply that if there is any contamination of our sample by spurious variables as a result of less confident flux density measurements obtained from the \textit{Quick Look} images, then that contamination is likely minimal and has no impact on our conclusions.

A final comment on the flux density measurements in the \textit{Quick Look} images is that \citet{Lacy2019} demonstrate that sources brighter than $1\,$Jy in Epoch 1.1 are subject to very unreliable flux density measurements.
Their Figure 3 shows the \textit{Quick Look} measurements being as low as $\sim 75\,$\% of their true value.
Only three of the variables in our sample are brighter than $1\,$Jy in Epoch 1, all of which were observed in Epoch 1.2 where this effect does not exist.
The brightest Epoch 1.1 measurement in our sample of variables is $664\pm1\,$mJy belonging to ICRF J085348.1$+$065447, which fades to $342\pm1\,$mJy.
This object is a previously known variable radio source, exhibiting strong variability between previous radio surveys, e.g., having $S_{1.4\,\text{GHz}}=790\pm28$mJy in NVSS and $S_{1.4\,\text{GHz}}=135.70\pm0.14$mJy in FIRST.
Moreover, if this source were an underestimate of a source that was actually brighter than $1\,$Jy at the time of the VLASS Epoch 1 observation resulting from the \textit{Quick Look} image quality issues, we would expect it to have $S_{\text{VLASS 1}}\gtrsim 750\,$mJy.
Given that all the other Epoch 1.1 measurements are fainter than for this object, we can be confident that the issue of unreliable fluxes above $1\,$Jy in Epoch 1.1 is not impacting our selection of radio variables.

\subsection{Future Prospects}

The major population traced by the radio variables sampled in this work appears to be blazars.
Many of these objects were not previously known as blazars, being fainter and at higher redshift than blazars in the \textit{Roma BZCat}.
Deep radio time-domain surveys like VLASS thus have the potential to identify new blazars for in-depth study and monitoring that can provide key insights into how AGN jets behave at lower radio luminosities, and/or earlier in the history of the Universe.
This will be made easier still by the forthcoming deep optical time-domain surveys.
For instance, the Vera C. Rubin Observatory Legacy Survey of Space and Time \citep[LSST,][]{Ivezic2019} is conducting a wide-fast-deep time-domain survey of the southern sky, including $\sim300$ visits in each of the $ugrizy$ bands over a ten year period. 
LSST will have $14,000\,\text{deg}^{2}$ overlap with VLASS, with $\sim10^{6}$ objects expected to be detected in both surveys \citep{Tellez2024}. 
Moreover, VLASS Epoch 4 \citep{Nyland2023} has temporal overlap with with the first year of observations from the Vera C. Rubin Observatory, allowing the combination of VLASS and LSST to be a powerful tool in studying the physics of AGN.

We have used the first two epochs of VLASS in this work, and as such are limited to two-point variability statistics in our source selection.
VLASS Epoch 3 was completed while this work was in-prep, and the majority of the \textit{Quick Look} images for the third epoch are now available.
The availability of an Epoch 3 source catalog in the near future will allow the work presented in this paper to be built upon by assessing the three-point variability statistics of sources in VLASS, an approach that will have two clear benefits.
First, the two-point variability is likely to miss many stochastically variable radio sources such as blazars that may not have had an outburst of activity timed such that it would be detected by comparing the Epoch 1 and Epoch 2 VLASS measurements.
Having information from a third epoch of observations will double the opportunities of identifying such random variability.
Second, a third epoch of flux density measurements provides the opportunity to identify sources that are consistently fading or flaring over a $5$-$6\,$year time period.
These objects may include phenomena such as newly switched on jets from supermassive black holes that can be studied to provide key insights into the AGN duty cycle \citep{Nyland2020}.

A secondary, but important, consideration from identifying radio variable sources is that it also identifies faint point sources that don't show variability between the observations used.
Such objects have the potential to act as phase calibrator sources for the next generation of radio telescopes such as the next-generation VLA, and the Square Kilometer Array, where calibrators of a few mJy in brightness will be required \citep{Wrobel2025}. 
Data from additional VLASS epochs can be used to select `quasi-stable' radio sources that consistently show no variability.
The most promising sources can then be targeted for long term monitoring in order to identify suitable faint phase calibrators.

\section{Summary and Conclusions} 
\label{sec:summary}

In this paper we have used data from the first two epochs of VLASS to identify and characterize the population of radio sources that vary on timescales of \text{$\sim2.5\,$years}.
We only consider sources that are detected in both Epoch 1 and Epoch 2 of VLASS here, with completely fading and rising sources that are only detected in one epoch being the purview of a future work.
The source measurements used in this work were based on the VLASS \textit{Quick Look} images, a data product known to suffer from quality limitations that are not homogeneous across the survey footprint.
In order to account for such issues, we analyze the usability of the \textit{Quick Look} data for identifying variability.
Using this information, we select $3,618$ variable radio sources (listed in Table \ref{tab:variables}), including a complete sample of $1,744$ variable source that have a mean flux density over the two epochs of $\mu_{S}>20\,$mJy.
The key characteristics of the $\mu_{S}>20\,$mJy radio variable population are as follows.
\begin{itemize}
    \item $\sim 5\,$\% of point sources with $\mu_{S}>20\,$mJy vary in their flux density at $3\,$GHz by $\geq30\,$\% over $\sim2.5\,$years, rising to $9\,$\% at $\mu_{S}>300\,$mJy.
    \item While there are approximately the same number of flaring and fading radio sources between Epoch 1 and Epoch 2 of VLASS, the absolute change in flaring sources is typically slightly larger than in fading sources, with median $|\Delta S|$ values of $15.6\,$mJy and $13.4\,$mJy respectively.
    These differences, while small, are found to be statistically significant when the $|\Delta S|$ distributions for the flaring and fading populations are compared, with a KS test returning a $p$-value of $4\times10^{-4}$.
    Our results here may be indicative of the average radio light curve of our sample being skewed towards having a steeper brightening than fading.
    \item Blazars and quasars appear to be the dominant parent population of radio sources that vary on timescales of $\sim2.5\,$years at $3\,$GHz, based on their multiwavelength properties.
    Many of the objects classified as quasars may in fact be previously unidentified blazars.
    Supporting this conclusion, both our sample and the \textit{Roma BZcat} have a similar fraction ($\sim 1/5$) with a $\gamma$-ray counterparts when accounting for the redshift and WISE color distributions.
    \item In addition to extragalactic phenomena, we identify radio variables with a galactic origin, including the sources with the strongest fractional variability in our sample, highlighting the usefulness of radio variability surveys for identifying and studying radio stars and microquasars.
\end{itemize}

\section*{Acknowledgments}

The authors thank the anonymous referee for their constructive comments.
We also thank Keith Bechtol for helpful discussions.

In carrying out this work we made use of the following software packages and tools:
AstroPy \citep{Astropy2013, Astropy2018, Astropy2022},
Matplotlib \citep{Hunter2007},
NumPy \citep{Harris2020},
SAOImage DS9 \citep{Joye2003},
SciPy \citep{Virtanen2020},
and TOPCAT \citep{Taylor2005}.

This publication uses data generated via the \href{https://www.zooniverse.org/}{Zooniverse.org} platform, development of which is funded by generous support, including from the National Science Foundation, NASA, the Institute of Museum and Library Services, UKRI, a Global Impact Award from Google, and the Alfred P. Sloan Foundation.

This work made use of the cross-match service \citep{Boch2012, Pineau2020} and the VizieR catalogue access tool \citep{Ochsenbein2000} provided by CDS, Strasbourg, France.
Additionally used the facilities of the Canadian Astronomy Data Centre (CADC), an organisation operated by the National Research Council of Canada with the support of the Canadian Space Agency.

We used data produced by The Canadian Initiative for Radio Astronomy Data
Analysis (CIRADA) in the preparation of this work.
CIRADA is funded by a grant from the Canada Foundation for Innovation 2017 Innovation Fund (Project 35999) and by the Provinces of Ontario, British Columbia, Alberta, Manitoba, and Quebec, in collaboration with the National Research Council of Canada, the US National Radio Astronomy Observatory and Australia's Commonwealth Scientific and Industrial Research Organisation.

The VLA is operated by NRAO, a facility of the National Science Foundation operated under cooperative agreement by Associated Universities, Inc.

WISE is a joint project of the University of California, Los Angeles, and the Jet Propulsion Laboratory/California Institute of Technology, and NEOWISE, which is a project of the Jet Propulsion Laboratory/California Institute of Technology. WISE and NEOWISE are funded by the National Aeronautics and Space Administration.

Funding for the SDSS IV has been provided by the Alfred P. Sloan Foundation, the U.S. Department of Energy Office of Science, and the Participating Institutions.
SDSS acknowledges support and resources from the Center for High-Performance Computing at the University of Utah. The SDSS web site is \url{www.sdss.org}.
SDSS is managed by the Astrophysical Research Consortium for the Participating Institutions of the SDSS Collaboration including the Brazilian Participation Group, the Carnegie Institution for Science, Carnegie Mellon University, Center for Astrophysics | Harvard \& Smithsonian (CfA), the Chilean Participation Group, the French Participation Group, Instituto de Astrofísica de Canarias, The Johns Hopkins University, Kavli Institute for the Physics and Mathematics of the Universe (IPMU) / University of Tokyo, the Korean Participation Group, Lawrence Berkeley National Laboratory, Leibniz Institut für Astrophysik Potsdam (AIP), Max-Planck-Institut für Astronomie (MPIA Heidelberg), Max-Planck-Institut für Astrophysik (MPA Garching), Max-Planck-Institut für Extraterrestrische Physik (MPE), National Astronomical Observatories of China, New Mexico State University, New York University, University of Notre Dame, Observatório Nacional / MCTI, The Ohio State University, Pennsylvania State University, Shanghai Astronomical Observatory, United Kingdom Participation Group, Universidad Nacional Autónoma de México, University of Arizona, University of Colorado Boulder, University of Oxford, University of Portsmouth, University of Utah, University of Virginia, University of Washington, University of Wisconsin, Vanderbilt University, and Yale University.

The Legacy Surveys consist of three individual and complementary projects: the Dark Energy Camera Legacy Survey (DECaLS; Proposal ID \#2014B-0404; PIs: David Schlegel and Arjun Dey), the Beijing-Arizona Sky Survey (BASS; NOAO Prop. ID \#2015A-0801; PIs: Zhou Xu and Xiaohui Fan), and the Mayall z-band Legacy Survey (MzLS; Prop. ID \#2016A-0453; PI: Arjun Dey). 
DECaLS, BASS and MzLS together include data obtained, respectively, at the Blanco telescope, Cerro Tololo Inter-American Observatory, NSF’s NOIRLab; the Bok telescope, Steward Observatory, University of Arizona; and the Mayall telescope, Kitt Peak National Observatory, NOIRLab. 
Pipeline processing and analyses of the data were supported by NOIRLab and the Lawrence Berkeley National Laboratory (LBNL). 
The Legacy Surveys project is honored to be permitted to conduct astronomical research on Iolkam Du’ag (Kitt Peak), a mountain with particular significance to the Tohono O’odham Nation.

NOIRLab is operated by the Association of Universities for Research in Astronomy (AURA) under a cooperative agreement with the National Science Foundation. 
LBNL is managed by the Regents of the University of California under contract to the U.S. Department of Energy.

This project used data obtained with the Dark Energy Camera (DECam), which was constructed by the Dark Energy Survey (DES) collaboration. 
Funding for the DES Projects has been provided by the U.S. Department of Energy, the U.S. National Science Foundation, the Ministry of Science and Education of Spain, the Science and Technology Facilities Council of the United Kingdom, the Higher Education Funding Council for England, the National Center for Supercomputing Applications at the University of Illinois at Urbana-Champaign, the Kavli Institute of Cosmological Physics at the University of Chicago, Center for Cosmology and Astro-Particle Physics at the Ohio State University, the Mitchell Institute for Fundamental Physics and Astronomy at Texas A\&M University, Financiadora de Estudos e Projetos, Fundacao Carlos Chagas Filho de Amparo, Financiadora de Estudos e Projetos, Fundacao Carlos Chagas Filho de Amparo a Pesquisa do Estado do Rio de Janeiro, Conselho Nacional de Desenvolvimento Cientifico e Tecnologico and the Ministerio da Ciencia, Tecnologia e Inovacao, the Deutsche Forschungsgemeinschaft and the Collaborating Institutions in the Dark Energy Survey. 
The Collaborating Institutions are Argonne National Laboratory, the University of California at Santa Cruz, the University of Cambridge, Centro de Investigaciones Energeticas, Medioambientales y Tecnologicas-Madrid, the University of Chicago, University College London, the DES-Brazil Consortium, the University of Edinburgh, the Eidgenossische Technische Hochschule (ETH) Zurich, Fermi National Accelerator Laboratory, the University of Illinois at Urbana-Champaign, the Institut de Ciencies de l’Espai (IEEC/CSIC), the Institut de Fisica d’Altes Energies, Lawrence Berkeley National Laboratory, the Ludwig Maximilians Universitat Munchen and the associated Excellence Cluster Universe, the University of Michigan, NSF’s NOIRLab, the University of Nottingham, the Ohio State University, the University of Pennsylvania, the University of Portsmouth, SLAC National Accelerator Laboratory, Stanford University, the University of Sussex, and Texas A\&M University.

BASS is a key project of the Telescope Access Program (TAP), which has been funded by the National Astronomical Observatories of China, the Chinese Academy of Sciences (the Strategic Priority Research Program “The Emergence of Cosmological Structures” Grant \# XDB09000000), and the Special Fund for Astronomy from the Ministry of Finance. 
The BASS is also supported by the External Cooperation Program of Chinese Academy of Sciences (Grant \# 114A11KYSB20160057), and Chinese National Natural Science Foundation (Grant \# 12120101003, \# 11433005).

The Legacy Survey team makes use of data products from the Near-Earth Object Wide-field Infrared Survey Explorer (NEOWISE), which is a project of the Jet Propulsion Laboratory/California Institute of Technology. 
NEOWISE is funded by the National Aeronautics and Space Administration.

The Legacy Surveys imaging of the DESI footprint is supported by the Director, Office of Science, Office of High Energy Physics of the U.S. Department of Energy under Contract No. DE-AC02-05CH1123, by the National Energy Research Scientific Computing Center, a DOE Office of Science User Facility under the same contract; and by the U.S. 
National Science Foundation, Division of Astronomical Sciences under Contract No. AST-0950945 to NOAO.

This work has made use of data from the European Space Agency (ESA) mission
{\it Gaia} (\url{https://www.cosmos.esa.int/gaia}), processed by the {\it Gaia}
Data Processing and Analysis Consortium (DPAC,
\url{https://www.cosmos.esa.int/web/gaia/dpac/consortium}). Funding for the DPAC
has been provided by national institutions, in particular the institutions
participating in the {\it Gaia} Multilateral Agreement.


\bibliographystyle{aasjournal.bst}

\bibliography{vlass_variables}

\begin{appendix}

\section{Supplementary Data}
\label{ap:supdata}

As part of this work we have created a catalog of variable radio sources (Table \ref{tab:variables}), produced png images comparing the brightness of likely variable sources in the first two VLASS epochs (e.g., see Figure \ref{fig:image-qa}), and identified multiwavelength counterparts for VLASS point sources (variables and non-variables). 
These data products have been collated in a Zenodo repository where they may be readily accessed by the community: \url{https://zenodo.org/records/18010746}. 
The contents of the repository are:
\begin{itemize}
    \item \texttt{Table1\_variables.fits}, a machine readable version of Table \ref{tab:variables} cataloging the variable radio sources identified in Section \ref{sec:finding-variables}.
    \item \texttt{Table5\_xIDs.fits}, a machine readable table providing the multiwavelength information found in \ref{sec:multiwavelength} for VLASS point sources with $\mu_{S}>20\,$mJy.
    Contains:
    \begin{itemize}
        \item the AllWISE ID and magnitudes in the W1-, W2-, and W3-bands for sources with $S/N > 2$ and no photometric quality flags in all three of these bands;
        \item the 4FGL ID, $\gamma$-ray flux, and $\gamma$-ray variability index for Fermi sources with \texttt{Flags}==0;
        \item the optical ID and redshift of sources whose redshift uncertainty satisfies $\sigma_{z}/(1+z) < 0.2$.
    \end{itemize}
    An example of the table format is provided in Table \ref{tab:xid}.
    \item \texttt{variable\_qa\_images.tar.gz}, a tarball of $4,124$ png three-panel images.
    Each file\textemdash one for each of the candidate variables visually inspected in Section \ref{sssec:qa}\textemdash has the filename format: \texttt{Jhhmmss.ss$\pm$ddmmss.s\_qa.png}, where ``\emph{Jhhmmss.ss$\pm$ddmmss.s}'' corresponds to the \texttt{VLASS1QLCIR} name given in Table \ref{tab:variables}.
\end{itemize}
Machine readable versions of Table \ref{tab:variables} and Table \ref{tab:xid} will also be made available on CDS after final publication of this paper.

\renewcommand{\arraystretch}{1.5}
\begin{sidewaystable}
\scriptsize
\caption{Multiwavelength cross identifications for VLASS point sources with $\mu_{S}>20\,$mJy \label{tab:xid}}
\hspace{-3em}
\begin{tabular}{cccccccccc}
    \hline
    \vspace{0.1em}\\
    VLASS1QLCIR$^{a}$ & AllWISE$^{b}$ & W1mag & W2mag & W3mag & 4FGL$^{c}$ & Flux1000 & VariabilityIndex & zID$^{d}$ & z \\ 
     &  & [mag] & [mag] & [mag] &  & [$10^{10}\,\text{cm}^{-2}\,\text{s}^{-1}$] &  &  &  \\
    (1) & (2) & (3) & (4) & (5) & (6) & (7) & (8) & (9) & (10) \\
    \vspace{0.05em}\\
    \hline
    \vspace{0.05em}\\
        J000113.38$+$581945.7 & -- & -- & -- & -- & -- & -- & -- & -- & -- \\
        J000113.77$-$082034.5 & J000113.84$-$082034.1 & $16.05\pm0.06$ & $15.04\pm0.11$ & $12.16\pm0.47$ & -- & -- & -- & -- & --\\
        J000114.84$+$235810.8 & J000114.86$+$235810.7 & $13.07\pm0.02$ & $12.71\pm0.03$ & $10.63\pm0.09$ & --  & -- & -- & 8000464815001453 & $0.112\pm0.045$ \\
        J000115.57$+$114556.3 & -- & -- & -- &  & -- & -- & -- &  8000397645002557 & $1.204\pm0.216$ \\
        J000117.48$+$095452.6 & J000117.47$+$095452.2 & $16.49\pm0.08$ & $15.87\pm0.15$ & $12.17\pm0.35$ & -- & -- & -- & SDSS J000117.48$+$095452.5 & $2.54511\pm0.00064$ \\
        J000118.02$-$074626.8 & J000118.01$-$074626.9 & $12.70\pm0.02$ & $11.75\pm0.02$ & $9.16\pm0.04$ & J0001.2$-$0747 & $7.02\pm0.47$ & 51.74 & -- & -- \\
        J000119.04$+$474200.7 & -- & -- & -- & -- & J0001.2$+$4741 & $1.22\pm0.27$ & 28.92 & -- & -- \\
        J000121.45$-$001140.0 & J000121.50$-$001140.0 & $14.98\pm0.04$ & $14.46\pm0.06$ & $12.10\pm0.37$ & J0001.4$-$0010 & $1.18\pm0.27$ & 9.37 & -- & -- \\
        J000121.66$+$252655.5 & J000121.67$+$252655.3 & $15.34\pm0.04$ & $14.35\pm0.05$ & $12.28\pm0.49$ & -- & -- & -- & 8000472683001511 & $1.052\pm0.320$ \\
        J000121.87$-$170324.6 & J000121.84$-$170324.7 & $14.70\pm0.03$ & $13.99\pm0.04$ & $11.01\pm0.13$ & -- & -- & -- & -- & -- \\
        ... & ... & ... & ... & ... & ... & ... & ... & ... & ... \\
        ... & ... & ... & ... & ... & ... & ... & ... & ... & ... \\
    \vspace{0.05em}\\
    \hline
    \vspace{0.05em}\\
    \multicolumn{10}{l}{\hspace{2em}\textsc{Note--}VLASS source name in Epoch 1 catalog (1); source name in AllWISE (2); brightness in WISE W1 band (3); brightness in WISE W2  band (4); brightness in WISE W3 band (5);}\\
    \multicolumn{10}{l}{\hspace{2em} source name in the Fermi 4FGL catalog (6); integrated flux density for energies in the range $1\,\text{GeV}\leq E \leq 100\,\text{GeV}$ (7); $\gamma$-ray variability index (8); ID of the optical source a redshift}\\
    \multicolumn{10}{l}{\hspace{2em} was obtained from (9); redshift of the source (10). Ten example rows of the table are shown here, with the full table consisting of $35,781$ rows available in machine readable format at}\\
    \multicolumn{10}{l}{\hspace{2em} \url{https://zenodo.org/records/18010746} and will be available on CDS after final publication of the paper.}\\
    \vspace{0.1em}\\
    \multicolumn{10}{l}{$^{b}$ Corresponds to the \texttt{VLASS1QLCIR} column in Table \ref{tab:variables} and the \texttt{Component\_name} column in version 3 of the CIRADA VLASS Epoch 1 \emph{Quick Look} catalog \citep{Gordon2021}.}\\
    \multicolumn{10}{l}{$^{b}$ Corresponds to the \texttt{AllWISE} column in the CDS catalog II/328 \citep{Cutri2014}.}\\
    \multicolumn{10}{l}{$^{c}$ Corresponds to the \texttt{Source\_Name} column in the 4FGL DR4 catalog \citep{Ballet2023}.}\\
    \multicolumn{10}{l}{$^{d}$ Corresponds to \texttt{SDSS16} in the CDS catalog V/154 \citep{Ahumada2020} if obtained from SDSS, or \texttt{ID} in the CDS catalog VII/292 if obtained from LS DR8 \citep{Duncan2022}.}
\end{tabular}
\end{sidewaystable}

\end{appendix}

\end{document}